\documentclass[prb,twocolumn,showpacs]{revtex4}

\usepackage{graphicx} 
\usepackage{dcolumn} 
\usepackage{bm}

\begin{document}

\title{Quantum Hall Phase Diagram  of Half-filled
Bilayers in the Lowest and the Second Orbital Landau Levels: Abelian versus Non-Abelian Incompressible Fractional Quantum Hall States}

\author{Michael R. Peterson and S. Das Sarma} 
\affiliation{Condensed Matter Theory Center, Department of Physics, University of Maryland,
College Park, MD 20742}

\date{\today}

\begin{abstract}
We examine the quantum phase diagram of the fractional quantum Hall
effect (FQHE) in the lowest two Landau levels in half-filled bilayer structures as
a function of tunneling strength and layer separation, i.e., we revisit the lowest Landau level filling 
factor 1/2 bilayer problem and make predictions involving bilayers in the 
half-filled second Landau level (i.e., filling factor 5/2).  Using
numerical exact diagonalization we investigate the important question
of whether this system supports a FQHE
described by the non-Abelian Moore-Read Pfaffian state in the strong
tunneling regime.  In the lowest Landau level, we find that although in principle, increasing
(decreasing) tunneling strength (layer separation) could lead to a
transition from the Abelian two-component Halperin 331 to non-Abelian one-component 
Moore-Read Pfaffian state, the
FQHE excitation gap is relatively small in the lowest Landau level Pfaffian regime, and we establish 
that all so far observed FQHE states in half-filled
lowest Landau level bilayers are most likely described by the Abelian
Halperin 331 state.  In the second Landau level we make the prediction that bilayer structures 
would manifest two distinct
branches of incompressible FQHE
corresponding to the Abelian 331 state (at moderate to low tunneling
and large layer separation) and the non-Abelian Moore-Read Pfaffian state (at
large tunneling and small layer separation).  The observation of these
two FQHE branches and the possible quantum phase transition between
them will be compelling evidence supporting the existence of the
non-Abelian Moore-Read Pfaffian state in the second Landau level.  We discuss our results in the context of existing experiments and theoretical works.
\end{abstract}


\maketitle

\section{Introduction}
\label{sec-intro}

Two important developments have rekindled interest in the phenomena of
even-denominator incompressible fractional quantum Hall  states
in two-dimensional high-mobility bilayer semiconductor structures.  The first is the recent
intriguing experimental observation by Luhman \textit{et
al}.~\cite{luhman-2008} of two distinct even-denominator 
fractional quantum Hall effect (FQHE) states at filling factors
$\nu=1/2$ and 1/4 in a very wide ($\sim 600 \AA$) single
quantum well (wide-quantum-well) structure at very high ($>40 T$) magnetic fields.
The second development, motivated by implications for fault-tolerant
topological quantum computation~\cite{sds-tqc,tqc-rmp}, is the great deal of
recent theoretical and experimental interest in the possible
non-Abelian nature of the  $\nu=5/2$ second Landau level (SLL) FQHE
state first observed by Willett \emph{et al.}~\cite{willett} in the single-layer 
system in 1987 with subsequent confirming 
observations~\cite{52-tilt,gammel,pan,eisenstein-cooper,xia,csathy,choi} 
over the years since.    The FQHE  at filling factor
$\nu=5/2$ presents an amazing confluence of ideas from
condensed matter physics, conformal field theory, topology, and
quantum computation.~\cite{tqc-rmp}  In particular, the fundamental nature of the
experimentally observed 5/2 FQHE, whether an exotic spin-polarized
non-Abelian incompressible paired state or a more common Abelian
incompressible paired state, has remained an enigma for more than 20
years.  Although theoretical and numerical work indicates that the 5/2
FQHE belongs to a non-Abelian Moore-Read Pfaffian~\cite{mr-pf} universality
class, there is scant experimental evidence supporting
this conclusion~\cite{dean} (see Ref.~\onlinecite{tqc-rmp} for a comprehensive 
review of non-Abelian physics and topological quantum computation).

These two developments lead to important and 
interesting questions.  One question is whether a non-Abelian
$\nu=1/2$ FQHE, i.e., the analog of the possibly non-Abelian
$\nu=5/2=2+1/2$ SLL single-layer state~\cite{foot-note}, can exist in the lowest Landau
level (LLL) under experimentally observable conditions.  This question
has a long history~\cite{he,greiter,he-1,nomura} in the theoretical literature
going back to the early 1990s and recent FQHE experiments at $\nu=1/2$
 make it imperative that a
theoretical analysis be carried out to achieve a proper qualitative
understanding of current experiments~\cite{luhman-2008,shayegan-new}.

Another important question concerns the non-Abelian-ness
(or not) of the experimental 5/2 state which is of profound importance
beyond quantum Hall physics~\cite{sds-tqc,tqc-rmp}.  Hence, it is useful to
contemplate novel situations where the nature of the 5/2 state will
manifest itself in a dramatic, but hitherto unexplored, manner.  
The current work addresses the dichotomy (i.e., the 1/2 FQHE
being an Abelian Halperin 331 bilayer state~\cite{halperin-331} versus the 5/2 FQHE being a
non-Abelian Moore-Read  Pfaffian~\cite{mr-pf,read-rezayi} 
 single-layer state) between single-layer SLL physics versus bilayer
LLL physics by studying the \textit{bilayer} FQHE 
in both the \emph{lowest} and \emph{second} LLs as a
function of layer separation and tunneling.  We ask 
 whether bilayers can support  \textit{both} Pf
and 331 FQHE.  

The proposed non-Abelian Moore-Read Pfaffian (Pf) state is a
weak-pairing single-layer (or one-component) FQHE state for 
half-filled LLs which, in principle,
applies to any orbital LL (i.e., LLL as well as SLL).  Thus, as
a matter of principle a $\nu=1/2$ LLL single-layer Pf FQHE is
certainly a possibility~\cite{he-1,storni,papic,nomura,greiter,shabani}
 although it has never been observed
experimentally.  Theoretically, due to the differences in the
electron-electron Coulomb interaction pseudo-potentials in the LLL
compared to the SLL the two situations (i.e., 1/2 and 5/2) are
quantitatively very different, and it is possible for the Pf FQHE to  
exist in the SLL, but not in the LLL (and vice versa).  The best existing numerical
work~\cite{mrp-tj-sds-prb,papic,papic-1,storni} indicates that either the
single-layer $\nu=1/2$ LLL Pf state does not exist in nature or if it exists,
does so only in rather thick 2D layers with an extremely
small FQHE excitation gap, making it impossible or very difficult to
observe experimentally.  By contrast, the single-layer $\nu=5/2$ SLL
FQHE is observed routinely, albeit at low temperatures ($\lesssim 100
mK$), in high mobility ($\gtrsim 10^7 cm^2/V s$) samples, and with a
rather small (but experimentally accessible) activation gap ($\sim
100$-$500 mK$).  In fact, it has been pointed out that the
experimental $\nu=5/2$ FQHE is always among the strongest  
observed FQHE states in the SLL.

Instead of studying a single-layer 2D system, we concentrate on the
spin-polarized and density balanced bilayer system assuming an 
arbitrary tunneling strength $t$ and an arbitrary
layer separation $d$ in the lowest and second LLs.  We confine our study to 
spin-polarized systems because all experimentally realized half-filled FQHE states appear to
be fully spin-polarized, consistent with the theoretical expectation~\cite{morf,feiguin}--whether 
bilayer two-component FQHE states at $\nu=1/2$ or one-component 
single-layer FQHE states at $\nu=5/2$.    By density ``balanced" we mean that each 
layer in the bilayer system has the same number of electrons.  Furthermore, it 
should be noted that  the tunneling strength $t$ is proportional to the symmetric-antisymmetric 
energy gap.  The experimentally realizable system we have in mind could
either be a true bilayer double-quantum-well structure or a single
wide-quantum-well (WQW) which manifests effective bilayer behavior where the
self-consistent field from the electrons produces effective
two-component behavior~\cite{luhman-2008,shayegan-new} (see Fig.~\ref{fig-bilayer-dqw-wqw} in Sec.~\ref{sec-app}). 

We numerically obtain the approximate quantum phase diagram
for a 1/2 filled bilayer system in either the lowest or second Landau 
level in the $t$-$d$ space using the spherical system finite size exact 
diagonalization (Lanczos) technique,
concentrating entirely on the Pf and the 331 FQHE phases.   

For the LLL, we revisit the $\nu=1/2$ bilayer FQHE and 
carry out an extensive comparison with all
existing bilayer $\nu=1/2$ FQHE experimental observations to ascertain
any hint of the existence of a non-Abelian Moore-Read 
Pfaffian state for large values of $t$.  No strictly
single-layer system, e.g., a heterostructure or a not-too-thick
quantum well, has ever demonstrated an incompressible FQHE at
$\nu=1/2$, instead manifesting only the compressible composite 
fermion~\cite{jain-prl,cf-book} Fermi
sea~\cite{kalmeyer,hlr}.   It is intuitive that our model system is an effective bilayer,  or single-layer,
$\nu=1/2$ system for small, or large, values of $t$, and therefore by
studying the quantum phase diagram as a function of $t$ and $d$ we
hope to shed light on the possible existence of a single-layer
$\nu=1/2$ FQHE in real systems.    

One of the results of the current work is that for the 
lowest Landau level bilayer FQHE system, (i) the recently observed WQW $\nu=1/2$
FQHEs~\cite{luhman-2008,shayegan-new} are strong-pairing Abelian Halperin
331 FQHE states~\cite{halperin-331,halperin-surfsci} which, however, sit close to the
boundary between the Abelian 331 and the weak-pairing non-Abelian
Pfaffian FQHE state~\cite{mr-pf}, and (ii) it may be conceivable,
as a matter of principle, to realize the LLL $\nu=1/2$ Pfaffian
non-Abelian FQHE in very thick bilayers, but as a matter of practice,
this is unlikely since the $\nu=1/2$ FQHE gap is extremely
small (perhaps zero) in the parameter regime where the Pf is more
stable than the 331 phase.  Our findings about the fragility of the
LLL $\nu=1/2$ non-Abelian Pf state are consistent with recent
conclusions~\cite{mrp-tj-sds-prb,papic,papic-1,storni}, but our main focus in the
current work is in understanding the $\nu=1/2$ \emph{bilayer} quantum phase
diagram treating tunneling $t$, layer separation $d$, 
and individual layer width $w$ of the 2D system as independent tuning parameters of
the Hamiltonian.

For the second Landau level, we predict that bilayers can support both the 
non-Abelian Moore-Read Pfaffian and the Halperin Abelian 331 FQHE
and that there could be a novel quantum phase transition, both as a function of
tunneling strength (at constant layer separation) and of layer
separation (at constant tunneling), between the two-component 331
Abelian and the one-component Pf non-Abelian states in a bilayer SLL 
system.  We show
that tuning the inter-layer tunneling and/or layer separation would
lead to a transition between the Abelian and the non-Abelian SLL
FQHE, which should be observable experimentally in
standard FQHE transport experiments.  In particular, we predict that
in realistic systems with finite single-layer width, the SLL bilayer state
would manifest two distinct FQHE phases separated by a region of
finite inter-layer separation and tunneling.  Existence of two
distinct incompressible FQHE bilayer states at total filling 
factor $\nu=5/2$, connected possibly by
a quantum phase transition, is a clear (experimentally testable)
prediction of our theory.  The observation of such a quantum phase transition,
originally predicted as possible by Read and Green~\cite{RG2000} (Fig. 1 
in Ref.~\onlinecite{RG2000}) under
general theoretical considerations (but never before demonstrated to
be feasible under realistic conditions), would strongly suggest the
existence of a non-Abelian 5/2 state since the two 
distinct FQHE phases in the same sample both cannot conceivably be Abelian 331 states.

We first present a background for our work in Section~\ref{sec-background} and then describe
our theoretical model in terms of a Hamiltonian and introduce all relevant 
parameters in Section~\ref{sec-model}.  Next we revisit the lowest Landau level problem in 
Section~\ref{sec-LLL} \emph{before} 
tackling the second Landau level problem in Section~\ref{sec-SLL}, 
since the LLL problem, due to its long history, is easier to understand and will 
provide a proper context and atmosphere when discussing our SLL results.  In 
Section~\ref{subsec-gap-LLL} we discuss in detail how we connect our results with current 
and previous experimental FQHE bilayer LLL results.  Furthermore, in Section~
\ref{subsec-bilayer-SLL} we discuss am important issue regarding bilayer FQHE systems in higher 
Landau levels--a difficulty or ambiguity that is quite subtle and has not been discussed previously 
as far as we 
know.  Lastly we present our conclusions in Section~\ref{sec-conc}.

\section{Background}
\label{sec-background}
Before describing and presenting our work, it is useful to provide a brief background 
of bilayer FQHE, a subject with a long history, in order to set a context for 
our work.

The two candidate wavefunctions we consider and compare in this work, with 
respect to $\nu=1/2$ and 5/2 bilayer incompressible states, the Halperin 331 
state~\cite{halperin-331,halperin-surfsci} and the Moore-Read Pfaffian state~\cite{mr-pf} were 
proposed in 1983 and 1991, respectively.   The 331 wavefunction is a two-component 
strongly paired Abelian state of two Laughlin~\cite{laughlin} phases in each layer whereas 
the Pfaffian is a one-component weak pairing non-Abelian superconducting 
state of chiral $p$-wave symmetry.  Both of these states are allowed incompressible 
fractional quantum Hall (FQH) phases at half-filled Landau levels which for our purpose could be 
either $\nu=1/2$ or 5/2($=2+1/2$).  The qualitative difference between these two 
incompressible FQH states is that the 331 is an Abelian two-component state 
whereas the Pfaffian is a non-Abelian one-component state.

It was first explicitly shown by Yoshioka, Girvin, and 
MacDonald~\cite{yoshioka-1988,yoshioka-1989} that a 
bilayer system at (total) half-filling could support a 331 incompressible state for a 
layer separation $d\sim l$ (where $l$ is a characteristic length scale 
called the magnetic length, defined below), i.e., when intra- and inter-layer correlations 
are comparable.  Later, He \textit{et al.}~\cite{he,he-1} carried out a detailed 
quantitative analysis of the possible existence of 331 FQHE in $\nu=1/2$ bilayer 
structures, making specific predictions for layer separation ($d$) and layer width ($w$) 
values where experimentally observable incompressible states may exist.  This lead 
to the observation of $\nu=1/2$ bilayer 331 FQHE by 
Eisenstein \textit{et al.}~\cite{eisenstein}.  Parallel to the Eisenstein \textit{et al.} observation 
of the 331 FQHE in bilayers, Shayegan \textit{et al.}~\cite{suen,shayegan-prl} 
observed a well-defined $\nu=1/2$ FQHE 
in wide-quantum-wells.  This WQW observation of $\nu=1/2$ FQHE was at first 
attributed to the one-component Moore-Read 
Pfaffian state by Greiter, Wen, and Wilczek~\cite{greiter}, 
who argued that the occurance of a $\nu=1/2$ FQHE in a single quantum well, rather 
than in double quantum wells, implies a one-component rather than two-component 
nature of the underlying incompressible state.  Later theoretical work by 
He \textit{et al.}~\cite{he-1} and further experimental work by 
Shayegan \textit{et al.}~\cite{shayegan-prl,suen-1} 
decisively established that the wide-well $\nu=1/2$ FQHE is, in fact, a manifestation 
of the two-component 331 rather than the one-component Pfaffian state.  The existence 
of a two-component 331 FQHE in wide single wells becomes possible by 
virtue of the self-consistent Hartree electric field arising from the electrons themselves which, 
for sufficiently large quantum well width ($W$) and carrier density ($n$), could lead to the single 
wide-quantum-well acting effectively as two distinct electron layers localized near the well 
boundaries (with density $n/2$ each) with a potential barrier separating them in 
the middle.  Such a two-component 331 FQHE description of observed $\nu=1/2$ 
incompressibility has become well-accepted, and the fact that no $\nu=1/2$ FQHE has ever 
been observed in relatively thin single quantum well structures or in 
single heterostructures has further reinforced the idea that the $\nu=1/2$ FQHE 
is due to the formation of the two-component Halperin 331 state in wide-quantum-wells~\cite{r1}.

From the perspective of $\nu=1/2$ FQHE, wide-quantum-wells should be considered 
as two-component bilayer systems with non-zero interlayer tunneling.  For relatively 
weak, or strong, tunneling between the layers, the system behaves as a 
two, or one, component system, and the question of the existence of Abelian 
331 or non-Abelian Pfaffian $\nu=1/2$ FQHE then, in some sense, boils 
down to the existence of incompressible FQHE in the weak 
or strong tunneling limit.  Unfortunately, how strong an interlayer tunneling is strong enough 
to render the system into a one-component Pfaffian FQHE is a quantitative question, 
which can only be addressed through detailed numerical study.  Such a numerical 
study is the main goal of this paper.  In the process, we also ask the question of 
one-component versus two-component (i.e., Pfaffian versus 331) FQHE in bilayer 
$\nu=5/2$ case, where the existence of the one-component Pfaffian $\nu=5/2$ FQHE 
is reasonably well-established 
theoretically~\cite{morf,rezayi-haldane-prl,park-1,scarola,scarola-nature,mrp-tj-sds-prl,mrp-tj-sds-prb,moller-simon-prb}, although 
actively debated~\cite{toke-jain-prl,toke-prl,wojs}. The 
bilayer $\nu=5/2$ case was never studied before in the literature whereas there 
were only two studies of the bilayer $\nu=1/2$ case comparing 331 versus Pfaffian state.  
The first is by He \textit{et al.}~\cite{he-1} and the second is by Nomura and 
Yoshioka~\cite{nomura}.  Our work for $\nu=5/2$ is distinct, and our work for $\nu=1/2$ 
transcends that of the earlier work in being much more complete.  Very recently, 
Papi\'c \textit{et al.}~\cite{papic} investigated this problem using a somewhat different model.

As mentioned in the Introduction (Sec.~\ref{sec-intro}), our work is partially motivated 
by the recent experimental observation of Luhman \textit{et al.}~\cite{luhman-2008} who found 
a $\nu=1/2$ FQHE in side single wells with relatively strong tunneling (i.e., 
large symmetric-antisymmetric energy splitting).  An interesting question is 
whether the $\nu=1/2$ FQHE observed in the Luhman \textit{et al.} experiment is a 
two-component 331 state or a one-component Pfaffian state.  We do not study 
the $\nu=1/4$ FQHE observed by Luhman \textit{et al.} which has recently been 
discussed by Papi\'c \textit{et al.}~\cite{papic}

We mention that all our theoretical work assumes complete spin-polarization 
of the electrons and considers only the balanced case where the average electron 
density is the same in each layer.  We also neglect all Landau level coupling 
effects, and as such our work does not distinguish between 
the non-Abelian Pfaffian and the non-Abelian 
anti-Pfaffian~\cite{levin,lee,peterson-apf,hao} states at 
$\nu=1/2$ or 5/2.  (Note that the neglect of LL mixing effects 
may not be a particularly good assumption for the 5/2 FQHE~\cite{dean,bishara}.)  
The reason for our considering full spin-polarization is that prior 
theoretical work by Morf~\cite{morf} and by Feiguin \textit{et al.}~\cite{feiguin} indicates 
that the $\nu=1/2$ or 5/2 state is likely to be fully spin-polarized.  If the bilayer 
incompressible states turn our to be unpolarized or partially polarized, our work 
simply would not be valid.

With this background, we study the stability of the 331 and the Pfaffian state in 
$\nu=1/2$ and 5/2 bilayers in the presence of finite interlayer tunneling ($t$), 
interlayer separation ($d$), and layer width ($w$).  Our goal is to obtain an appropriate 
zeroth-order quantum phase diagram in the $t$-$d$-$w$ space for $\nu=1/2$ 
and 5/2 bilayer FQHE.  Given that we calculate and compare the exact many-body 
ground state for many individual values of $t$, $d$, and $w$, we are forced 
to use a rather modest system size for $N=8$ electrons (i.e., 4 in each layer) 
in all our exact diagonalization work.  Past experience shows that a 8-electron 
system is quit adequate for qualitative understanding as long as precise thermodynamic 
values of excitation gap or ground state energy are not desired.

\section{Theoretical Model}
\label{sec-model}

We use the simplest model Hamiltonian $\hat{H}$ incorporating both 
finite tunneling $t$ and  finite layer separation $d$ (as well as finite layer width $w$--or 
finite thickness) for our bilayer FQHE system:
\begin{widetext}
\begin{eqnarray}
\hat{H}=\sum_{i<j}^N [V_\mathrm{intra}(|\mathbf{r}_i-\mathbf{r}_j|) + 
V_\mathrm{intra}(|\tilde{\mathbf{r}}_i-\tilde{\mathbf{r}}_j|)
+ V_\mathrm{inter}(|\mathbf{r}_i-\tilde{\mathbf{r}}_j|)] - t (\hat{S}_x)_\mathrm{layer} \;,
\label{eq-1}
\end{eqnarray}
$\mathbf{r}_i$ and $\tilde{\mathbf{r}}_i$ are the position of the
$i$-th electron in the right and left layer, respectively.  In Eq.~(\ref{eq-1}),
$V_\mathrm{intra}(r)=e^2/(\kappa\sqrt{r^2+w^2})$ (we
use the Zhang-Das Sarma~\cite{zds} potential to model the
single layer quasi-2D interaction) and
$V_\mathrm{inter}(r)=e^2/(\kappa\sqrt{r^2+d^2})$ are the intralayer
and interlayer Coulomb interaction incorporating a finite layer width
$w$ and a center-to-center interlayer separation $d$ ($>w$ by
definition).  The $x$-component of the pseudo-spin operator
$(\hat{S}_x)_\mathrm{layer}$ (written in the layer basis representation) controls the tunneling 
between the two quantum wells with large $t$ denoting strong tunneling.  

Note that in the bilayer problem when the electron density is balanced in each layer, i.e., total 
number of particles in each layer is $N/2$, there are essentially two natural Hilbert space 
representations.   The layer basis (in which Eq.~\ref{eq-1} is written) or the 
symmetric-antisymmetric basis where $c_{mS}=(c_{mR}+c_{mL})/\sqrt{2}$ and
$c_{mA}=(c_{mR}-c_{mL})/\sqrt{2}$ destroy an electron in the symmetric
($S$) and antisymmetric ($A$) superposition states, respectively,
where $m$ is angular momentum and $c_{mR}$ and $c_{mL}$ destroy an electron in
the right and left quantum well, respectively.  $S(A)$ can be considered to be an
effective pseudo-spin index for the bilayer system.  In the symmetric-antisymmetric basis, the
Hamiltonian is written as
\begin{eqnarray}
\hat{H}=\frac{1}{2}\sum_{\{m_i,\sigma_i=A,S\}}\langle
m_1\sigma_1,m_2\sigma_2|V|m_4\sigma_4,m_3\sigma_3\rangle
 c^\dagger_{m_1\sigma_1}c^\dagger_{m_2\sigma_2}c_{m_4\sigma_4}c_{m_3\sigma_3}
-\frac{t}{2}\sum_{m}(c^\dagger_{mS}c_{mS} -
c^\dagger_{mA}c_{mA})\;.
\end{eqnarray}
\end{widetext}
As written above in Eq.~\ref{eq-1}, the intra-layer Coulomb potential energy
between two electrons is $V=V_\mathrm{intra}$  and the inter-layer potential
energy is $V=V_\mathrm{inter}$.  
A difference between the two representations is that the tunneling operator in the layer-basis can 
be written as $(\hat{S}_x)_\mathrm{layer}$ while, in the symmetric-antisymmetric basis, the 
tunneling operator is $(\hat{S}_z)_{SAS}=\frac{1}{2}\sum_{m}(c^\dagger_
{mS}c_{mS} -
c^\dagger_{mA}c_{mA})=\frac{1}{2}\sum_{m}(c^\dagger_{mR}c_{mL} + c^\dagger_{mL}c_{mR})= 
(\hat{S}_x)_\mathrm{layer}$.  The layer- and symmetric-antisymmetric-bases are related through a 
pseudo-spin rotation.   Of course, the choice is a matter of personal taste and 
convenience and one should really 
appeal to a physical explanation to understand the tunneling:  
no matter the basis choice, the tunneling term controls 
the probability that electrons jump back and forth between the two layers keeping the total number 
of electrons in each layer fixed at $N/2$, i.e., keeping the bilayer system density balanced.

We numerically diagonalize $\hat{H}$ for finite $N$ assuming 
specific values of $w$, $d$, and $t$ (each expressed throughout in dimensionless units using
the magnetic length $l=(c\hbar/eB)^{1/2}$ as the length unit and the
Coulomb energy $e^2/(\kappa l)$, where $\kappa$ is the background
dielectric constant of the host semiconductor, 
as the energy unit; $e$ is electron charge, $c$ is the speed of light in vacuum, 
and $B$ is the magnetic field strength).  We utilize 
the spherical geometry where the electrons are confined to
a spherical surface of radius $R=\sqrt{N_\phi/2}$, $N_\phi$ is the
total magnetic flux piercing the surface ($N_\phi$ is an integer
according to Dirac), and the filling factor in the partially 
occupied Landau level (whether LLL or SLL) is $\lim_{N\rightarrow\infty}N/N_\phi$ .  To fully consider tunneling, the Hilbert space 
must contain basis states with $\tilde N$ and $N-\tilde N$ electrons in the right and 
left layers for $\tilde{N}\in[0,N]$.  

Following the standard
well-tested procedures~\cite{he,greiter,he-1,nomura,mrp-tj-sds-prb,papic,papic-1,storni} used
extensively in the FQHE literature, we calculate the overlap between
the exact numerical $N$ electron ground state wavefunction of the
Coulomb Hamiltonian defined by Eq.~(\ref{eq-1}) and the candidate $N$
electron variational states which are the Abelian Halperin 331
strong-pairing~\cite{halperin-331} and the non-Abelian Moore-Read Pfaffian
weak-pairing~\cite{mr-pf} wavefunctions:
\begin{eqnarray}
\Psi_{331}&=&\prod_{i<j}^{N/2}(z_i-z_j)^3\prod_{i<j}^{N/2}(\tilde{z}_i-\tilde{z}_j)^3
\prod_{i,j}^N (z_i-\tilde{z}_j)\\
\Psi_\mathrm{Pf}&=&\mathrm{Pf}\left\{\frac{1}{z_i-z_j}\right\}\prod_{i<j}^N(z_i-z_j)^2
\;,
\end{eqnarray}
respectively, where $z=x-iy$ is the electron coordinate in the $x$-$y$ plane.  
Intuitively, the ``331" in $\Psi_{331}$ originates from the exponents for 
$(z_i-z_j)$, $(\tilde{z}_i-\tilde{z}_j)$, and $(z_i-\tilde{z}_j)$, respectively, i.e., 
3, 3, and 1.  (Note that other Halperin 331-type wavefunctions~\cite{moller-1,moller-2} have been 
considered for other filling factors, e.g., total $\nu=1$, and recently the total $\nu=5/2+5/2=5$ bilayer FQHE has been considered~\cite{shi}, however, none of these 
states directly apply to the situation that we are investigating of a half-filled LLL or SLL.)
Both $\Psi_{331}$ and $\Psi_\mathrm{Pf}$ are written above in the layer 
representation, where the 331 state pairs electrons between layers and 
$\Psi_\mathrm{Pf}$ pairs electrons in a single-layer, since writing the 
wavefunctions in real space and appealing to the layer basis 
representation is much more intuitive.  However, we are working 
within the ``balanced" density situation so the Pf wavefunction, in particular, 
should be thought of as a wavefunction that describes pairing among electrons in the symmetric 
state (($c_{mR} + c_{mL})/\sqrt{2}$) since the electrons are never physically 
completely in a single layer, right or left.

We emphasize that we only consider the above two candidate 
wavefunctions.  It is well known~\cite{kalmeyer,hlr}
that a composite fermion liquid phase (or composite fermion Fermi sea) exists for 
$\nu=1/2$ one-component systems both from extensive theoretical analysis and 
experimental observations~\cite{cf-book}.  On the other hand, it is strongly suspected 
that for one-component systems at $\nu=5/2$, the ground state is the Moore-Read 
Pfaffian phase.  Again, this is known from theoretical and experimental 
works~\cite{morf,rezayi-haldane-prl,scarola,mrp-tj-sds-prl,mrp-tj-sds-prb,moller-simon-prb}.   
There are other candidate wavefunctions that can be written down besides 
$\Psi_{331}$ and $\Psi_\mathrm{Pf}$.  Namely, one could consider $\Psi_{222}$ 
and $\Psi_{440}$ (both obvious generalizations of naming convention 
used for $\Psi_{331}$) that describe a pseudo-spin unpolarized composite 
fermion Fermi sea and two uncorrelated one-quarter-filled composite  
fermion Fermi seas, respectively~\cite{cf-book}.  However, those states have different ``shifts" 
(see below) in the spherical geometry than one another 
and $\Psi_{331}$ and $\Psi_\mathrm{Pf}$ and cannot be compared on 
an equal footing~\cite{foot-note-1} and comparing them each to the 
exact state $\Psi_0$ would require many additional calculations that are
simply beyond the scope of this work.

After diagonalizing $\hat{H}$ we ensure that (i) the ground state is homogeneous, i.e., has
total orbital angular momentum $L=0$ and 
is therefore an incompressible state, and (ii) there is a gap, 
the FQHE
excitation gap, separating the ground state from all excited
states.  We are concerned with the Pf and
331 variational states for total filling factor 1/2 in either the lowest or second Landau level, so $N_
\phi=2N-3$, the ``-3" is known as the ``shift" and is a consequence of the 
curvature of the spherical geometry in which we work.  Throughout, 
we consider $(N,N_\phi)=(8,13)$ and mention that due to the 
pseudo-spin component present in the bilayer problem the Hilbert space is 
large (more than $10^5$ states for $N=8$ and over
$7\times10^6$ for $N=10$).  Usually when one is utilizing exact diagonalization one 
wishes to consider many different system sizes and extrapolate the finite 
size results to the infinite system size through some sort of finite-size 
scaling.  The computational difficulty of our problem does not allow this.  $N=6$ electrons 
at $N_\phi=2N-3$ is aliased with a one-component FQHE corresponding 
to $\nu=2/3$--remember the filling factor $\nu=\lim_{N\rightarrow\infty}N/N_\phi$, so 
for certain combinations of $N$ and $N_\phi$ there can be correspondence between 
the finite systems and two distinct filling factors (the so-called aliasing problem).  Hence, 
the $N=6$ case will produce ambiguous results (and, in fact, 
this aspect raises some questions about the work of Nomura and Yoshioka~\cite{nomura} 
who used $N=6$ to do a similar investigation for $\nu=1/2$ bilayers).  $N=10$ is too big for us 
to exactly diagonalize (mentioned above) so we are left with only being 
able to consider $N=8$ and no finite size scaling is possible.  Although this is a 
drawback, it is not uncommon in theoretical FQHE studies to use a modest system 
size, but many different sets of physical parameters (i.e., $t$, $d$, $w$ for our case) to 
bring out qualitative features.  Furthermore, 
since the theoretical techniques are standard, we do not
give the details, concentrating instead on the results and their
implications for bilayer FQHE experiments.

We investigate this system with the usual probes used in theoretical
FQHE studies by calculating: (i) wavefunction overlap between
variational ansatz ($\Psi_\mathrm{Pf}$  and $\Psi_{331}$ 
states) and the exact ground state $\Psi_0$ of $\hat{H}$--an overlap of
unity or zero indicates that the physics is or is not described by the ansatz;
(ii) expectation value of the exact ground state 
of $(N_S-N_A)/2$ , where $N_S$ and $N_A$ are the 
expectation values of the number of electrons in the symmetric or the antisymmetric 
states, respectively--a value of zero or $N/2$
indicating the ground state to be two- or one-component; and (iii)
energy gap (provided the ground state is a uniform state with
$L=0$)--a non-zero excitation gap indicating a possible FQHE state.
In other words, we ask: (i) What is the physics (i.e., 331 or Pf)?;
(ii) Is the system one- or two-component?; (iii) Will the system
display FQHE (i.e., is the system compressible or incompressible)?

\begin{figure*}[t]
\begin{center}
\mbox{(a)}
\mbox{\includegraphics[width=7.cm,angle=0]{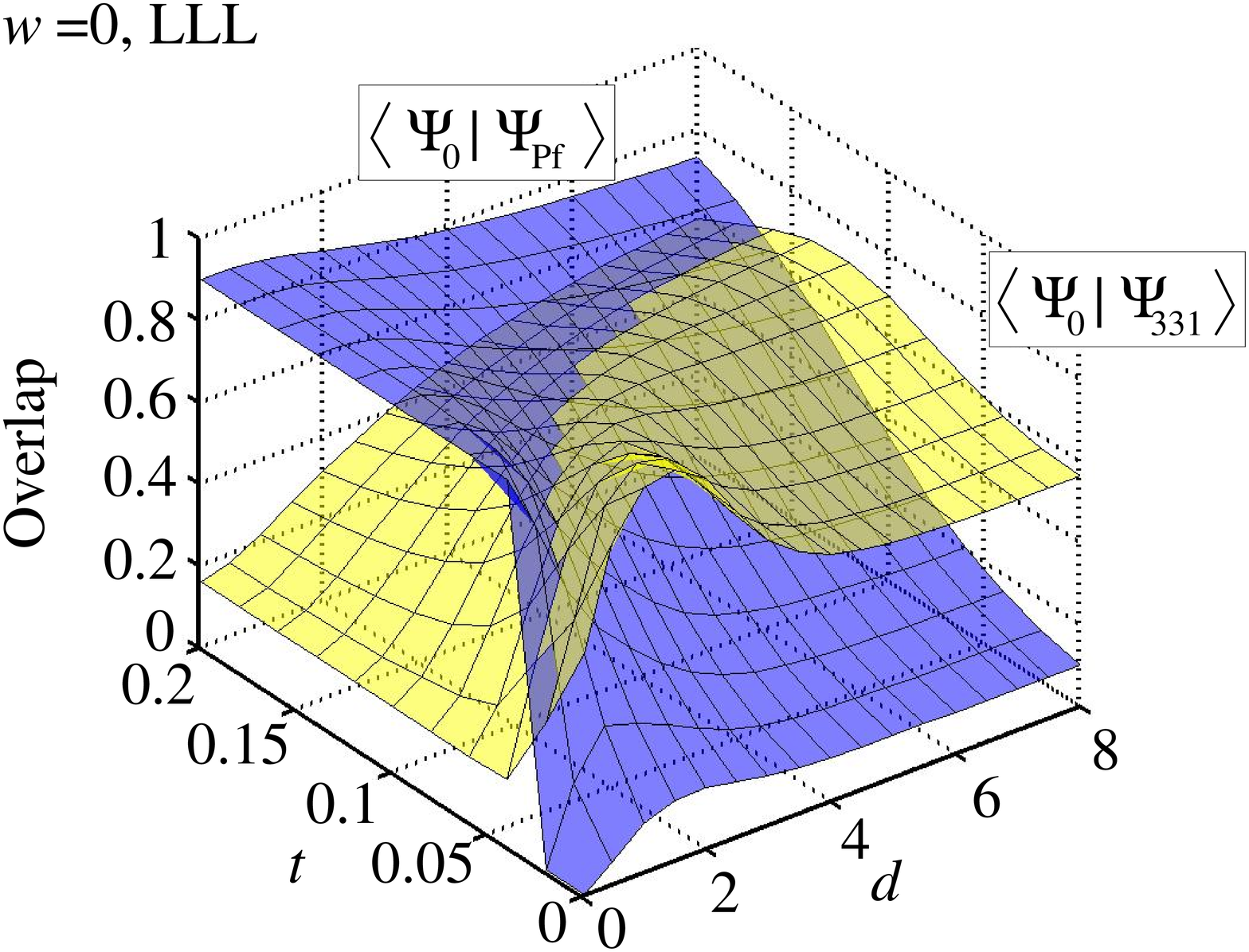}}
\mbox{(b)}
\mbox{\includegraphics[width=7.cm,angle=0]{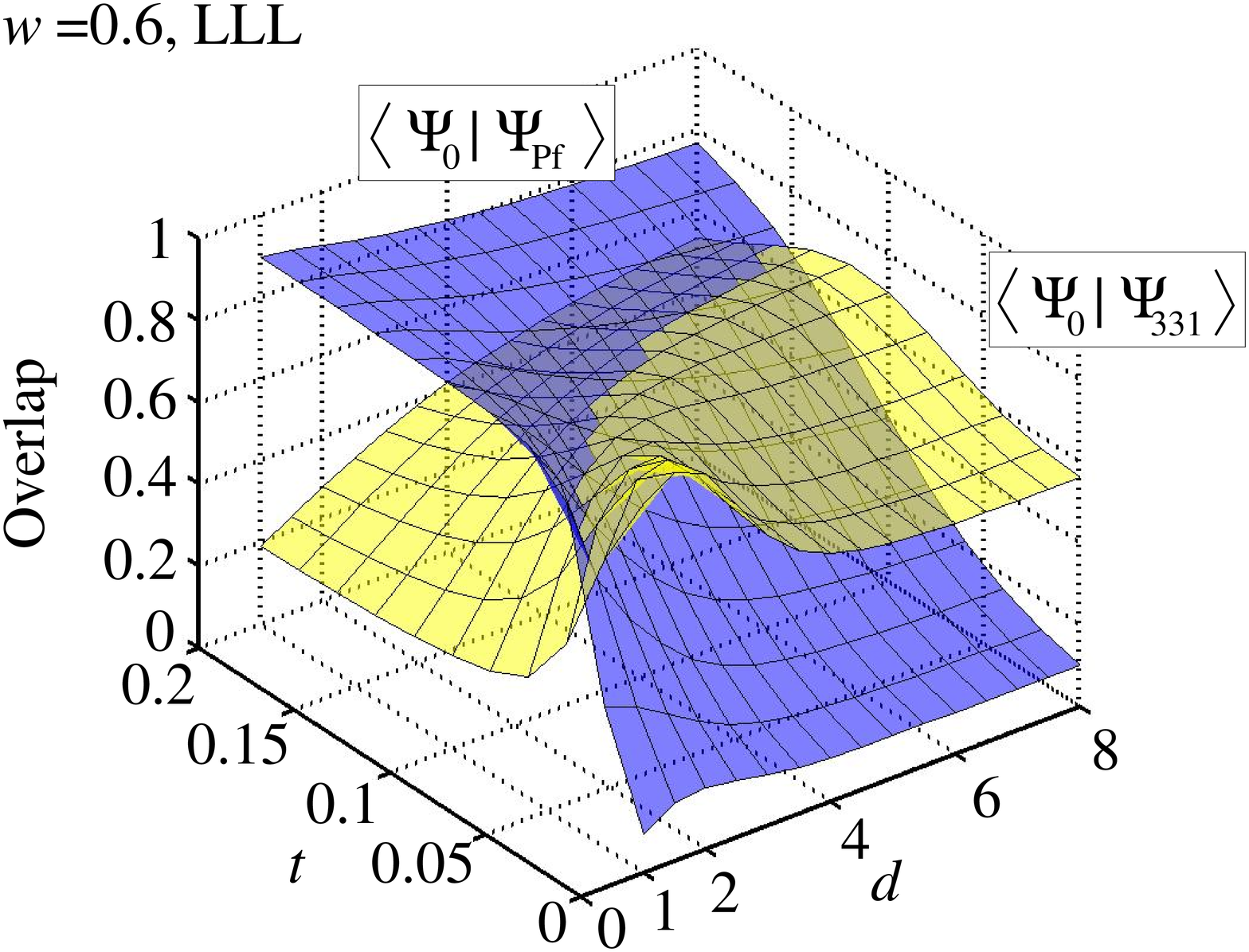}}\\
\mbox{(c)}
\mbox{\includegraphics[width=7.cm,angle=0]{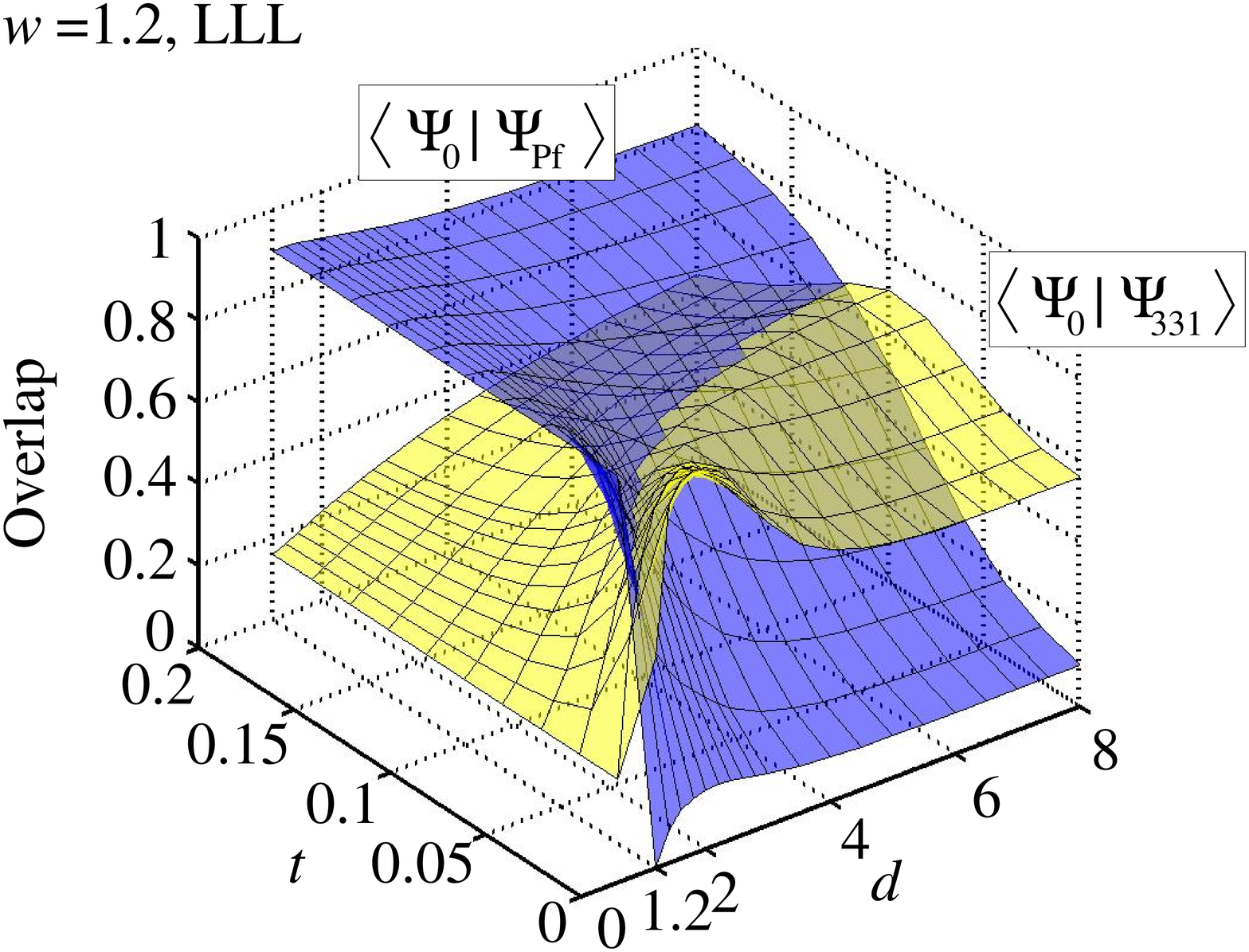}}
\mbox{(d)}
\mbox{\includegraphics[width=7.cm,angle=0]{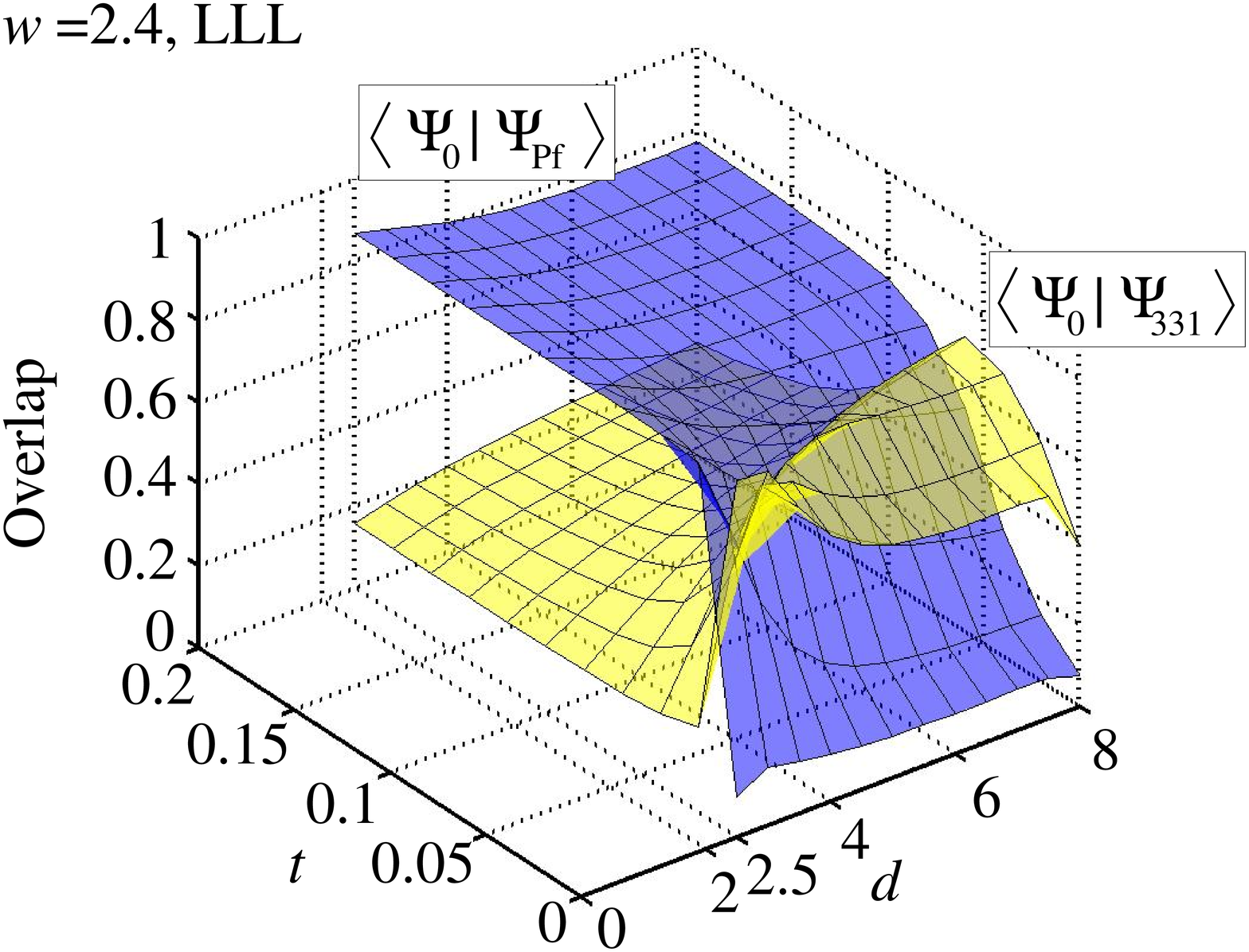}}
\end{center}
\caption{(Color online) Wavefunction overlap between the exact ground
state, $\Psi_0$, and the Pf state, $\Psi_{\mathrm{Pf}}$ (blue (dark
gray)), and 331 state, $\Psi_{331}$ (yellow (light gray)), as a function
of layer separation $d$ and tunneling amplitude $t$ for the $\nu=1/2$
lowest LL with $N=8$ electrons and single layer width (a) $w=0$, (b) $w=0.6$, (c) $w=1.2$, 
and (d) $w=2.4$ ($d\geq w$ necessarily).}
\label{fig-o-LLL}
\end{figure*}

The calculated overlap and gap determine the nature of the FQHE and
its strength in our theory.  We operationally define the
system to be in the 331 or Pf phase depending on whether the overlap
between the exact ground state and the 331 or Pf wavefunctions is larger, i.e., 
if $\langle \Psi_0|\Psi_{331}\rangle > \langle \Psi_0|\Psi_\mathrm{Pf}\rangle$ then 
the system is said to be in the 331 phase and vice versa.  We emphasize that
our work is a comparison between these two incompressible states only,
and we cannot comment on the possibility of some other state
(i.e., neither 331 nor Pf) being the ground state.  We
do, however, believe that if the system is incompressible at a
particular set of parameter values (i.e., $d$, $t$,
$w$, etc.), it is very likely to be described by one of
these two candidate states, 331 or Pf.  
We cannot, however, rule out the possibility that the real system has a
compressible ground state (without manifesting FQHE), e.g., a
composite fermion Fermi sea, not considered in our calculation.  This is more likely to 
happen when our calculated excitation gap is very small.

\section{Lowest Landau Level Results}
\label{sec-LLL}

\subsection{What is the physics?--lowest Landau level}
\label{subsec-o-LLL}

\begin{figure*}[]
\begin{center}
\mbox{(a)}
\mbox{\includegraphics[width=6.5cm,angle=0]{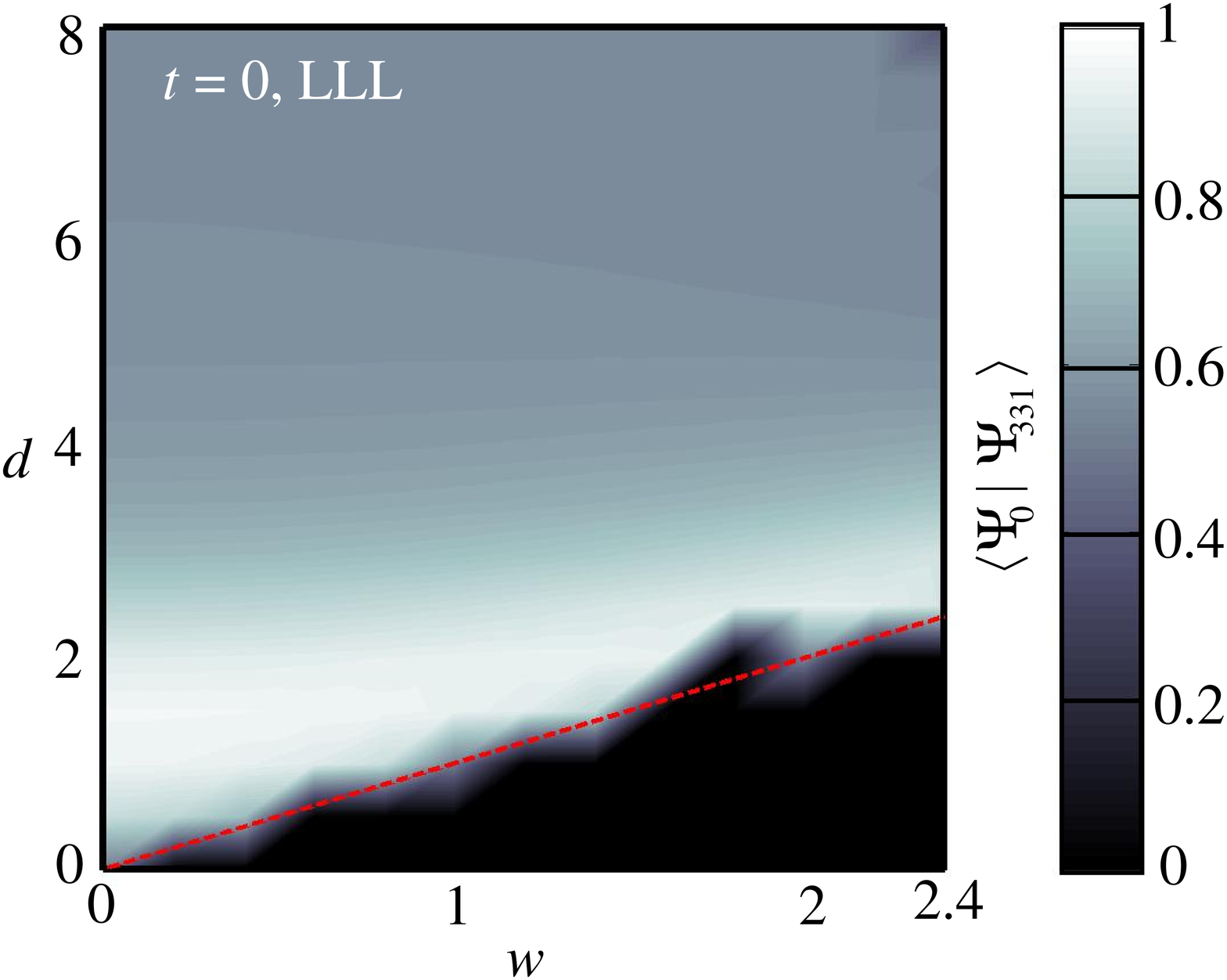}}
\mbox{(b)}
\mbox{\includegraphics[width=6.5cm,angle=0]{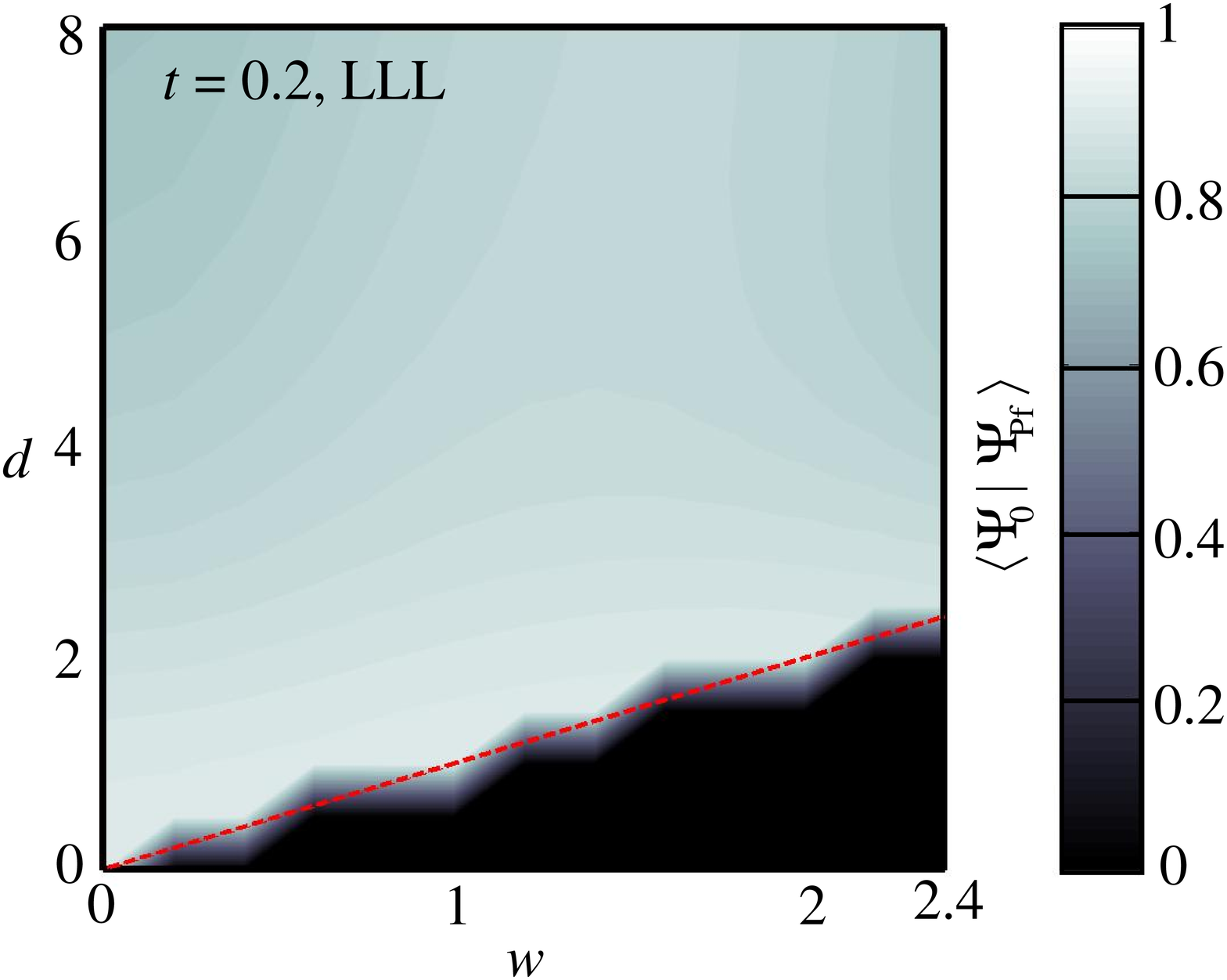}}
\end{center}
\caption{(Color online) Wavefunction overlap between the exact ground
state $\Psi_0$ and $\Psi_{\mathrm{Pf}}$ at strong
tunneling $t=0.2$ (a) and $\Psi_{331}$ at zero tunneling
$t=0$ (b) as a function of separation $d$ and single-layer well width
$w$ (where $d\geq w$) for the $\nu=1/2$ lowest LL with $N=8$
electrons.  White corresponds to an overlap of unity while black
corresponds to an overlap of zero.  The dashed red line is the condition $w=d$ and 
for $w>d$ the bilayer system is undefined, i.e., the single-layer width cannot be larger than the 
layer separation.}
\label{fig-o-LLL-w}
\end{figure*}

The calculated wavefunction overlap
between the exact ground state $\Psi_0$ and the two appropriate
candidate variational wavefunctions ($\Psi_{\mathrm{Pf}}$ and
$\Psi_{331}$) as a function of distance $d$ and tunneling energy
$t$ is shown in Fig.~\ref{fig-o-LLL}.  First we focus on the situation
with zero width $w=0$ (Fig.~\ref{fig-o-LLL}(a)) and concentrate on the
overlap with $\Psi_\mathrm{Pf}$.  In the limit of zero tunneling and
zero separation, the overlap with $\Psi_\mathrm{Pf}$ is zero and quickly jumps to 
approximately 0.9 when the tunneling is increased to only a small amount of approximately 0.025,    
 and for increasing tunneling, $\langle \Psi_0|\Psi_\mathrm{Pf}\rangle $ remains relatively 
constant.  In fact, this zero $d$ and large tunneling result should be compared to the single layer 
results we have given earlier (Fig. 9(a) in Ref.~\onlinecite{mrp-tj-sds-prb}).  
For $d\neq 0$, in the weak tunneling limit the overlap with $\Psi_\mathrm{Pf}$ remains very small.  
However, in the large tunneling limit, increasing $d$ only reduces the overlap marginally and the 
Pf description remains quite good even for large $d$.  This shows that
the strong-tunneling (one-component) regime is well described by the
Pf state.  It should be noted, however, that the overlap between $\Psi_0$ and $\Psi_\mathrm{Pf}$ 
is never much above 0.9--this should be compared to the SLL where 
we know~\cite{morf,rezayi-haldane-prl,scarola,mrp-tj-sds-prl,mrp-tj-sds-prb,moller-simon-prb,storni} 
that the Pf wavefunction is a good physical description in the small layer 
separation and large tunneling limit 
(see Section~\ref{subsec-o-SLL}).

Next we consider the overlap between $\Psi_0$ and $\Psi_{331}$.  In
the zero tunneling limit, as a function of $d$, we find the overlap
starts very small, increases to a moderate maximum of $\approx 0.80$ at
$d\sim1$ before achieving an essentially constant value of
$\approx0.6$.  For $d>4$ the overlap increases with increasing tunneling to a maximum of nearly 
0.8.  Thus, the weak-tunneling (two-component) regime is well described by $\Psi_{331}$ and the 
331 state is very robust to tunneling when layer separation is large.  For small layer separation 
($d<2$), non-zero tunneling very quickly suppresses the overlap between $\Psi_{331}$ and the 
exact state to a very small value.

\begin{figure*}[t]
\begin{center}
\mbox{(a)}
\mbox{\includegraphics[width=6.cm,angle=0]{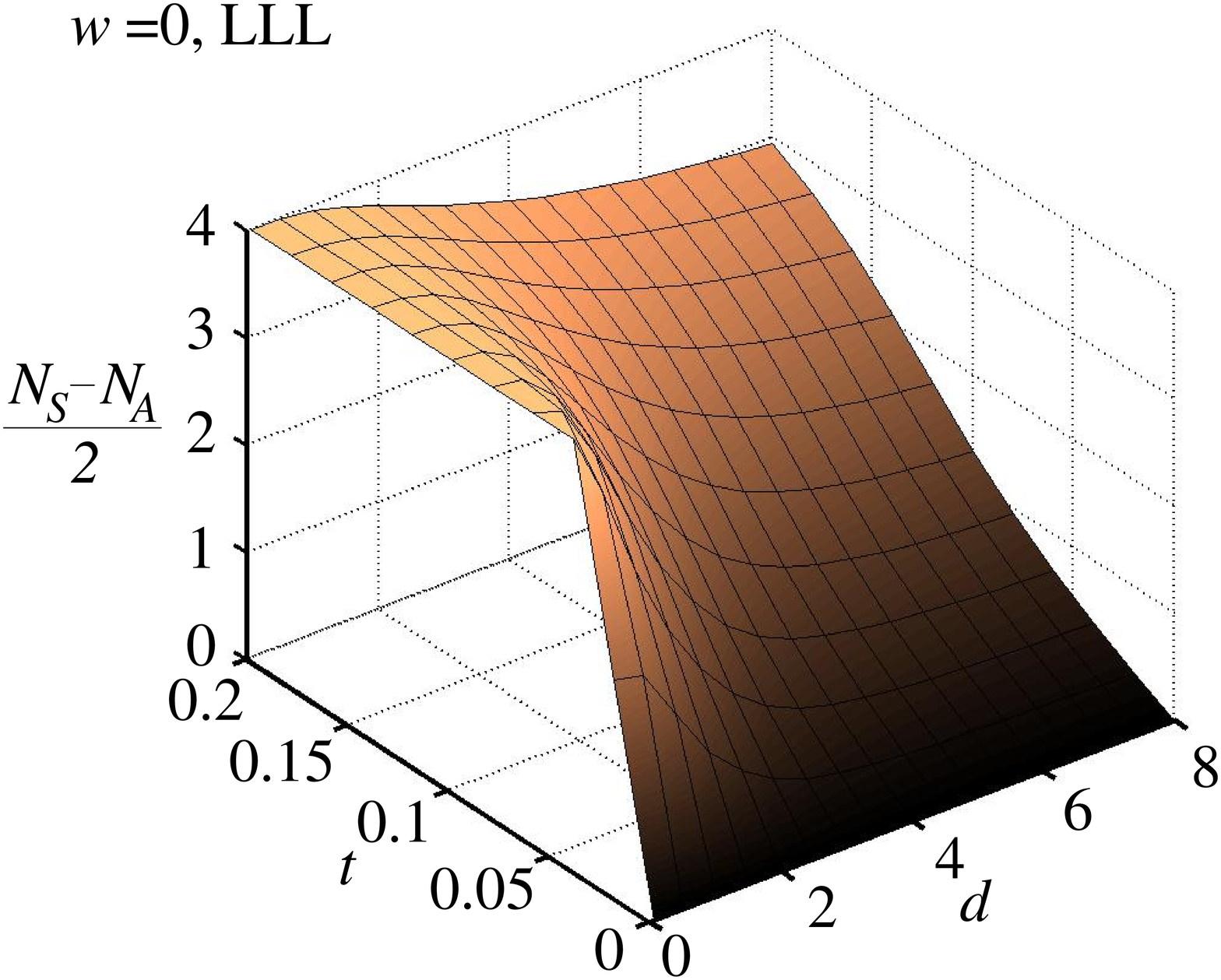}}
\mbox{(b)}
\mbox{\includegraphics[width=6.cm,angle=0]{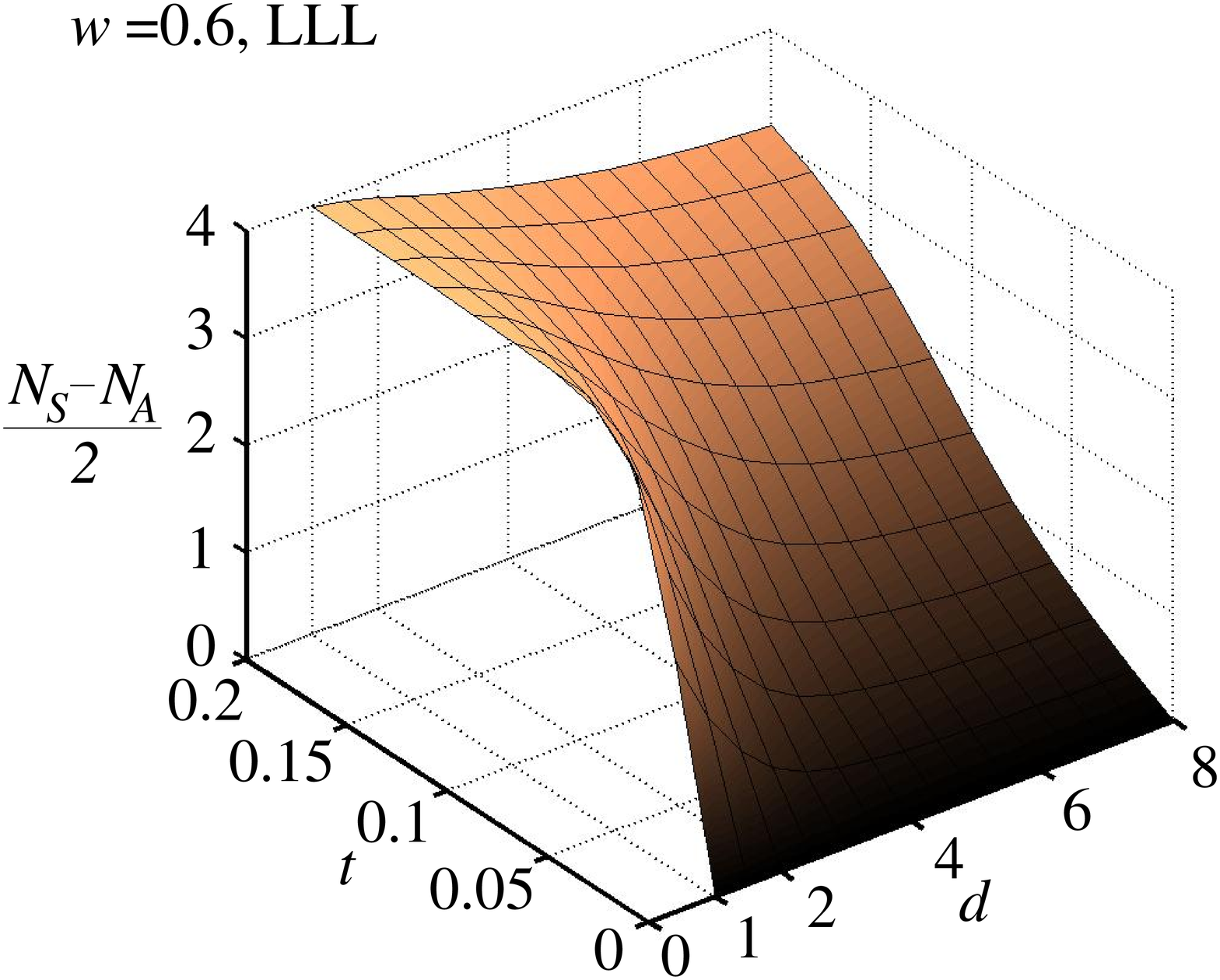}}\\
\mbox{(c)}
\mbox{\includegraphics[width=6.cm,angle=0]{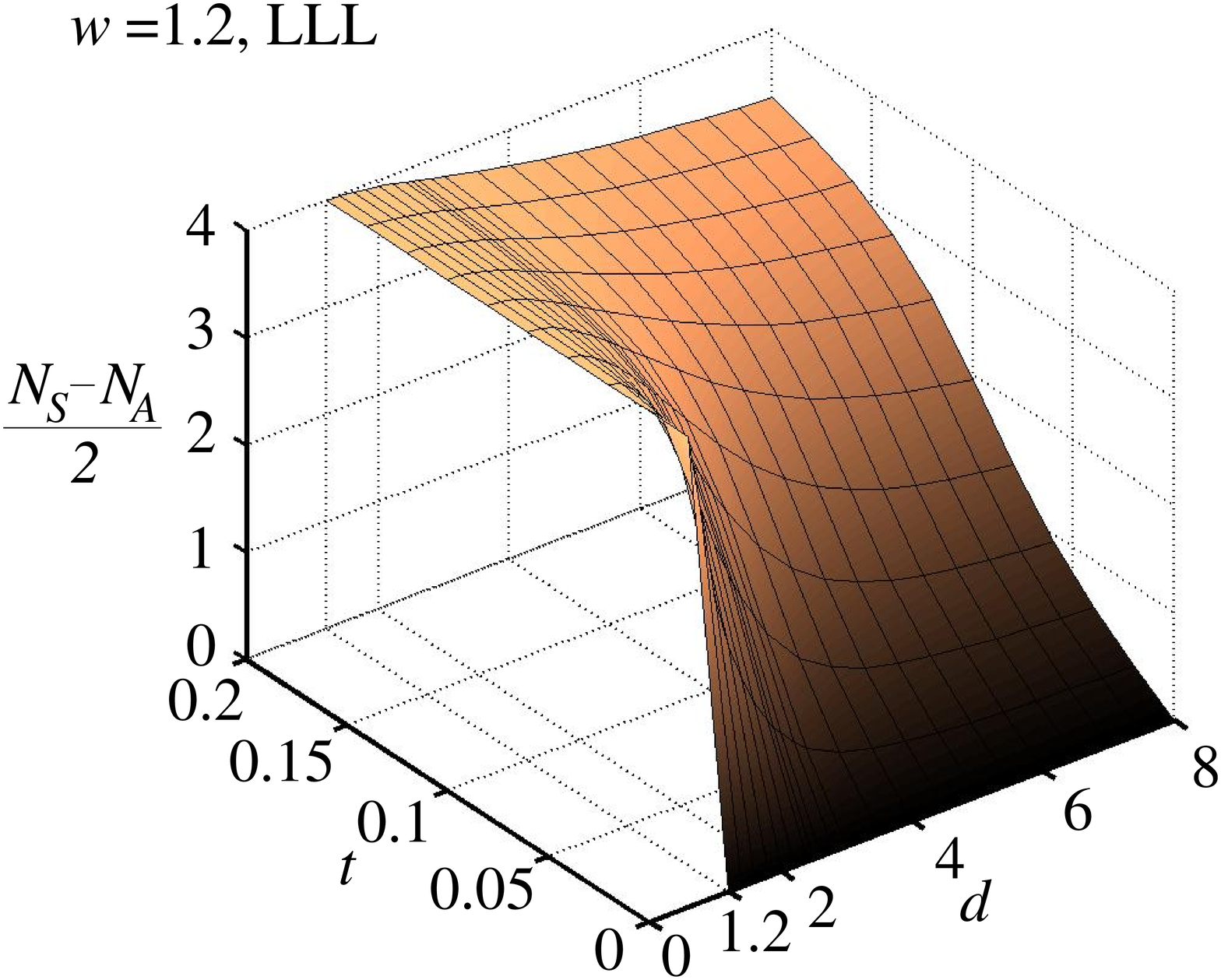}}
\mbox{(d)}
\mbox{\includegraphics[width=6.cm,angle=0]{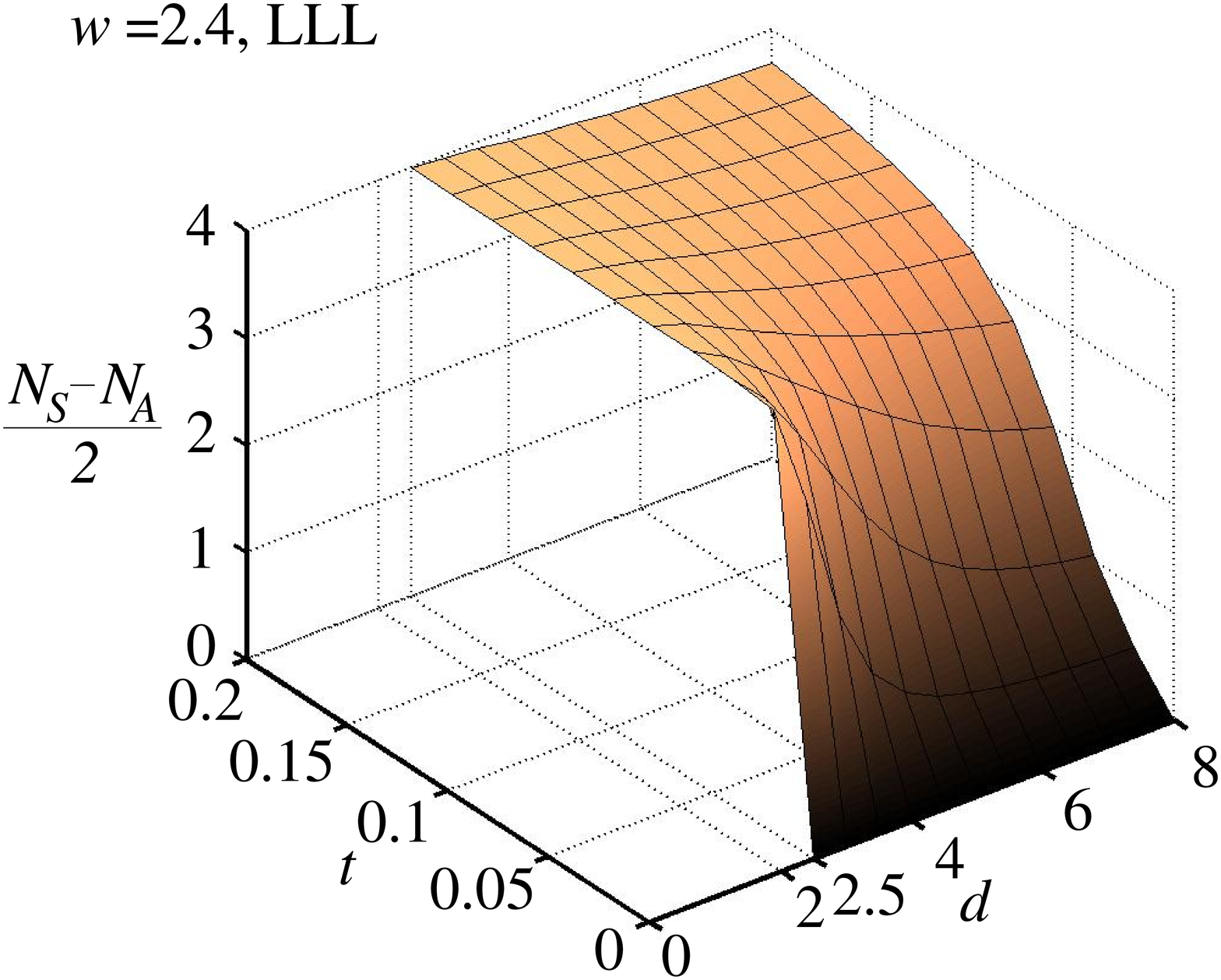}}
\end{center}
\caption{(Color online) $(N_S-N_A)/2$ (or $\langle (S_z)_{SAS}\rangle$ when the 
Hamiltonian $\hat{H}$ is 
written in the SAS-basis) as a function of layer separation $d$ and tunneling amplitude $t$ 
for $\nu=1/2$ and $N=8$ electrons for single-layer width (a) $w=0$, (b) $w=0.6$, (c) $w=1.2$ 
and $w=2.4$.}
\label{fig_Sz_LLL}
\end{figure*}

We know from earlier work~\cite{mrp-tj-sds-prl,mrp-tj-sds-prb} that finite layer width within 
a single quantum well, at least in the large tunneling and small layer separation limit, i.e., the 
one-component limit, may enhance the overlap between $\Psi_0$ and $\Psi_\mathrm{Pf}$.  In 
Fig.~\ref{fig-o-LLL-w}(a) and (b) we show $\langle \Psi_0|\Psi_{331}\rangle$  
and $\langle \Psi_0|\Psi_\mathrm{Pf}\rangle$, respectively, 
versus layer separation $d$ and single layer width $w$ for small tunneling 
$t=0$  and large tunneling $t=0.2$.   For the small 
tunneling situation where 331 is a good ansatz, we 
see clearly that for $w=0$ the overlap is maximum for a finite value of $d\sim 2$.  
However, for finite $w$ 
we find that the position of maximum overlap does not change much and, in fact, for large $w$ the 
maximum overlap obtains essentially for the $w\gtrsim d$ condition. 

In the case of $\Psi_\mathrm{Pf}$ (Fig.~\ref{fig-o-LLL-w}(b)) we find a result similar to the single-
layer finite-thickness results (cf. Ref.~\onlinecite{mrp-tj-sds-prb}) 
where the overlap with the exact state is 
nearly constant for increasing $w$ and $d$.   For $w=d$ we expect this behavior but it is 
interesting to note how robust the overlap is for finite $w$ and large $d$.  

In Fig.~\ref{fig-o-LLL}(b)-(d) we consider the overlap between the exact state and the 331 and Pf 
states as function of separation and tunneling for finite layer widths of $w=0.6$, 1.2, and 2.4.   If we 
consider first $\langle \Psi_0|\Psi_\mathrm{Pf}\rangle$ we see that not much changes for different 
values of $w$ as one would expect from examining Fig.~\ref{fig-o-LLL-w}(b).  
The only real qualitative change is that for $w=d$ for $w=0.6$ and 2.4 the overlap 
is non-zero, albeit very small, compared to the $w=0$ case where it is identically zero.  However, 
for $w=1.2=d$ the overlap is zero, thus, we suppose that the small but non-zero overlap for $d=w
$ is simply a finite size effect.  For the overlap between the exact state and the 331 variational 
state we find that finite single layer width has two effects.  First, it slightly lowers the maximum 
overlap obtained (this is evident in Fig.~\ref{fig-o-LLL-w}) and reduces the area in $d$-$t$ phase 
space where the 331 state is a good ansatz.   In essence, for finite $w$ the Pfaffian phase 
pushes out the 331 phase, at least as far as wavefunction overlap determines 
the phase.

\subsection{Is the system one- or two-component?--lowest Landau level}
\label{subsec-Sz-LLL}

Due to the potential confusion between whether the pseudo-spin operator $(\hat{S}_z)_{SAS}$ or 
$(\hat{S}_x)_\mathrm{layer}$ controls the tunneling we will report our results in terms of $(N_S-
N_A)/2$ where $N_S$ and $N_A$ are the expectation values of the total number of particles in 
the symmetric $S$ or antisymmetric $A$ state, respectively.  This is the more precise variable 
since whether we are calculating $(\hat{S}_z)_{SAS}$ or $(\hat{S}_x)_\mathrm{layer}$ the answer 
is always equal to $(N_S-N_A)/2$.

Our conclusion so
far, based on overlap calculations, is completely consistent with the
calculated expectation value of $(N_S-N_A)/2$ which is essentially the order
parameter describing the one-component to two-component transition,
i.e., $(N_S-N_A)/2\approx N/2$ describes a one-component 
phase since in that case $N=N_S$, 
whereas $(N_S-N_A)/2\approx 0$ describes a two-component 
phase since in that case $N_S=N_A=N/2$.  Note that in our $N=8$ 
system the maximum and minimum value of the pseudo-spin expectation value 
is 4 and 0, respectively.  Fig.~\ref{fig_Sz_LLL}(a)-(d) 
shows the calculated $(N_S-
N_A)/2$ as a
function of $d$ and $t$ for $w=0$, $w=0.6$, $w=1.2$, and $w=2.4$ .  First we consider $w=0$ 
(Fig.~\ref{fig_Sz_LLL}(a)).  For $d=0$ and $t=0$ the system is two-component and the 
system is SU(2) symmetric.   Only a small amount of tunneling is required to very quickly 
push the 
system to be one-component.  As $d$ is increased, more tunneling is required to make the system 
one-component.  Of course, this can be readily understood physically: for non-zero layer 
separation $d$ the electrons in the 
symmetric state pay a higher Coulomb energy price than electrons 
in the antisymmetric state since $1/r>1/\sqrt{r^2+d^2}$, for $d\neq 0$.

\begin{figure*}[t]
\begin{center}
\mbox{(a)}
\mbox{\includegraphics[width=7.cm,angle=0]{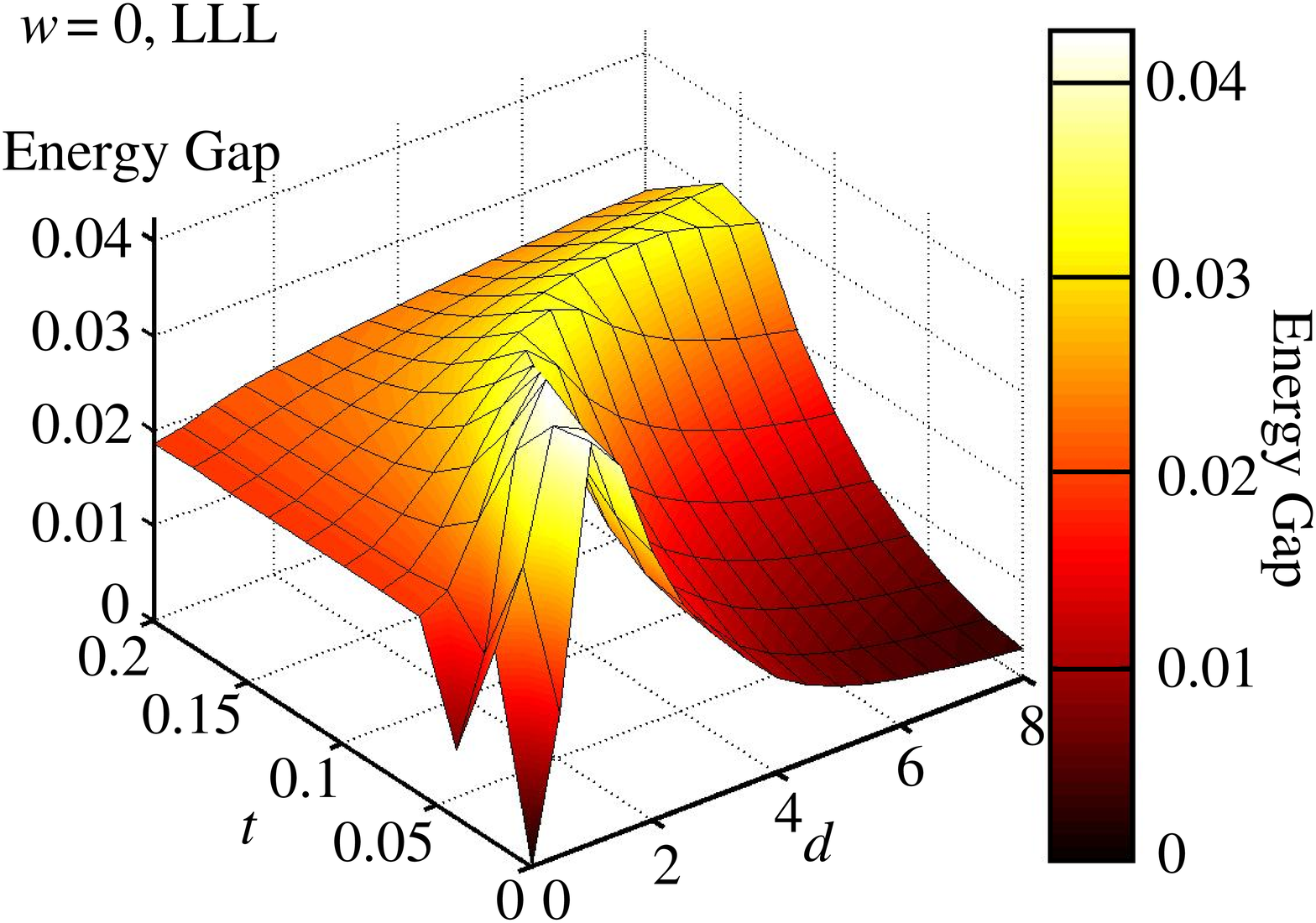}}
\mbox{(b)}
\mbox{\includegraphics[width=7.cm,angle=0]{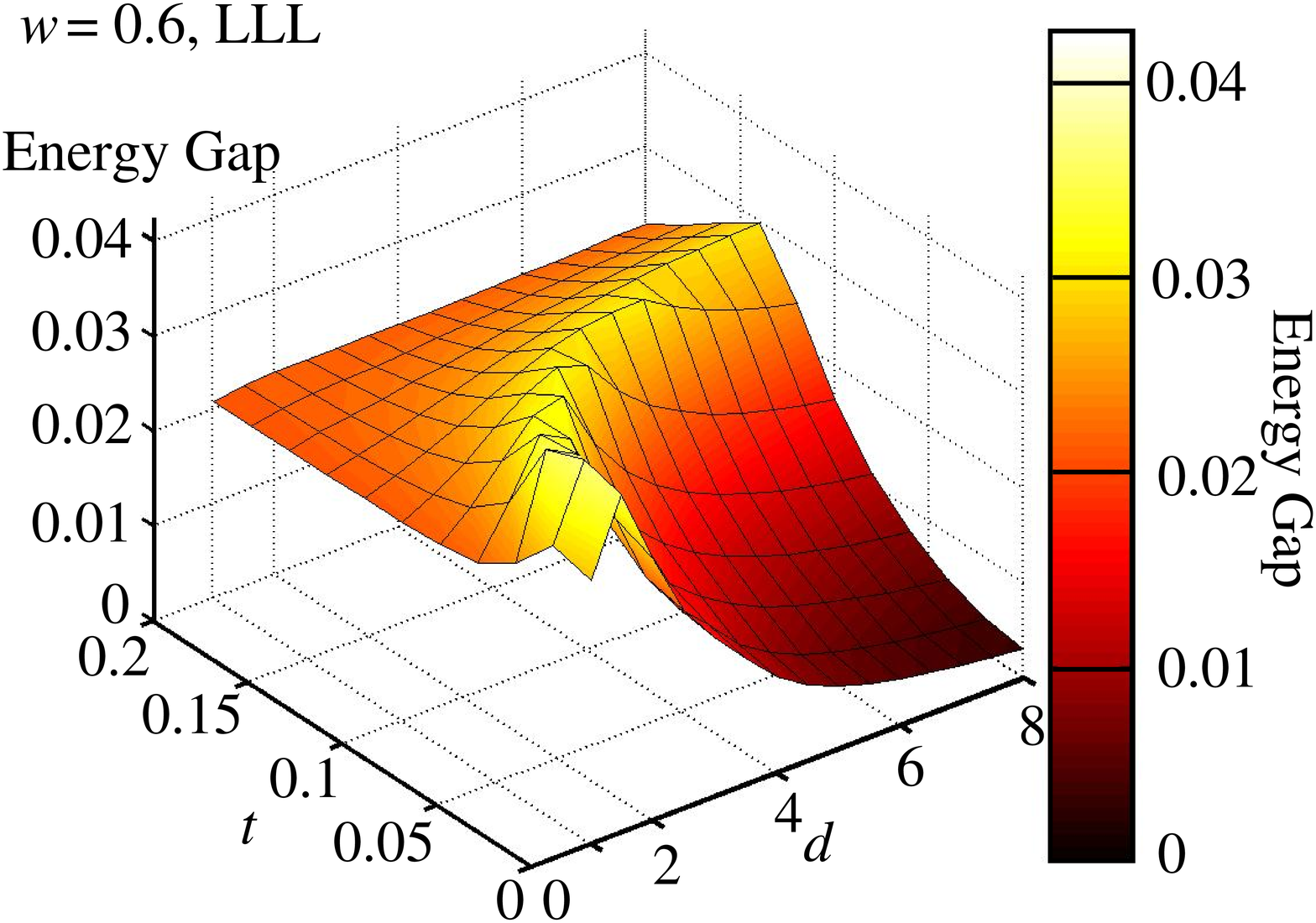}}\\
\mbox{(c)}
\mbox{\includegraphics[width=7.cm,angle=0]{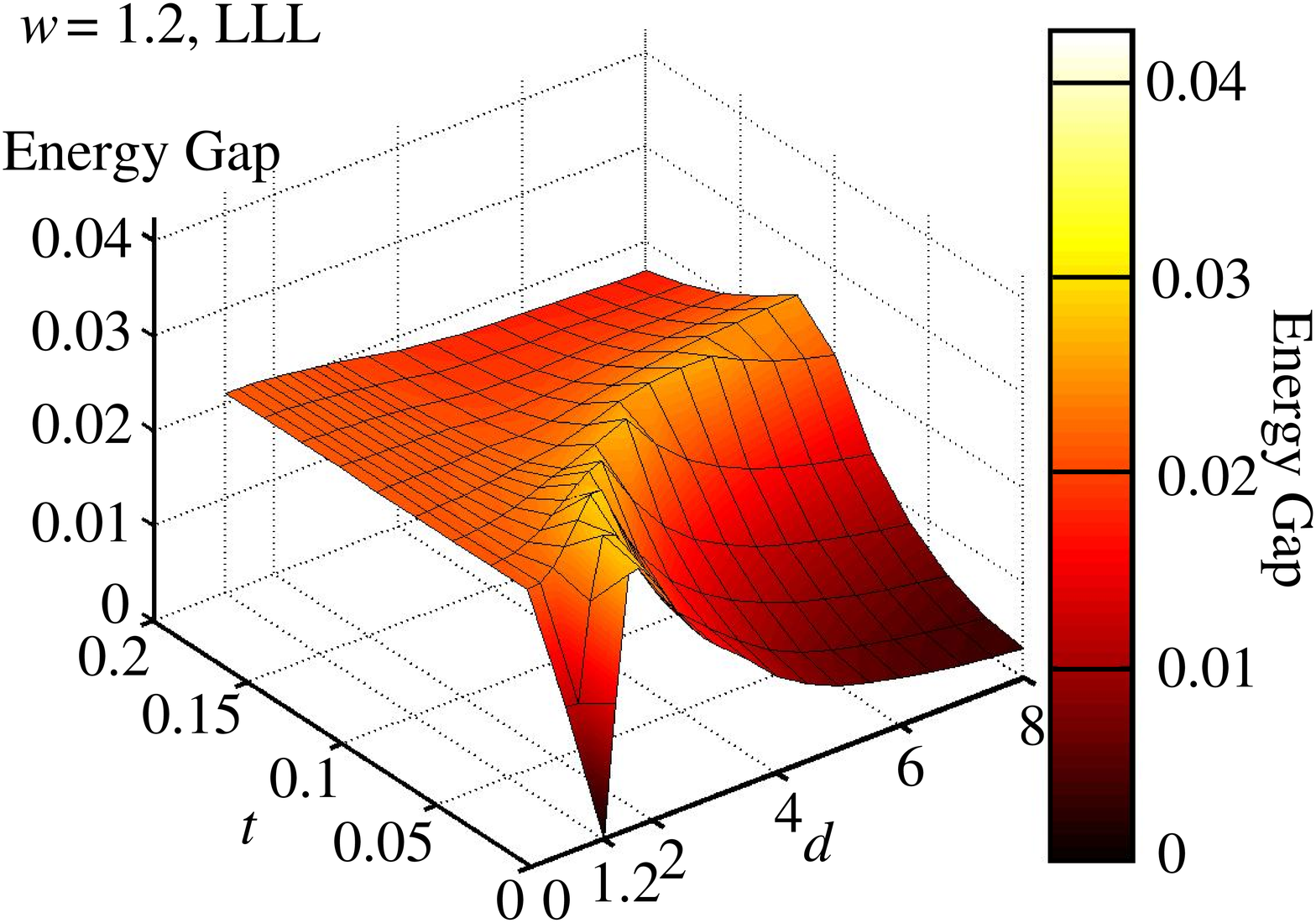}}
\mbox{(d)}
\mbox{\includegraphics[width=7.cm,angle=0]{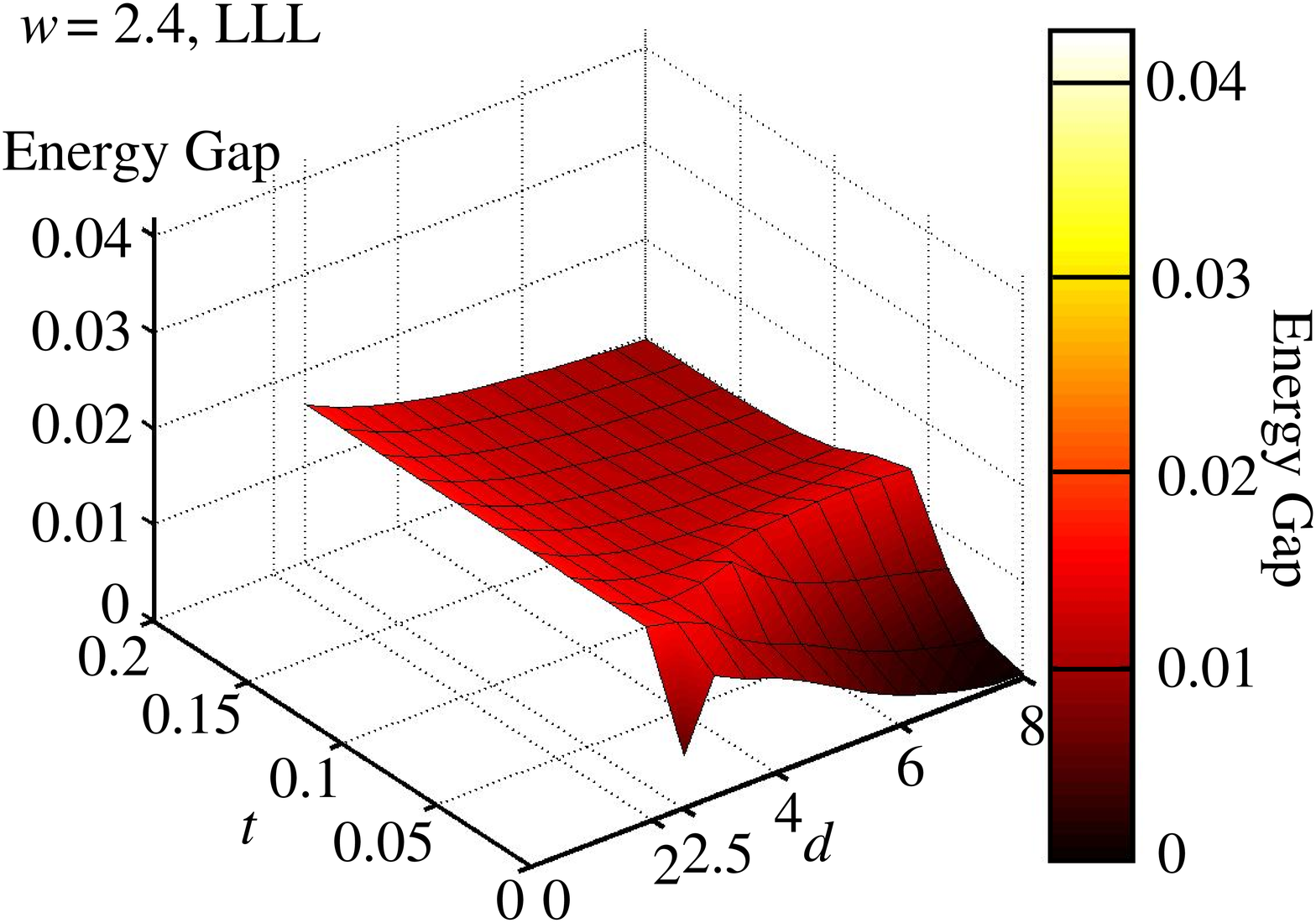}}
\end{center}
\caption{(Color online) FQHE energy gap as a function of layer separation $d$ 
and tunneling amplitude $t$ for the $\nu=1/2$ lowest LL system 
with $N=8$ electrons.  We consider 
single quantum well widths of (a) $w=0$, (b) $w=0.6$, (c) $w=1.2$, and (d) $w=2.4$.  The 
energy gap is also color-coded such that black is zero gap and white is maximum on a scale 
from zero to 0.0425 in Coulomb energy units, $e^2/(\kappa l)$.}
\label{fig_gap3D_LLL}
\end{figure*}

For non-zero single-layer width (finite $w$) the general behavior outlined above does not 
change appreciably, cf. Figs.~\ref{fig_Sz_LLL}(b)-(d).  The only real difference is that for 
increasing $w$, less tunneling is required, at the same value of layer separation $d$, to make the 
system one-component.  That is, it is easier to push the system into one-component behavior via 
tunneling when $w\neq 0$.

Comparing these results to our overlap results shown in Fig.~\ref{fig-o-LLL} we see that when the 
system is effectively two-component for weak tunneling the Halperin 331 state has a higher 
overlap with the exact state than the Moore-Read 
Pfaffian state.  This is expected since the Pf state describes a 
one-component state.  On the other hand, when the system is effectively one-component for strong 
tunneling the one-component Pf state has a higher overlap with the exact state 
$\Psi_0$ than $\Psi_{331}$ does.  Looking closer, one finds that 
when $(N_S-N_A)/2$ is increased from zero 
towards $(N_S-N_A)/2\sim 2.5$ the overlap with the exact state switches from being higher with 
the two-component 331 state to the one-component Pf state (this is true irrespective of $w$).  In 
other words, the two-component description of the exact state provided by $\Psi_{331}$ 
survives until nearly 60$\%$ more electrons are in the symmetric state than 
the antisymmetric state--the two-component 331 state is surprisingly robust to tunneling 
that drives the system towards one-component behavior.

\subsection{Will the system display the FQHE?--lowest Landau level}
\label{subsec-gap-LLL}

Wavefunction overlap and pseudo-spin are only two properties that elucidate the physics.
Another property is the energy gap above the $L=0$ ground state in the
excitation spectra; a crucial characteristic that determines the incompressibility or the 
robustness of the FQHE.  In Fig.~\ref{fig_gap3D_LLL}(a)-(d) we show the energy
gap, defined as the difference between the first excited and ground
state energies at constant $N_\phi$, as a function of $d$ and $t$
for the $\nu=1/2$ lowest LL system for (a) $w=0$, (b) $w=0.6$, (c) $w=1.2$, 
and (d) $w=2.4$.  It is clear that for 
finite $d$ and $t$ there is a 
1/2 FQHE with a finite gap.  At the SU(2) symmetric point the energy gap is 
vanishingly small.  Interestingly, the energy gap has a peak in $t$-$d$ space--a ridge along which 
the energy gap is maximum.  Generally, as $w$ is increased the value of the energy gap 
decreases as is expected since 
finite width of a single quantum well reduces the Coulomb energy 
by softening it.  For $w=2.4$, the maximum width considered, the energy gap retains the basic 
qualitative structure as for the other widths, however, its overall value is very small.  Note as well 
that the maximum energy gap ridge is approximately located at the point in the $d$-$t$ parameter 
space where the overlap between the exact state crosses over from being larger with the Pfaffian 
ansatz and the 331 ansatz (more on this below), i.e., the gap is maximum close to the 
transition line between 331 and Pfaffian.

\begin{figure*}[t]
\begin{center}
\mbox{(a)}
\mbox{\includegraphics[width=7.cm,angle=0]{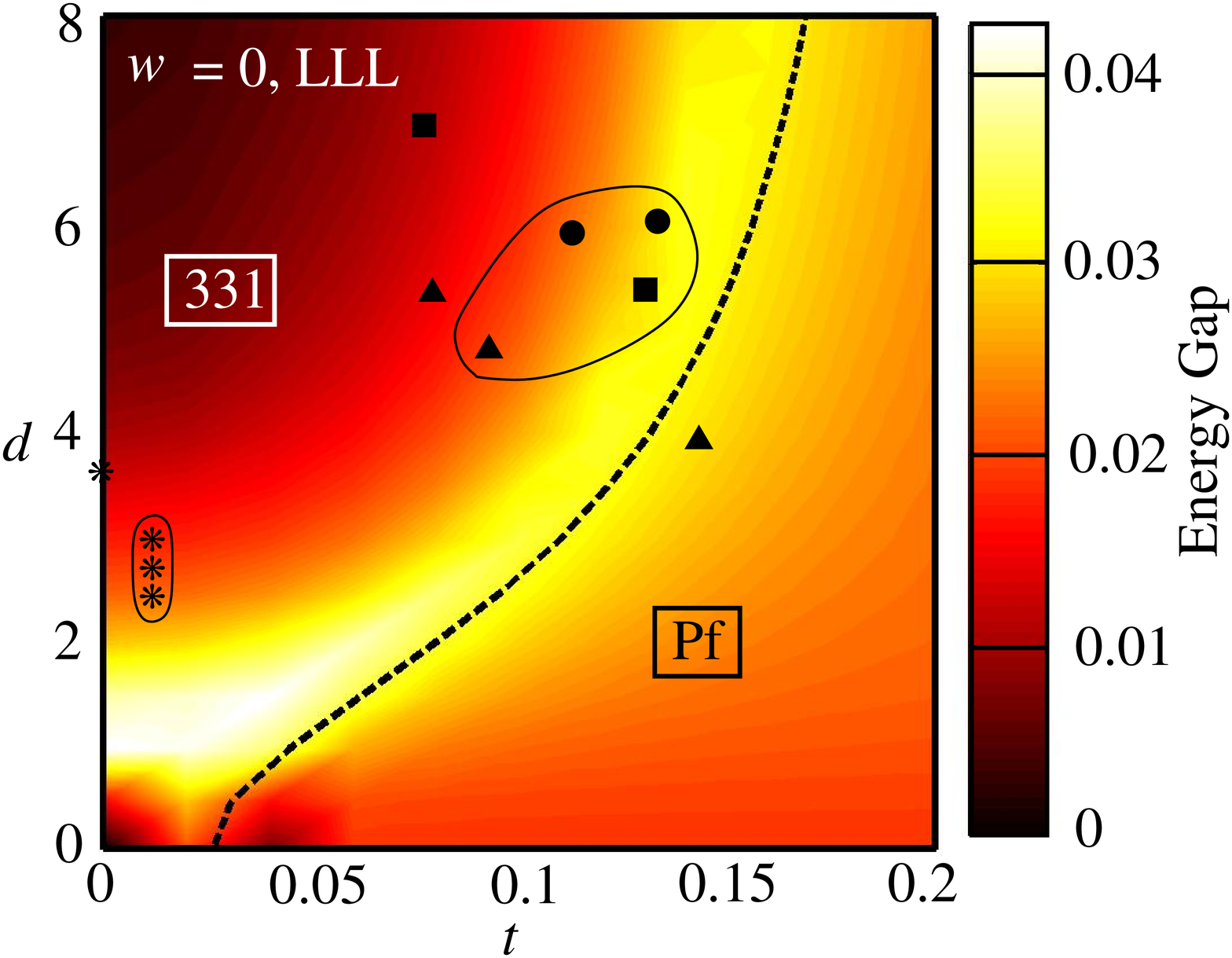}}
\mbox{(b)}
\mbox{\includegraphics[width=7.cm,angle=0]{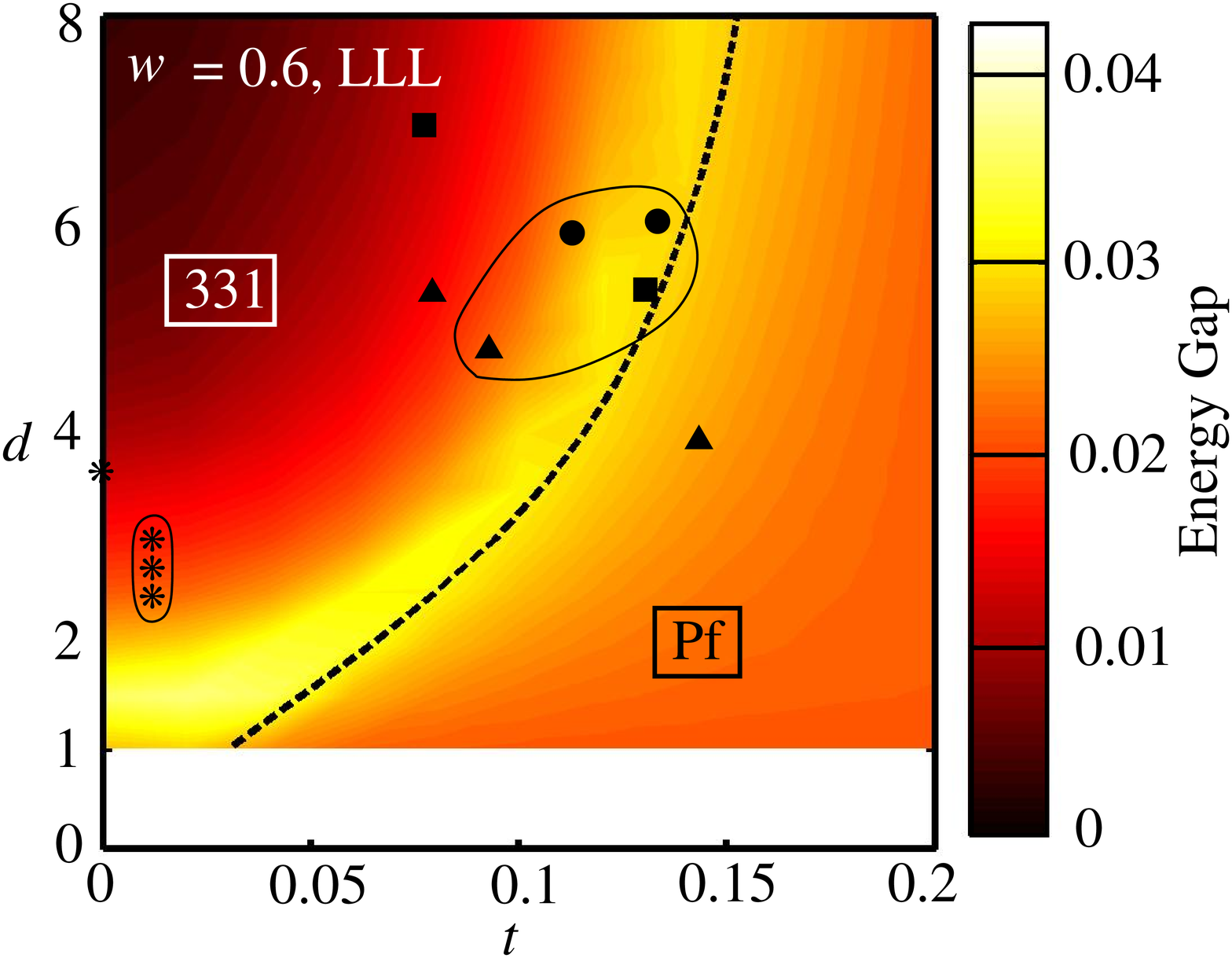}}\\
\mbox{(c)}
\mbox{\includegraphics[width=7.cm,angle=0]{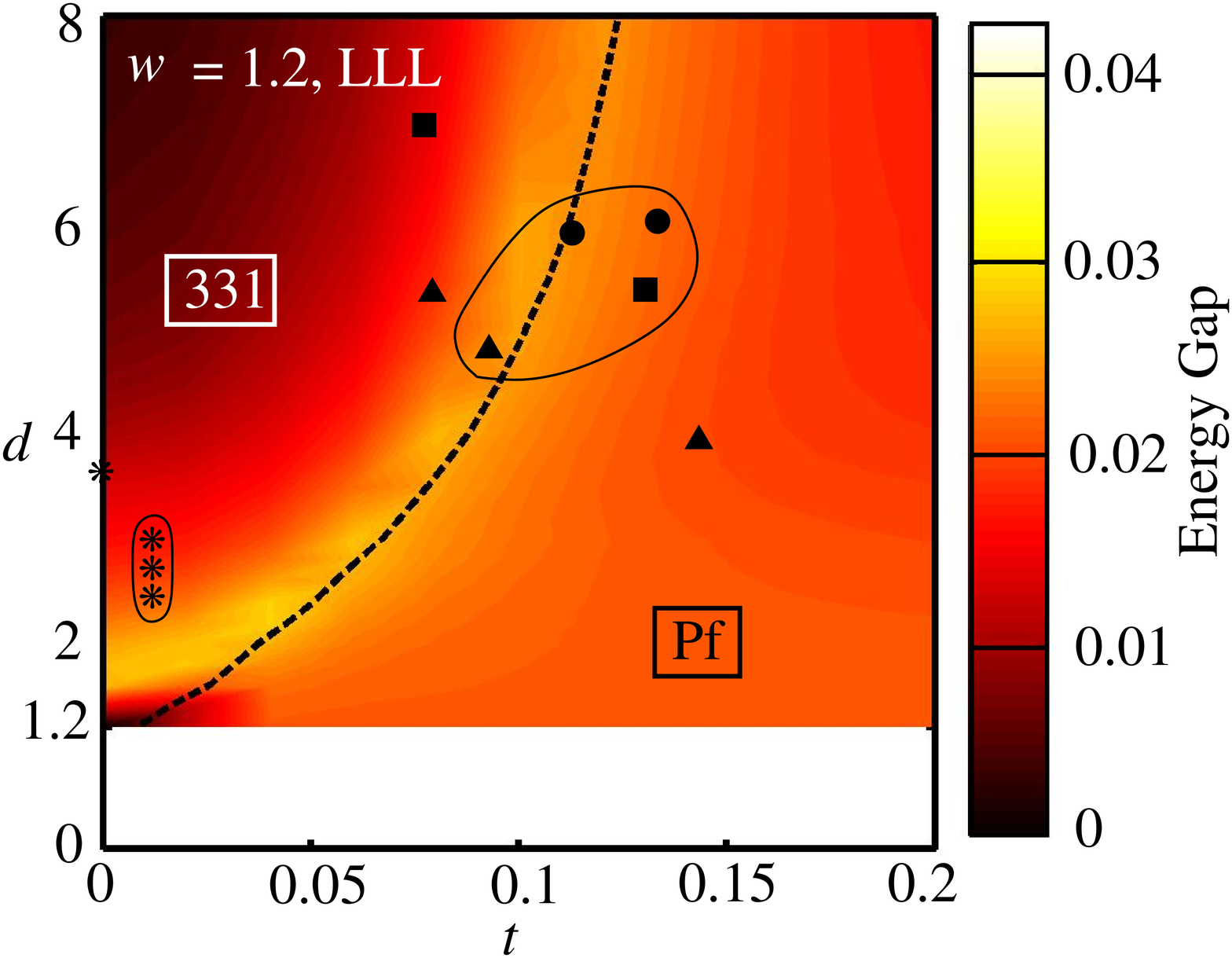}}
\mbox{(d)}
\mbox{\includegraphics[width=7.cm,angle=0]{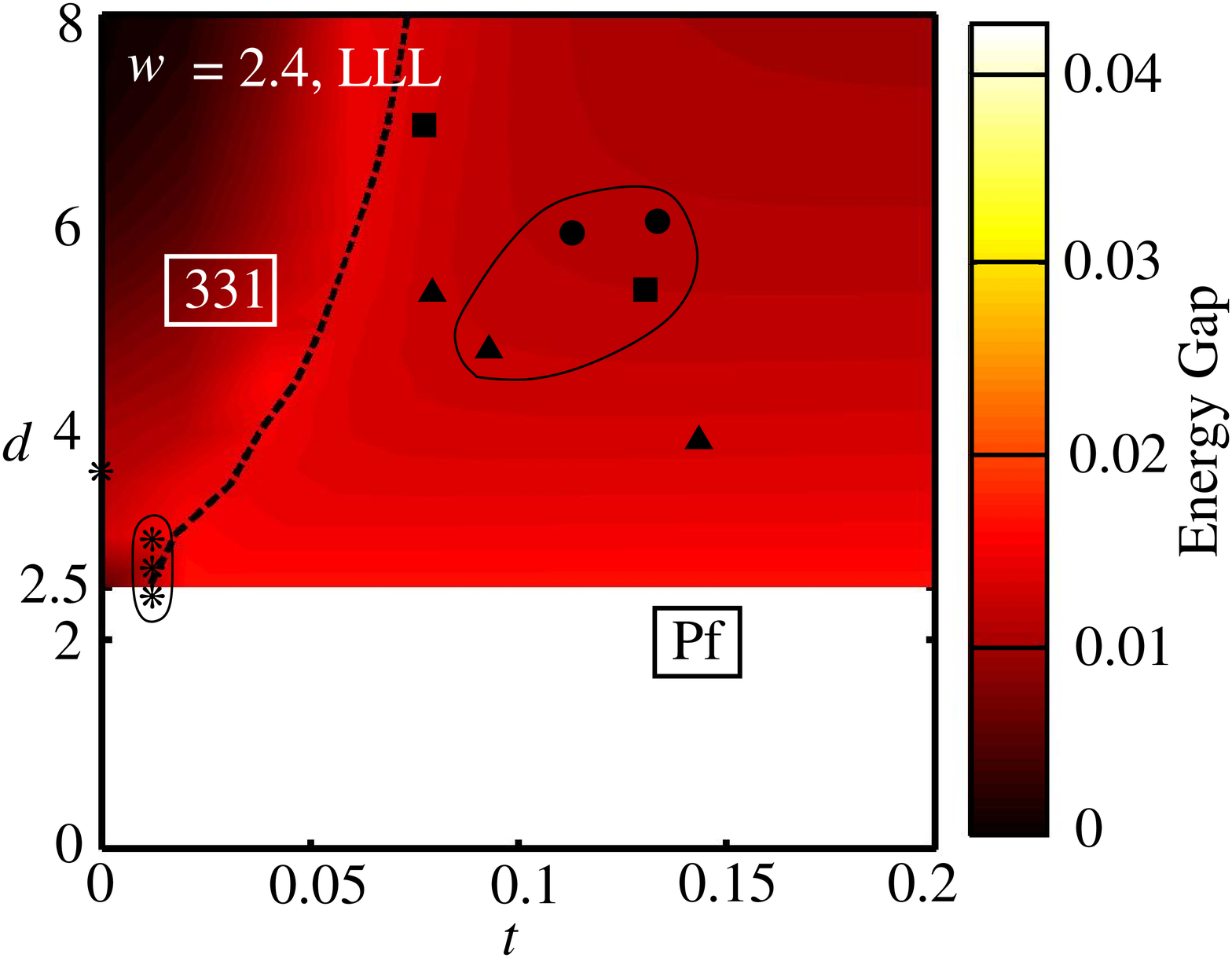}}
\end{center}
\caption{(color online) Quantum phase diagram (QPD) and FQHE gap
(color coded) versus layer separation $d$ and tunneling strength $t$
for widths (a) $w=0$, (b) $w=0.6$, (c) $w=1.2$, and (d) $w=2.4$.  For the QPD, the
331 and Pf phases (as discussed in the text) are separated by a dashed
black line and labeled appropriately.  The FQHE gap is given as a
contour plot with color coding given by the color-bar from dark to
light, i.e., white being a largest value of 0.0425 and black being value
of 0.  The asterisks, triangles, circles, and squares correspond to
the different experiments in
Refs.~\onlinecite{eisenstein},~\onlinecite{suen},~\onlinecite{suen-1},~\onlinecite{shayegan-new}
and~\onlinecite{luhman-2008}, respectively.  Only experimental points
showing FQHE are within the large solid circles in (a) with the lower
smaller circles and upper larger circles indicating experiments in
double-quantum-well structures and WQW structures, respectively.  We note that the single
triangle on the Pf side of the QPD does not manifest any experimental
FQHE indicating that the theoretical gap may be overestimated for the
Pf state.}
\label{fig-gap-LLL}
\end{figure*}

In Fig.~\ref{fig-gap-LLL} we project the energy gap onto the two-dimensional $t$-$d$ plane 
with the color coding indicating the numerical FQHE gap
strength on the same scale as in Fig.~\ref{fig_gap3D_LLL}.  Note that the Coulomb interaction is 
undefined for $d<w$ so no results are obtained in that case.  On this same plot we 
draw with a black dashed line the line that separates the 331 phase from the Pfaffian phase.  This 
phase boundary is determined by noting which state has a higher overlap with the exact state 
$\Psi_0$ in 
which region of parameter space, i.e., if $\langle \Psi_{331}|\Psi_0\rangle$ is 
larger than $\langle \Psi_\mathrm{Pf}|\Psi_0\rangle$ then we say the exact system is in the 
331 phase and vice-versa.  Note that the dashed line is only an operational phase
boundary within our calculation since all we know is that the 331 (Pf)
has higher (lower) overlap above (below) this line.  We cannot rule out the possibility that 
the in thermodynamic limit some other state (including possibly a compressible 
state) besides the $\Psi_{331}$ or $\Psi_\mathrm{Pf}$ dominates.

We show the approximate quantum 
phase diagram (QPD) for all four values of the layer width parameter (a) $w=0$,
(b) $w=0.6$, (c) $w=1.2$, and (d) $w=2.4$.  The zero- (Fig.~\ref{fig-gap-LLL}(a)) and the
intermediate-width (Fig.~\ref{fig-gap-LLL}(b)) results are of physical
relevance whereas the (unrealistically) large width results
(Fig.~\ref{fig-gap-LLL}(c) and (d)) are provided here only for completeness (since
this is the regime where the Pf state dominates over the 331 state in
the quantum phase diagram).  We note that we are using the simplistic Zhang-Das Sarma
(ZDS) model~\cite{zds} for describing the well width effect, and
crudely speaking $w=1$ in the ZDS model corresponds roughly to
$w_{QW}\approx 6$ where $w_{QW}$ is the corresponding physical (i.e.,
effective single-layer) quantum well width.  For a single WQW, where
the effective bilayer is created by the self-consistent potential of
the electrons themselves~\cite{luhman-2008,shayegan-new,suen,suen-1}, our
$w$ is typically much less than the total width $W$ of the
WQW--very roughly speaking $w\sim W/12$, and $d\sim W/2$.  As
emphasized above, we treat $t$, $d$, and $w(<d)$ as independent tuning
parameters.

\begin{figure*}[]
\begin{center}
\mbox{(a)}
\mbox{\includegraphics[width=6.cm,angle=0]{fig6a.eps}}
\mbox{(b)}
\mbox{\includegraphics[width=6.cm,angle=0]{fig6b.eps}}\\
\mbox{(c)}
\mbox{\includegraphics[width=6.cm,angle=0]{fig6c.eps}}
\mbox{(d)}
\mbox{\includegraphics[width=6.cm,angle=0]{fig6d.eps}}
\end{center}
\caption{(color online) FQHE energy gap versus tunneling strength $t$
for several values of layer separation $d$ for (a) $w=0$, (b) $w=0.6$, (c) $w=1.2$, 
and (d) $w=2.4$.  A dashed vertical line of the same color corresponds to the boundary
between the Pfaffian phase (right of the line) and the 331 phase (left of the line).}
\label{fig_gap_vs_D_LLL}
\end{figure*}

\begin{figure*}[]
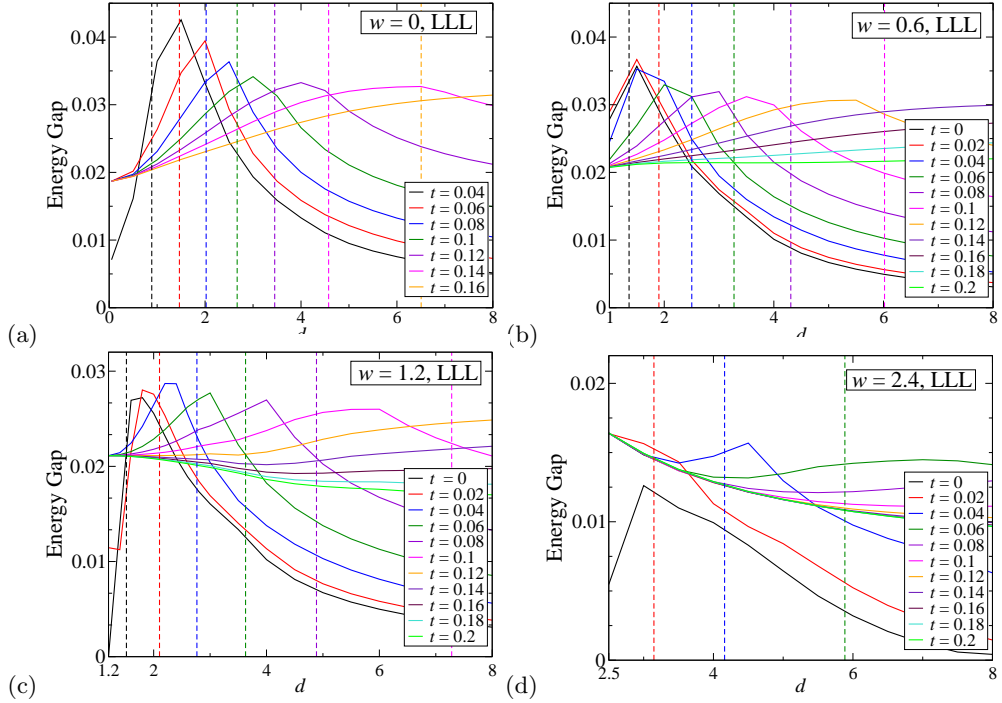

\begin{center}
\mbox{(a)}
\mbox{\includegraphics[width=6.cm,angle=0]{fig7a.eps}}
\mbox{(b)}
\mbox{\includegraphics[width=6.cm,angle=0]{fig7b.eps}}\\
\mbox{(c)}
\mbox{\includegraphics[width=6.cm,angle=0]{fig7c.eps}}
\mbox{(d)}
\mbox{\includegraphics[width=6.cm,angle=0]{fig7d.eps}}
\end{center}
\caption{(color online) FQHE energy gap versus layer separation $d$
for a few values of   tunneling strength $t$ for (a) $w=0$, (b) $w=0.6$, (c) $w=1.2$, 
and (d) $w=2.4$.  A dashed vertical corresponds to the phase boundary
between the Pfaffian phase (left of the line) and the 331 phase (right of the line).}
\label{fig_gap_vs_d_LLL}
\end{figure*}

In Fig.~\ref{fig-gap-LLL}, we have put as discrete symbols all existing
$\nu=1/2$ bilayer published FQHE experimental data (both for double quantum
well systems and single wide-quantum-wells) in the literature,
extracting the relevant parameter values (i.e., $d$ and $t$) from the
experimental works~\cite{eisenstein,suen,suen-1,luhman-2008,shayegan-new}.
Because of the ambiguity and uncertainty in the definition of $w$
(i.e., how to precisely relate our theoretical $w$ in the Zhang-Das
Sarma model to the experimental layer width in real samples), we have
put the data points on all four QPDs shown in Fig.~\ref{fig-gap-LLL}
although the actual experimental width values correspond to only
Figs.~\ref{fig-gap-LLL}(a) and (b).  See Sec.~\ref{sec-app} for a detailed description of exactly 
how the experimental points were determined.

Results shown in Fig.~\ref{fig-gap-LLL} bring out several important points of
physics not clearly appreciated earlier in spite of a great deal of
theoretical exact diagonalization work on $\nu=1/2$ bilayer FQHE: \\
(i) It is obvious that large, or small, $t$
and small, or large, $d$, in general, lead to a decisive preference for
the existence of $\nu=1/2$ Pf, or 331, FQHE.  The fact that large $t$
values would preferentially lead to the Pf state over the 331 state
is, of course, expected since the system becomes an effective
one-component system for large tunneling strength.  \\
(ii) What is,
however, not obvious, but apparent from the QPDs shown in
Fig.~\ref{fig-gap-LLL}, is that the FQHE gap (given in color coding in the
figures) is maximum near the phase boundary between  331 and Pf.  \\
(iii) Another non-obvious result is the persistence of the 331
state for very large (essentially arbitrarily large!)  values of the
tunneling strength $t$ as long as the layer separation $d$ is also
large--thus having a large $t$ by itself, as achieved in the Luhman
\textit{et al}. experiment~\cite{luhman-2008}, is not enough to realize
the single-layer $\nu=1/2$ Pf FQHE, one must also have a relatively
small value of layer separation $d$ so that one is below the phase
boundary (dashed line) in Fig.~\ref{fig-gap-LLL}.  The explanation for the
Luhman experimental $\nu=1/2$ FQHE being a 331 sate, as can be seen in
Fig.~\ref{fig-gap-LLL}, is indeed the fact that both $t$ and $d$ are large in
these samples making 331 a good variational state.  \\
(iv) An important
aspect of Fig.~\ref{fig-gap-LLL} is that the Pf FQHE gap tends to be very
small--this is particularly true for larger values of $w$, where the
Pf overlap is large.  This implies, as emphasized by Storni
\textit{et al}.~\cite{storni}, that the observation of a $\nu=1/2$ Pf
state is unlikely since the activation
gap would be extremely (perhaps even vanishingly) small.\\
(v) For larger values of $w$ (and large $t$), our calculated QPD is dominated
by the Pf state--particularly for the unrealistically large width
$w=2.4$ (corresponding to $w_{QW}\sim 14$!) where all the experimental
$d$ and $t$ values fall in the Pf regime of the phase diagram.  We
emphasize, however, that this Pf-dominated large-$w$ (and large-$t$)
regime will be difficult to access
experimentally since the FQHE gap would be likely extremely small 
as in Fig.~\ref{fig-gap-LLL}.  Our results however cannot decisively rule 
out the possibility of a $\nu=1/2$ Pfaffian state in the strong tunneling 
and small separation regime.

We now discuss the published experimental results in light of our
theoretical QPD.  First, we note that most of the existing
experimental points fall on the 331 side of the phase diagram which is
consistent with our QPD in Fig.~\ref{fig-gap-LLL}.  In particular, only 
samples on the 331 side of the QPD with reasonably large FQH
gaps, i.e., the data points close to the phase boundary, exhibit
experimental FQHE.  By contrast, the one data point (solid triangle in
Figs.~\ref{fig-gap-LLL}(a) and (b)) on the Pf side of the phase boundary does
not manifest any observable FQHE in spite of its location being in a
regime of reasonable FQHE excitation gap according to our phase
diagram.  This is consistent with the finding of Storni \textit{et
al.}~\cite{storni} that the $\nu=1/2$ FQHE gap in a single-layer
system is likely to be vanishingly small in the thermodynamic limit.
It is, therefore, possible that the Pf regime in our QPD has a
much smaller excitation gap than what we obtain on the basis of our
$N=8$ particle diagonalization calculation.  We refer to Storni
\textit{et al.}~\cite{storni} for more details on the theoretical
status of the single-layer LLL $\nu=1/2$ FQHE.

For a more detailed view of the $\nu=1/2$ bilayer FQHE, we show in
Figs.~\ref{fig_gap_vs_D_LLL}(a)-(d) and~\ref{fig_gap_vs_d_LLL}(a)-(d), 
respectively, our calculated 
FQHE gap as a function of $t$ (for a few fixed $d$ values) and as a function of
$d$ (for a few fixed $t$ values).   Note that similar results were obtained by Nomura and Yoshioka 
in Ref.~\onlinecite{nomura}, however, they only 
considered the energy gap versus tunneling for two 
values of separation $d$ and fixed $w=3.8$ for a $N=6$ electron system, which  
is aliased with a possible FQHE at $\nu=2/3$ and therefore suspect.   
In each figure, we also depict the
line separating the 331 (smaller $t$/larger $d$) and the Pf (larger
$t$/smaller $d$) regimes in the phase diagram.  The qualitatively
interesting point is, of course, the non-monotonicity in the FQHE
gap as a function of $t$ or $d$ with a maximum close (but always on
the 331 side) to the phase boundary.  The non-monotonicity in the FQHE
gap as a function of $t$ (but not $d$) was earlier pointed out, but
our finding that the peak lies \textit{always} on the 331 side of the
phase boundary is a new result.  We emphasize that this result is strong evidence
that the 331 phase is the dominant FQHE phase in $\nu=1/2$
systems.  We believe that the only chance of observing the $\nu=1/2$
Pf FQHE is to look on the Pf side of the phase boundary at
fairly large values of $d$ and $t$.  This is in sharp contrast to the
SLL $\nu=5/2$ bilayer FQHE where we show in Section~\ref{sec-SLL} that there are two
sharp ridges far away from each other in the $d$-$t$ space
corresponding to the $\nu=5/2$ Pf and 331 bilayer
phases~\cite{mrp-sds-sll-bilayer}. 

\begin{figure*}[t]
\begin{center}
\mbox{(a)}
\mbox{\includegraphics[width=7.cm,angle=0]{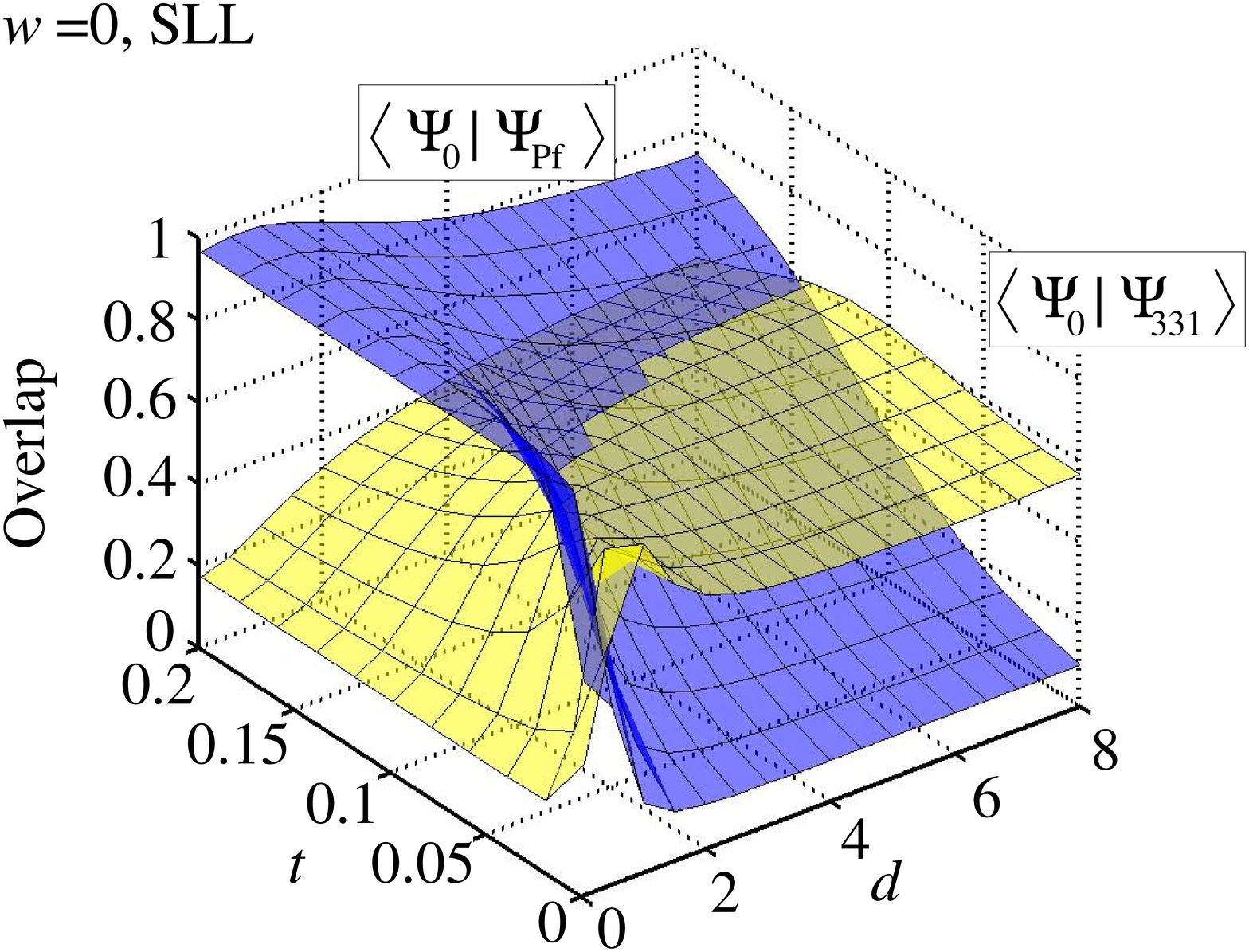}}
\mbox{(b)}
\mbox{\includegraphics[width=7.cm,angle=0]{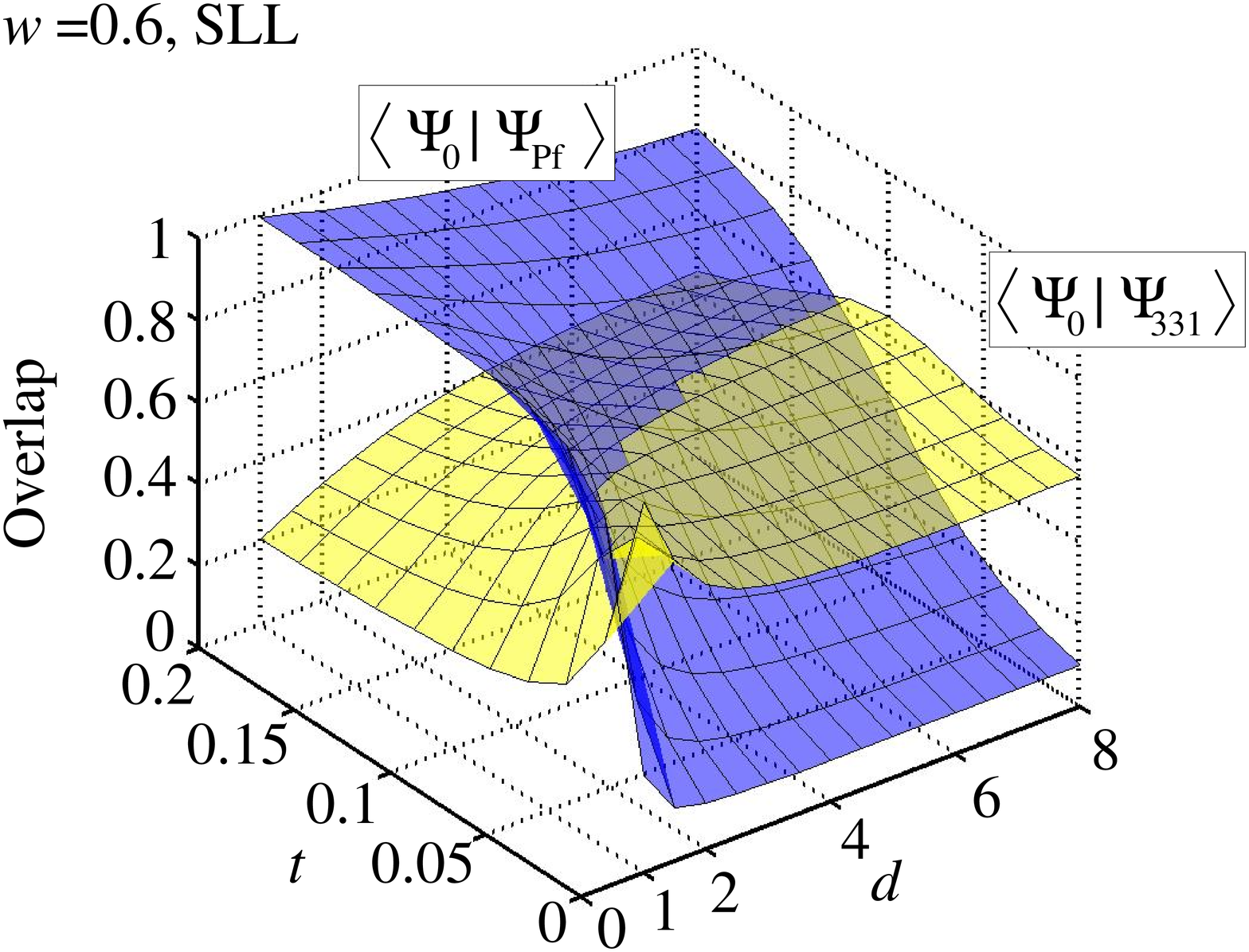}}\\
\mbox{(c)}
\mbox{\includegraphics[width=7.cm,angle=0]{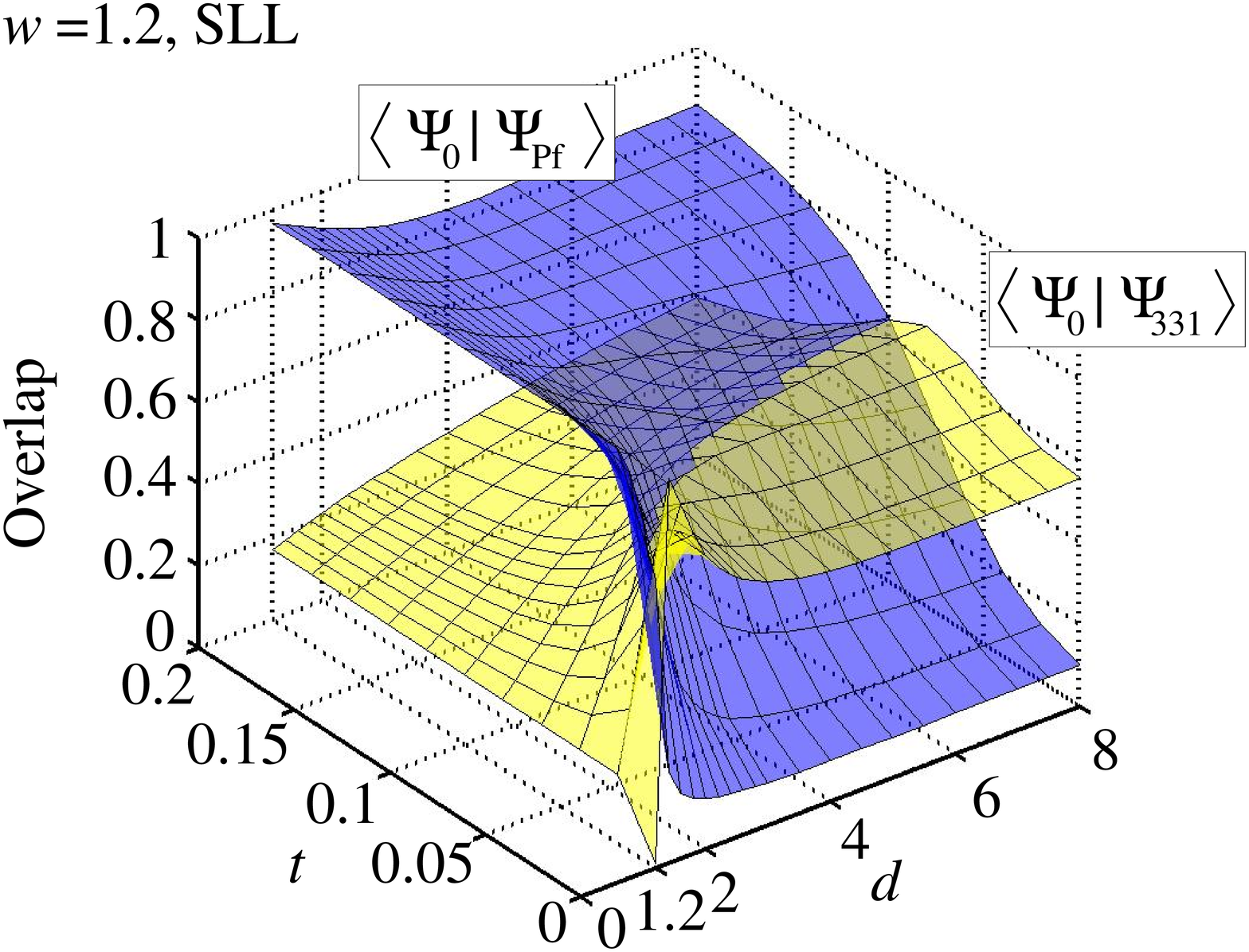}}
\mbox{(d)}
\mbox{\includegraphics[width=7.cm,angle=0]{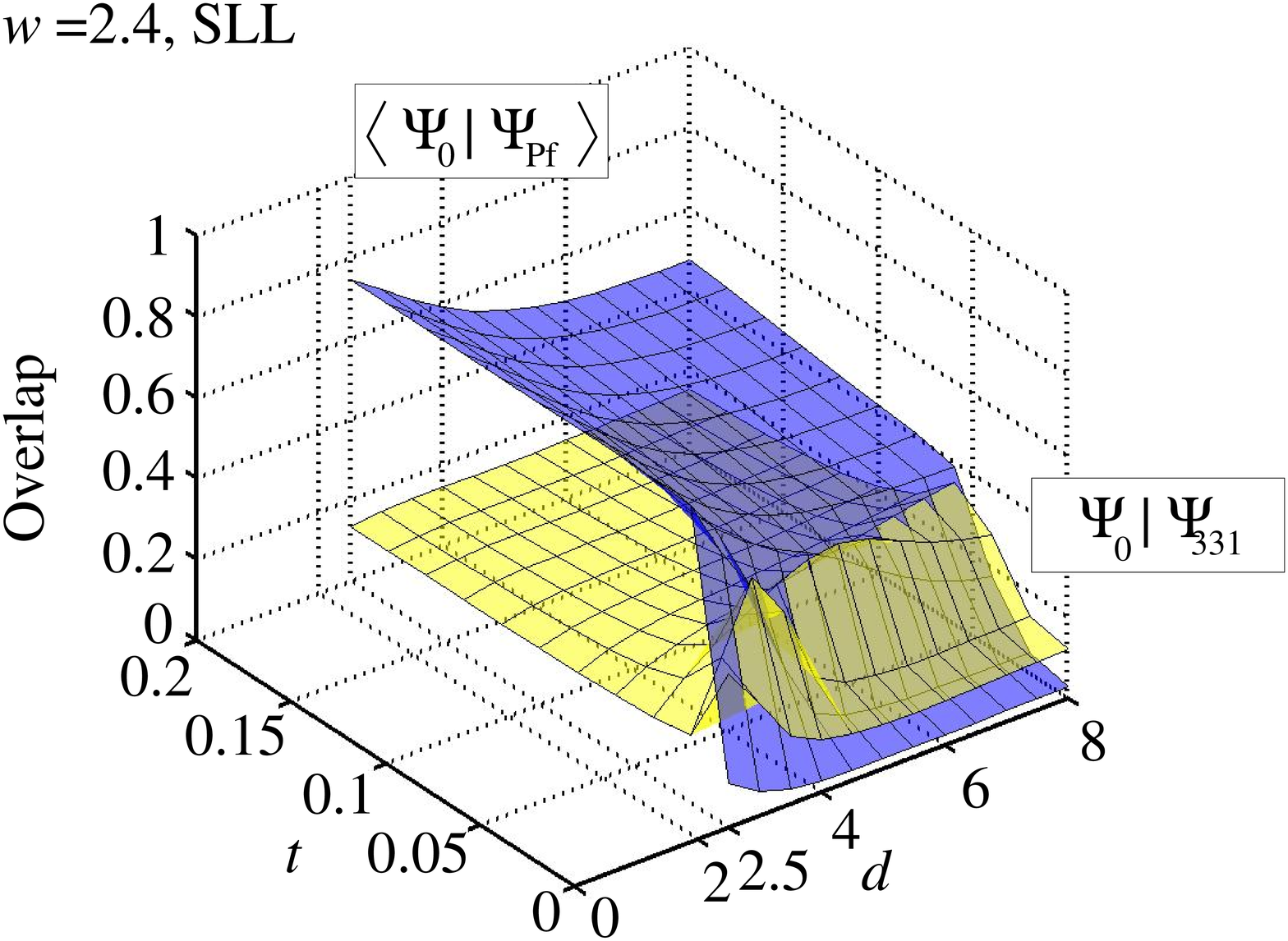}}
\end{center}
\caption{(Color online) Wavefunction overlap between the exact ground
state $\Psi_0$ and the Pf state $\Psi_{\mathrm{Pf}}$ (blue (dark
gray)) and 331 state $\Psi_{331}$ (yellow (light gray)) as a function
of distance $d$ and tunneling amplitude $t$ for the half-filled
second LL with $N=8$ electrons and width (a) $w=0$, (b) $w=0.6$, (c) $w=1.2$, 
and (d) $w=2.4$ ($d>w$ necessarily).}
\label{fig-o-SLL}
\end{figure*}

We conclude this section by commenting on the nature of the quantum phase
transition (QPT) between the 331 and the Pf phase in the $d$-$t$
space.  Our calculated QPD implies a continuous QPT from the
strong-pairing 331 to the weak-pairing Pf state with 
increasing $t$ and/or decreasing $d$ as predicted by Read and Green~\cite{RG2000}.
(Note this has also been very recently discussed by Dimov, Halperin, and 
Nayak~\cite{dimov}.)
In fact, our finite size diagonalization based QPD of Fig.~\ref{fig-gap-LLL}
is topologically equivalent to the phase diagram predicted by Read and
Green (Fig. 1 of Ref.~\onlinecite{RG2000}).  
We emphasize, however, that we cannot distinguish a quantum
phase transition from a crossover because of the limitations of finite
size calculations.  It is also possible that a different phase, e.g. a
compressible composite fermion Fermi liquid phase, has lower energy
and intervenes between the 331 and Pf phases so that the system goes
from 331 to Pf (or vice versa) through two first-order transitions.
What we have shown here is that if the $\nu=1/2$ bilayer Pf phase
exists at all, it would manifest most strongly in very wide samples
and close to the phase boundary with the 331 phase with an extremely
small FQHE excitation gap.  We have also shown, through an explicit
comparison with our $t$-$d$-$w$ phase diagram (Fig.~\ref{fig-gap-LLL}) that
all published $\nu=1/2$ FQHE
data~\cite{luhman-2008,shayegan-new,suen,suen-1,eisenstein} are consistent
with the existence of \textit{only} the 331 phase in the LLL.  Our work does not 
rule out the possibility of a weak Pf FQHE at $\nu=1/2$ for large values 
of tunneling.

\section{Second Landau Level}
\label{sec-SLL}

\begin{figure*}[t]
\begin{center}
\mbox{(a)}
\mbox{\includegraphics[width=7.cm,angle=0]{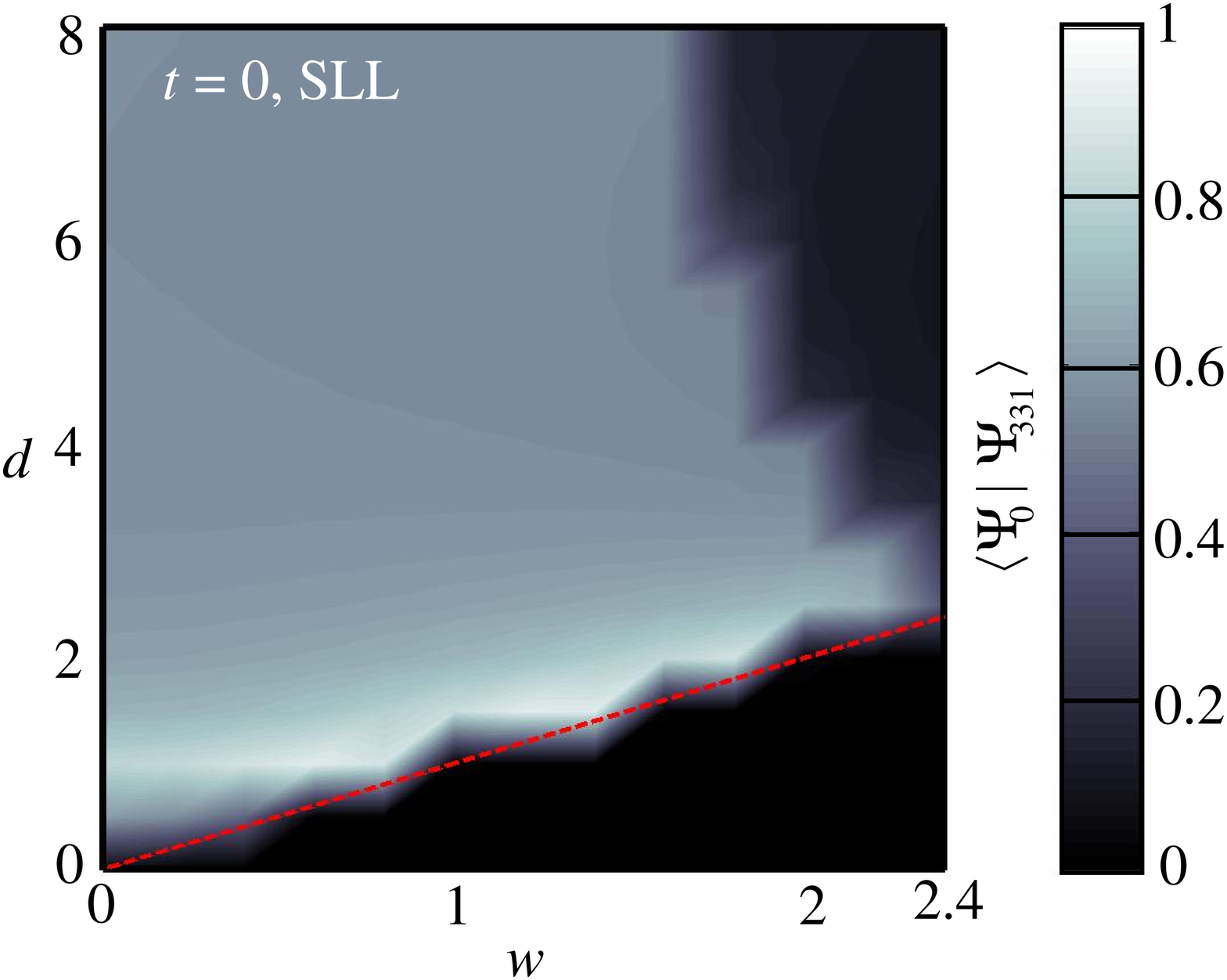}}
\mbox{(b)}
\mbox{\includegraphics[width=7.cm,angle=0]{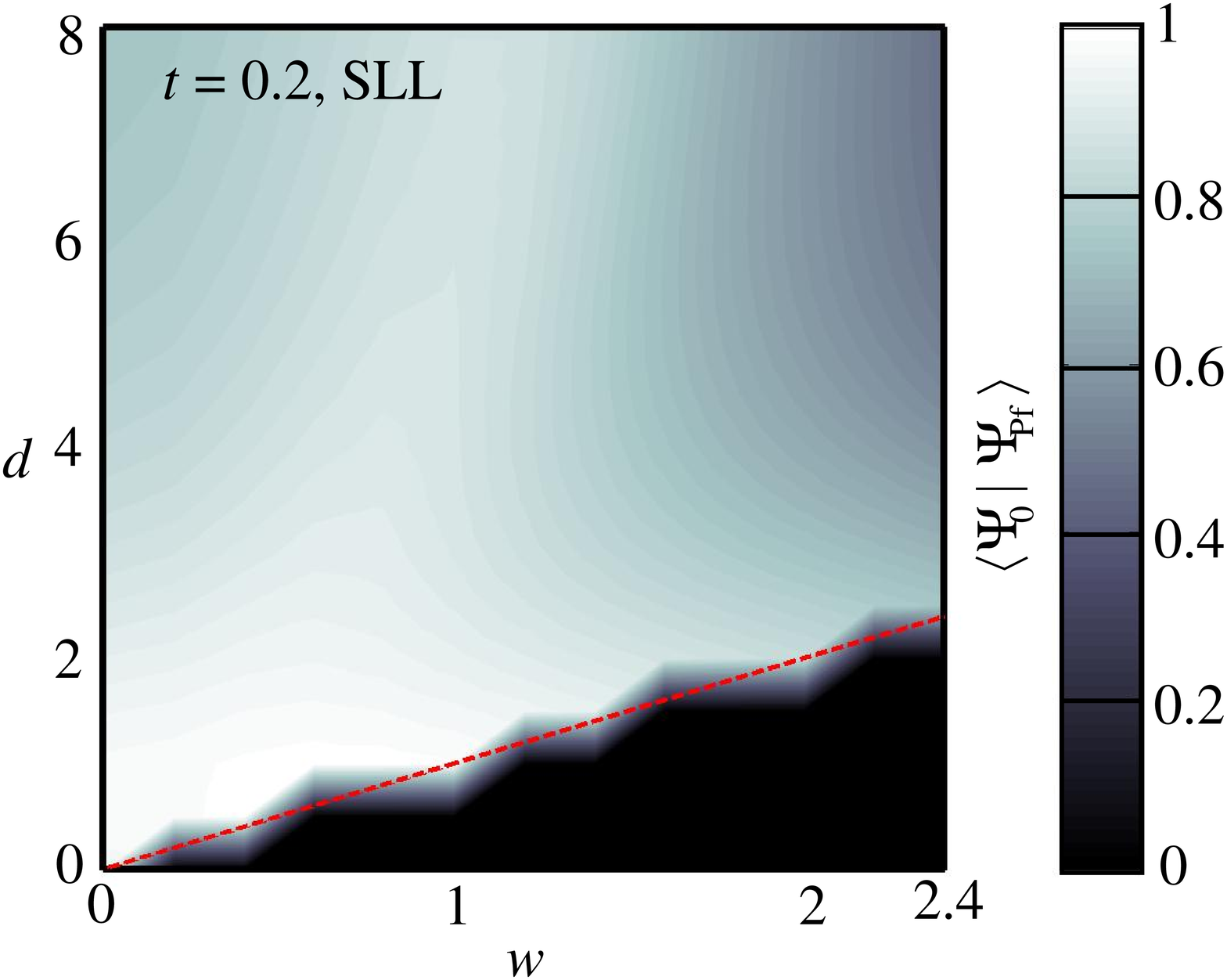}}
\end{center}
\caption{(Color online)  Wavefunction overlap between the exact ground
state $\Psi_0$ and $\Psi_{\mathrm{Pf}}$ at strong
tunneling $t=0.2$ (a) and $\Psi_{331}$ at zero tunneling
$t=0$ (b) as a function of separation $d$ and single-layer well width
$w$ (where $d\geq w$) for the SLL with $N=8$
electrons.  White corresponds to an overlap of unity while black
corresponds to an overlap of zero.  As in Fig.~\ref{fig-o-LLL-w}, the  
dashed red line is the condition $w=d$ and 
for $w>d$ the bilayer system is undefined, i.e., the single-layer width cannot be larger than the 
layer separation.}
\label{fig-o-SLL-w}
\end{figure*}

We now consider the same set of questions we asked regarding the physics 
of FQHE bilayers in the $\nu=1/2$ lowest LL (Sec.~\ref{sec-LLL}) about the bilayer FQHE 
systems in the second Landau level (SLL).  Our approach here is to theoretically consider 
a single-layer FQHE at a half-filled SLL system, i.e., $\nu=2+1/2=5/2$ where the 
first two lowest Landau levels of spin-up and spin-down are completely 
occupied and inert.  Thus, the half-filled SLL interacting electrons are projected 
into the LLL using the appropriate SLL Haldane pseudopotentials~\cite{hald}.  The 
exact nature of this procedure has been given many times in many places in 
great detail and will not be reiterated here (see Ref.~\onlinecite{cf-book} for a good 
description).  To consider a SLL bilayer system we then allow the 
single-layer to become a bilayer 
by reducing the tunneling and/or increasing the layer separation $d$.  There are actually 
quite subtle points in defining this procedure in the SLL and they are
discussed below in Sec.~\ref{subsec-bilayer-SLL} in detail.  However, 
our procedure is completely well defined theoretically, and is a direct analog of 
the $\nu=1/2$ LL calculation in the SLL.

\subsection{What is the physics?--second Landau level}
\label{subsec-o-SLL}

\begin{figure*}[t]
\begin{center}
\mbox{(a)}
\mbox{\includegraphics[width=6.cm,angle=0]{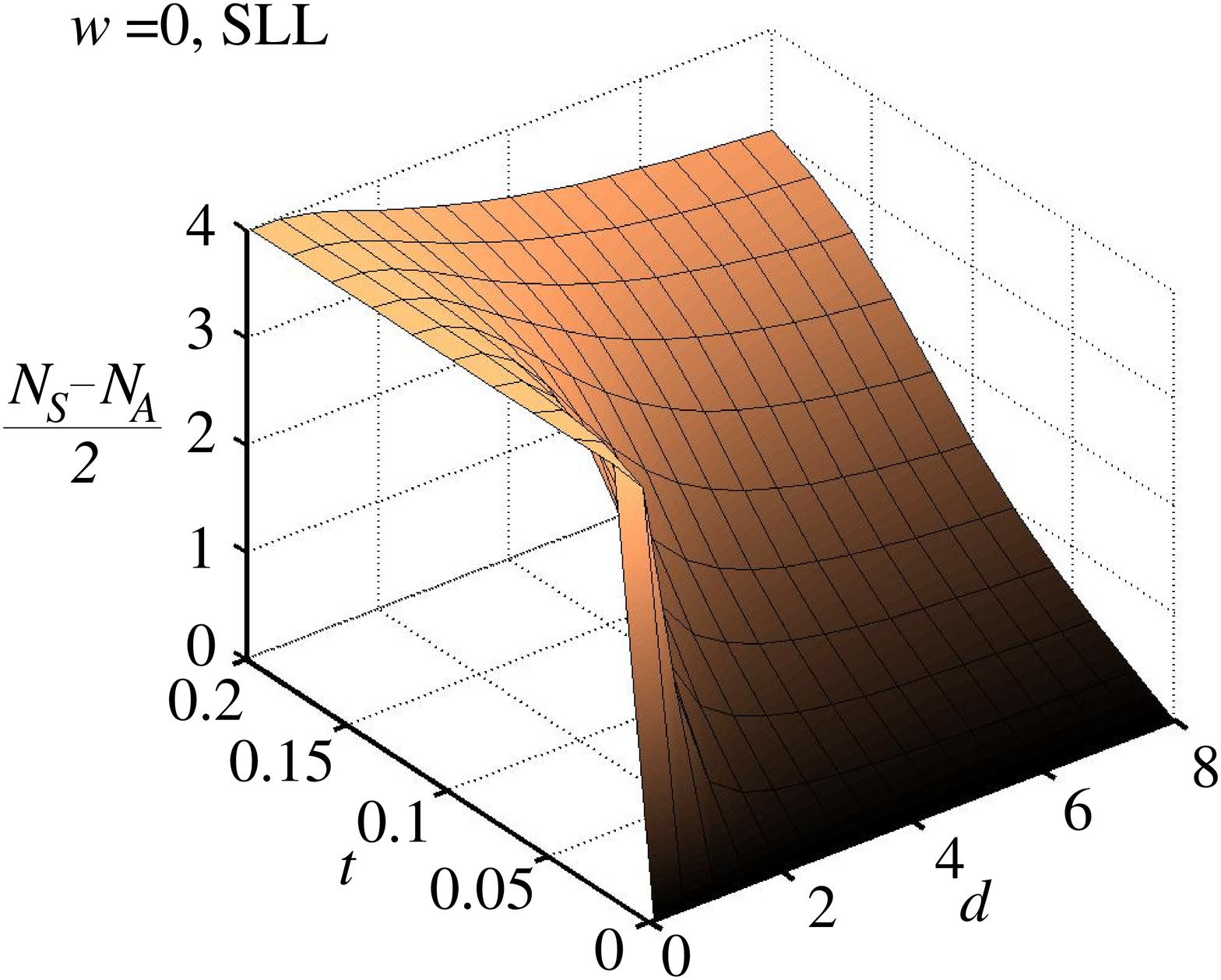}}
\mbox{(b)}
\mbox{\includegraphics[width=6.cm,angle=0]{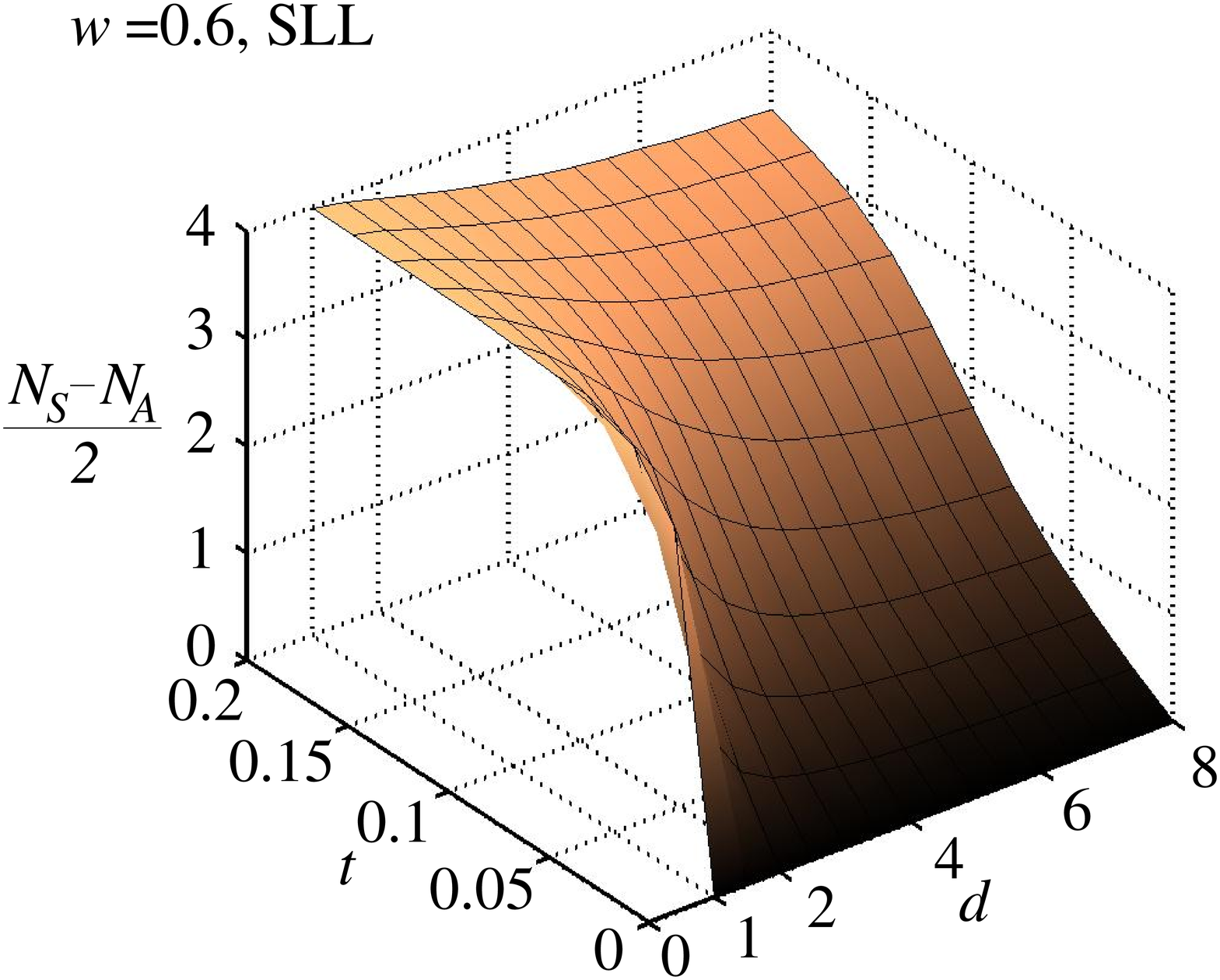}}\\
\mbox{(c)}
\mbox{\includegraphics[width=6.cm,angle=0]{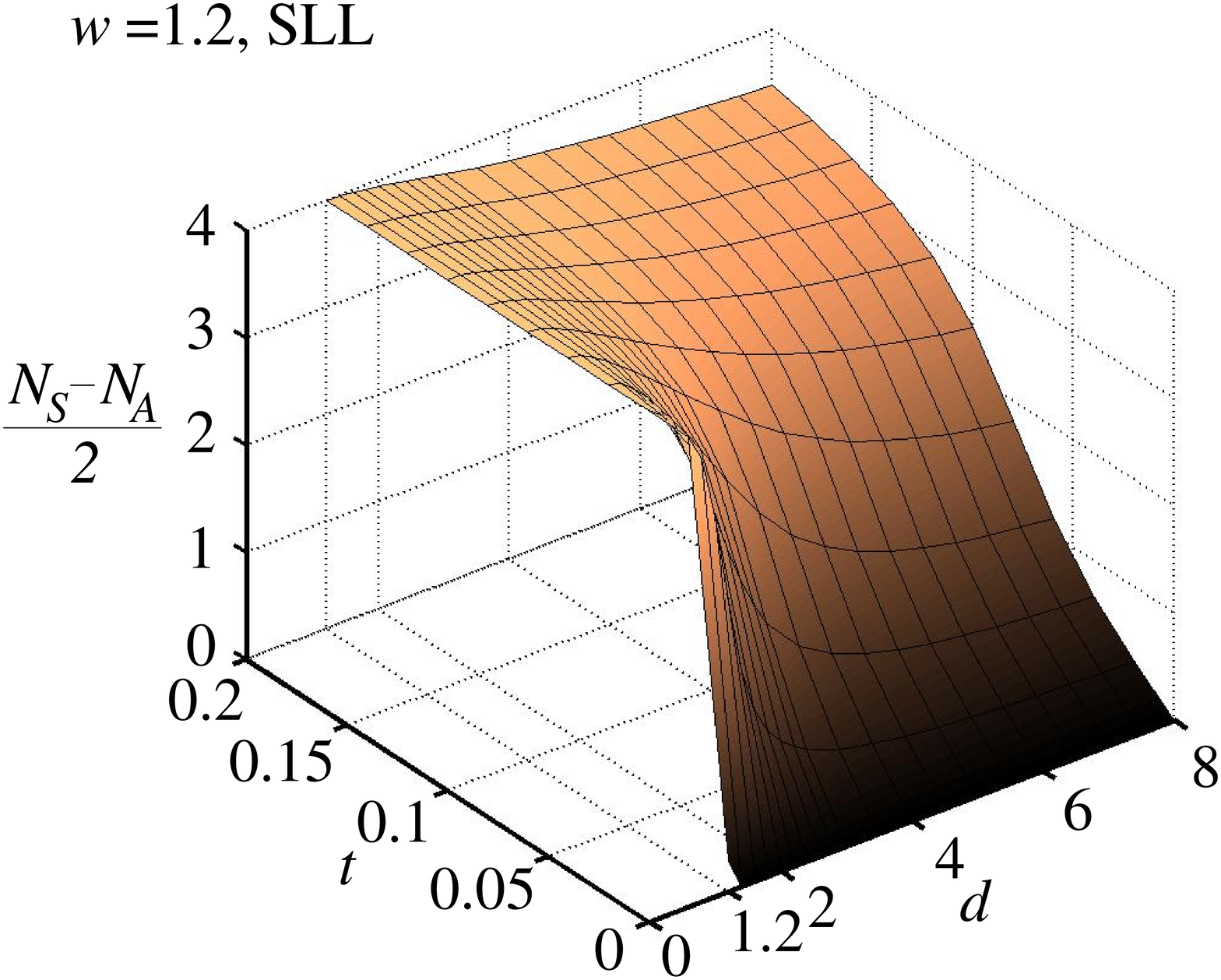}}
\mbox{(d)}
\mbox{\includegraphics[width=6.cm,angle=0]{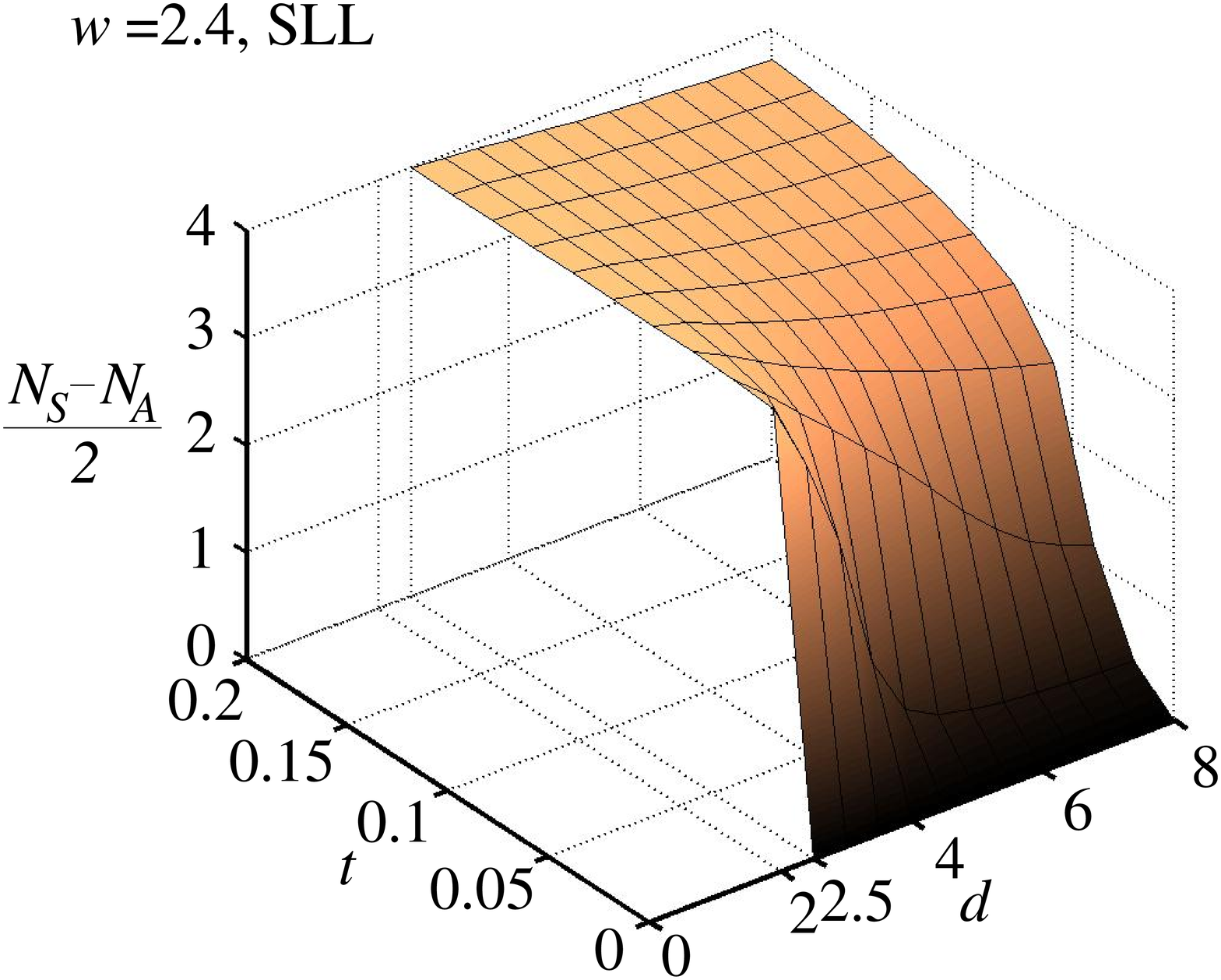}}
\end{center}
\caption{(Color online) $(N_S-N_A)/2$ as a function of distance
$d$ and tunneling amplitude $t$ for the half-filled second Landau level, $N=8$ electrons, and 
(a) $w=0$, (b) $w=0.6$, (c) $w=1.2$, and (d) $w=2.4$. }
\label{fig-Sz-SLL}
\end{figure*}

The calculated wavefunction overlap
between the exact ground state $\Psi_0$ and the two appropriate
candidate variational wavefunctions ($\Psi_{\mathrm{Pf}}$ and
$\Psi_{331}$) as a function of distance $d$ and tunneling energy
$t$ is shown in Fig.~\ref{fig-o-SLL} for single-layer widths of $w=0$, $w=0.6$, $w=1.2$, 
and $w=2.4$.  First we focus on the situation
with zero width $w=0$ (Fig.~\ref{fig-o-SLL}(a)) and concentrate on the
overlap with $\Psi_\mathrm{Pf}$.  In the limit of zero tunneling and
zero $d$ the overlap with $\Psi_\mathrm{Pf}$ is small--approximately
0.5 (unclear from the figure).  However, only weak tunneling is
required to produce a state with a sizable overlap of approximately
$\sim0.96$ and as tunneling increases this overlap remains
large and approximately constant.  For $d\neq 0$, in the weak
tunneling limit, the overlap with $\Psi_\mathrm{Pf}$ decreases
drastically.  Adding moderate to strong tunneling we obtain a sizable
overlap, decreasing gently in the large $d$ limit.  This shows that
the strong-tunneling (one-component) regime is well described by the
Pf state.  This strong tunneling Pf regime appears to be robust.

Next we consider the overlap between $\Psi_0$ and $\Psi_{331}$.  In
the zero tunneling limit, as a function of $d$, the overlap
starts small, increases to a moderate maximum of $\approx 0.80$ at
$d\sim1$ before achieving an essentially constant value of
$\approx0.56$.  For $d>4$ the overlap remains relatively constant and
slowly decreases as the tunneling is increased (in fact, there is a
slight increase in the overlap to $\sim0.6$ for a region of positive
$d>4$ and $0.1\lesssim t\lesssim 0.15$).  Thus, the weak-tunneling
(two-component) regime is well described by $\Psi_{331}$.

Similar to what we did in the LLL, we consider the effects of finite width, knowing that it 
may enhance the overlap~\cite{mrp-tj-sds-prl,mrp-tj-sds-prb}.  In Fig.~\ref{fig-o-SLL-w} (a) 
and (b) we 
show $\langle \Psi_0|\Psi_{331}\rangle$  and $\langle \Psi_0|\Psi_
\mathrm{Pf}\rangle$, respectively, versus layer separation $d$ and single layer width $w$ for 
small tunneling 
$t=0$ and large tunneling $t=0.2$.  
In the zero tunneling limit (Fig.~\ref{fig-o-SLL-w}(a)) we see that 
the overlap can be increased significantly ($\sim0.94$) by using $d\approx1$ and $w\approx d$. 
The overlap between the exact state and the 331 state remains high with increasing width as 
long as the layer separation $d\approx 1$--this compares with the LLL case where the overlap is 
maximum for a larger separation near 1.8.  However, the general feature that the maximum 
does not change appreciably with increasing $w$ is consistent with the LLL results.  The 
main difference is the overlap with 331 is lower, and increasing $w$ and $d$ eventually 
pushes the overlap to zero. 

For the strong tunneling limit ($t=0.2$), 
in the case of $\Psi_\mathrm{Pf}$ (Fig.~\ref{fig-o-LLL-w}(b)),
our results  are consistent with the single-
layer finite-thickness results (cf. Refs.~\onlinecite{mrp-tj-sds-prl} and~\onlinecite{mrp-tj-sds-prb}).  
Compared to the LLL, 
however, the overlap increases to a maximum for a finite $w$ and that behavior continues 
to hold when $d$ is increased.  Eventually, large single-layer width causes the overlap 
with the Pfaffian state to decrease for any $d$ significantly larger than $w$.  
Thus, finite layer width enhances the overlap with both 331 and Pf states in their
regimes of phase space.

\begin{figure*}[]
\begin{center}
\mbox{(a)}
\mbox{\includegraphics[width=7.cm,angle=0]{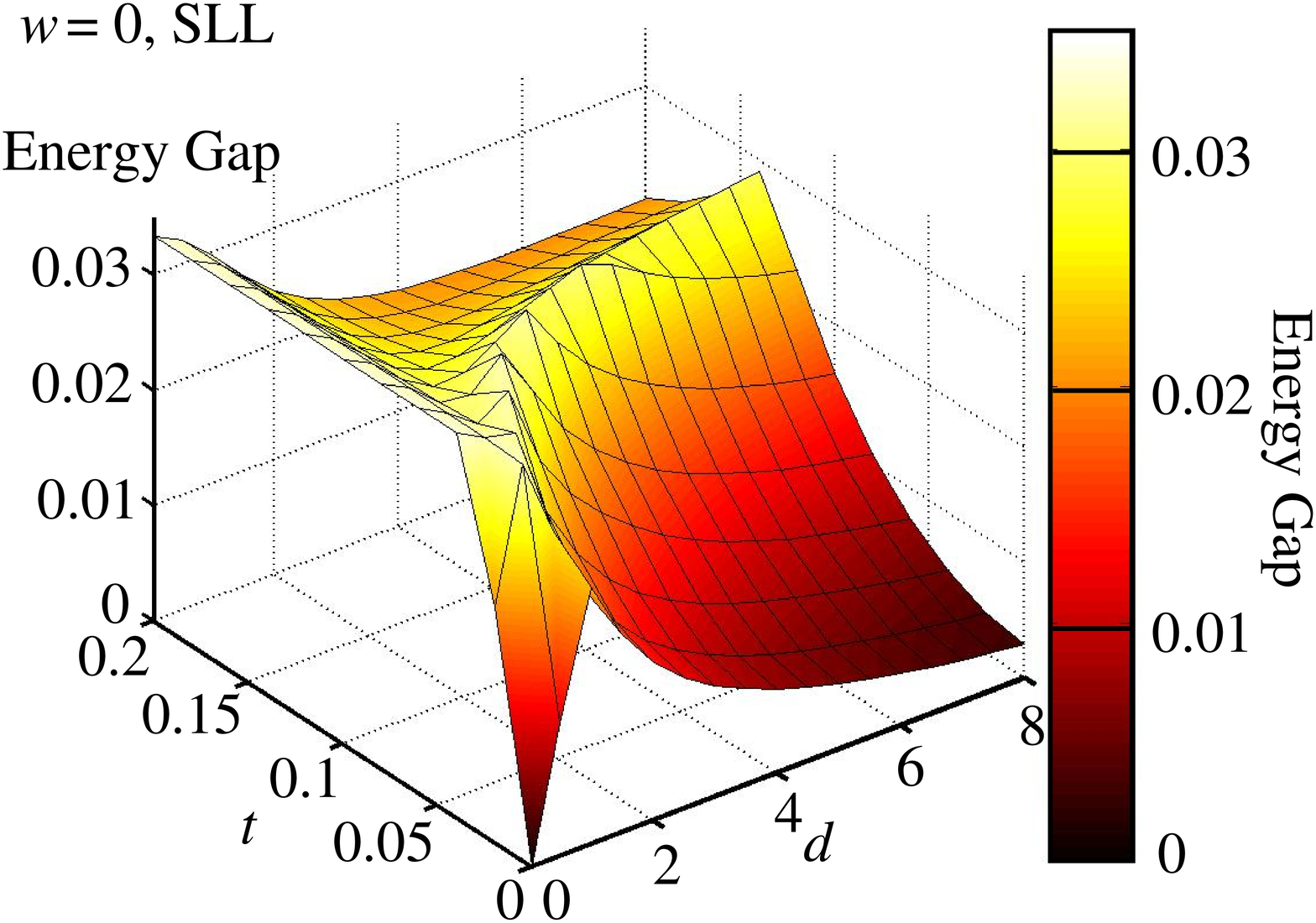}}
\mbox{(b)}
\mbox{\includegraphics[width=7.cm,angle=0]{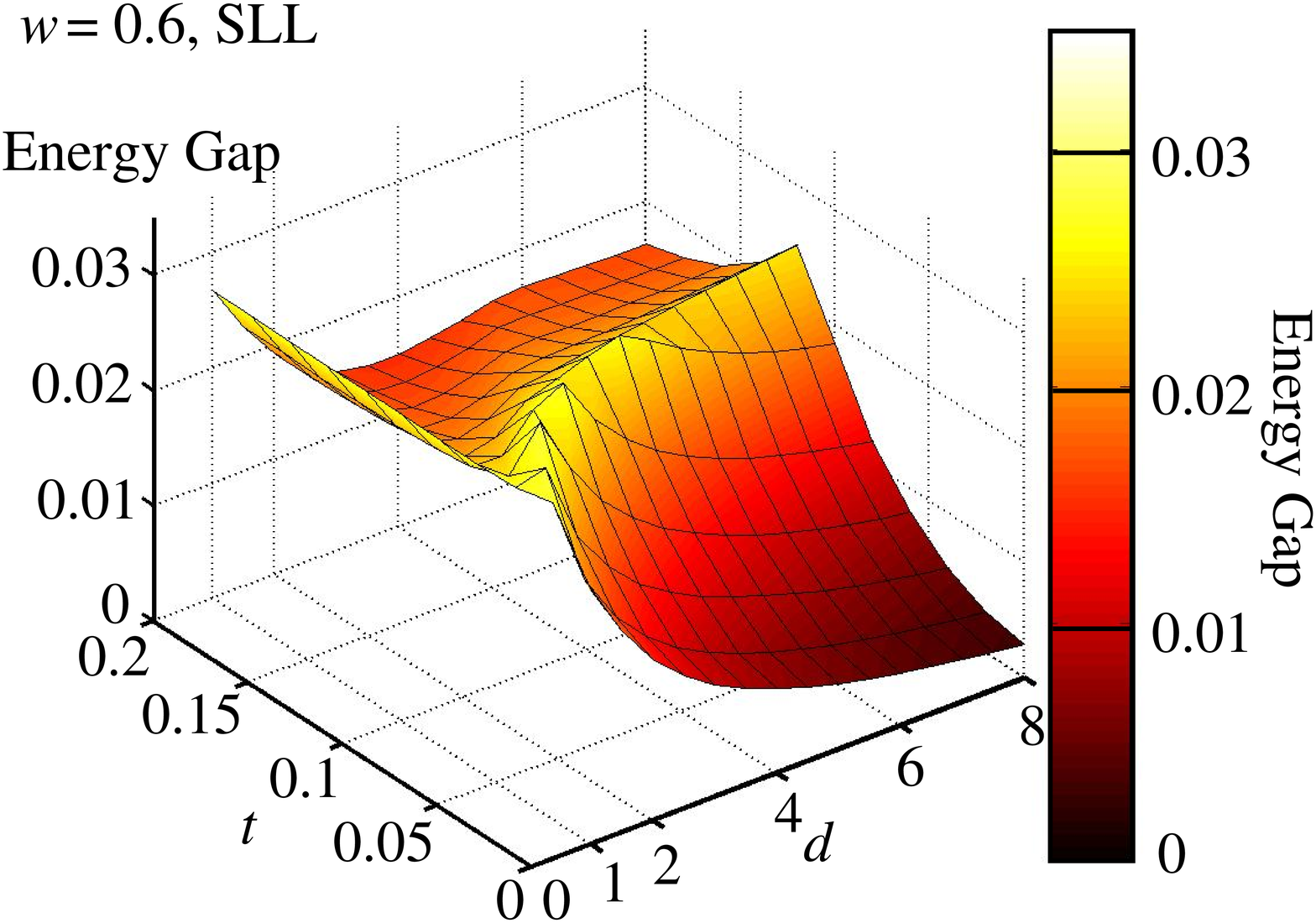}}\\
\mbox{(c)}
\mbox{\includegraphics[width=7.cm,angle=0]{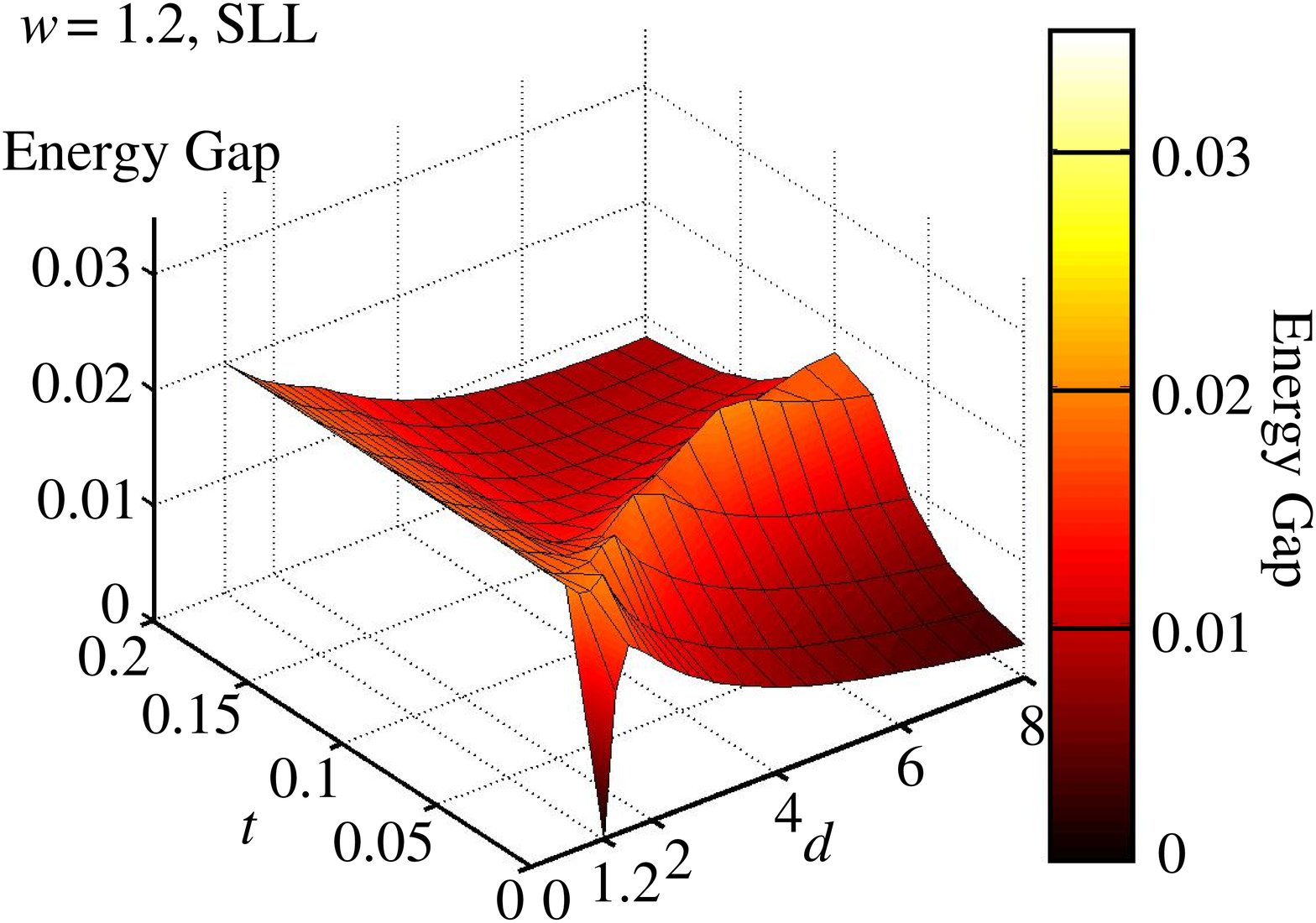}}
\mbox{(d)}
\mbox{\includegraphics[width=7.cm,angle=0]{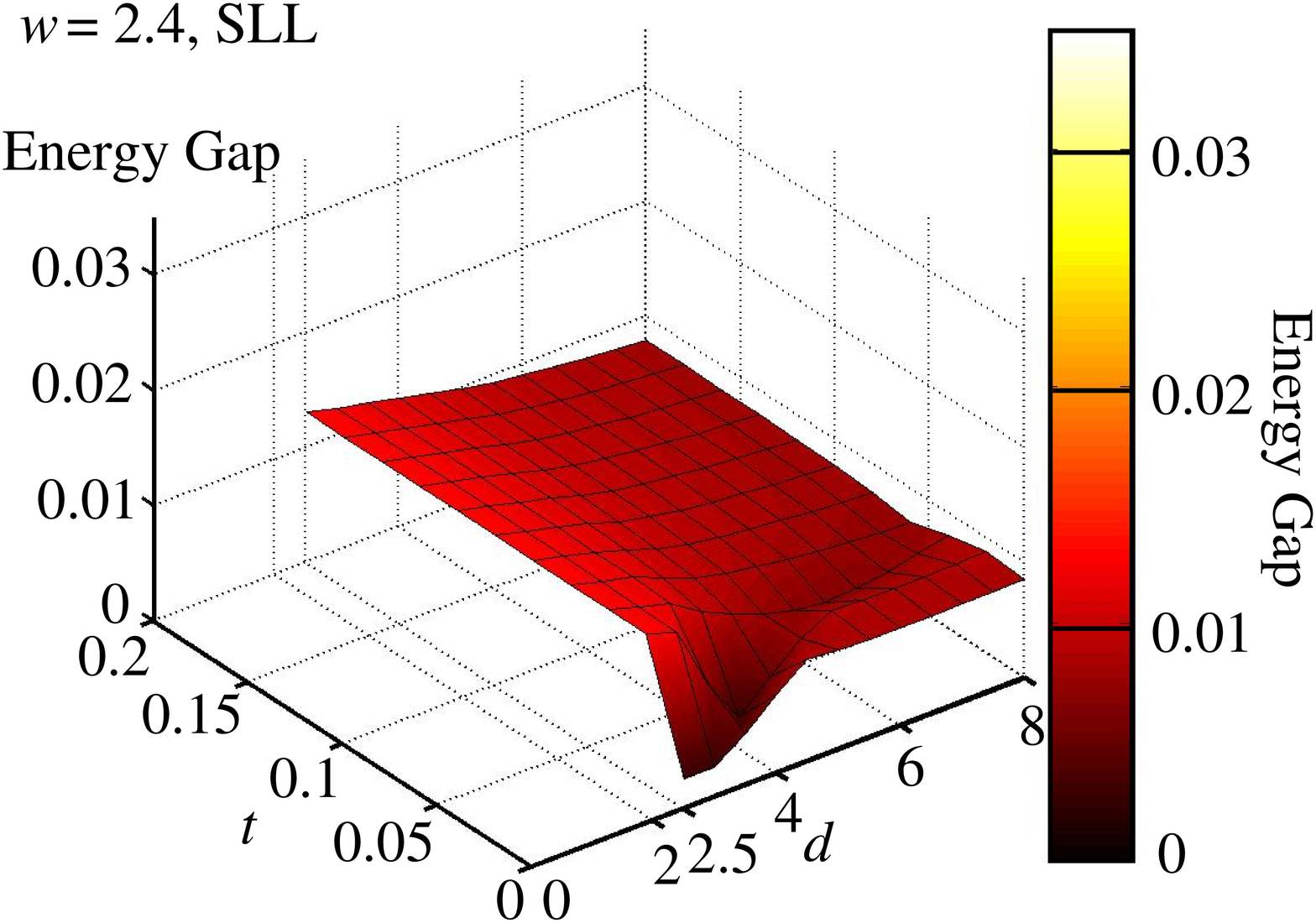}}
\end{center}
\caption{(Color online) FQHE energy gap as a function of layer separation $d$ 
and tunneling amplitude $t$ for the second LL system with $N=8$ electrons.  We consider 
single quantum well widths of (a) $w=0$, (b) $w=0.6$, (c) $w=1.2$, and (d) $w=2.4$.  The 
energy gap is also color-coded such that black is zero gap and white is maximum on a scale 
from zero to 0.035.}
\label{fig-gap3D-SLL}
\end{figure*}

By examining Fig.~\ref{fig-o-SLL}(b)-(d) we see results qualitatively similar to the LLL results 
shown in Fig.~\ref{fig-o-LLL}.  The real difference here is that the overlap between 
the exact state and the Pf state, in the region of phase space where it is 
better than the 331 state, is higher in the SLL than it is in the LLL.  This is 
expected behavior when one considers the single-layer 
results~\cite{mrp-tj-sds-prl,mrp-tj-sds-prb} where the Pf is 
known to be an excellent candidate 
for the $\nu=5/2$ single-layer FQHE.  The other difference complimentary to the Pfaffian 
behavior is that the overlap between the exact state and the 331 state is lower 
in the SLL than it is in the LLL.  In fact, for large layer width $w=2.4$ (Fig.~\ref{fig-o-SLL}(d)) 
the overlap with the 
331 is quite low and it would be unreasonable to assume that the exact state is 
adequately described by the Halperin 331 state.  The Pfaffian overlap is also 
lower for $w=2.4$ in the SLL than it is in the LLL (somewhat surprisingly).  In fact, after 
investigating the FQHE energy gap for $w=2.4$ (below in Sec.~\ref{subsec-gap-SLL}) it 
is clear that the bilayer SLL system most likely would not exhibit any FQHE for such 
large widths.  Again, 
similar to the LLL results (Fig.~\ref{fig-o-LLL}(b)-(d)), we note that for increasing $w$, the 
region in phase space described by the Pf phase increases at the 
expense of the 331 state.  

\subsection{Is the system one- or two-component?--second Landau level}
\label{subsec-Sz-SLL}

Fig.~\ref{fig-Sz-SLL} shows the calculated value of $(N_S-N_A)/2$ as a
function of $d$ and $t$ with (a) $w=0$, (b) $w=0.6$, (c) $w=1.2$, and $w=2.4$ 
and, as before, our overlap based conclusions are consistent 
with the value of $(N_S-N_A)/2$.  It 
is difficult to clearly tell the difference between the results in the SLL compared 
to those in the LLL.  The slight difference between the 
two is that in the large $d$ limit slightly more tunneling is required to drive the 
system to the one-component regime than in the LLL case.  Further, in the strong tunneling 
limit, the SLL system's one-component character is more robust to increasing 
layer separation $d$--but only just.  Again, similar to the LLL, we find that 
when $(N_S-N_A)/2\sim2.5$ the overlap with the exact ground state switches 
from being either higher with the Pf state or the 331 state. 

\subsection{Will the system display the FQHE?--second Landau level}
\label{subsec-gap-SLL}

\begin{figure*}[]
\begin{center}
\mbox{(a)}
\mbox{\includegraphics[width=7.cm,angle=0]{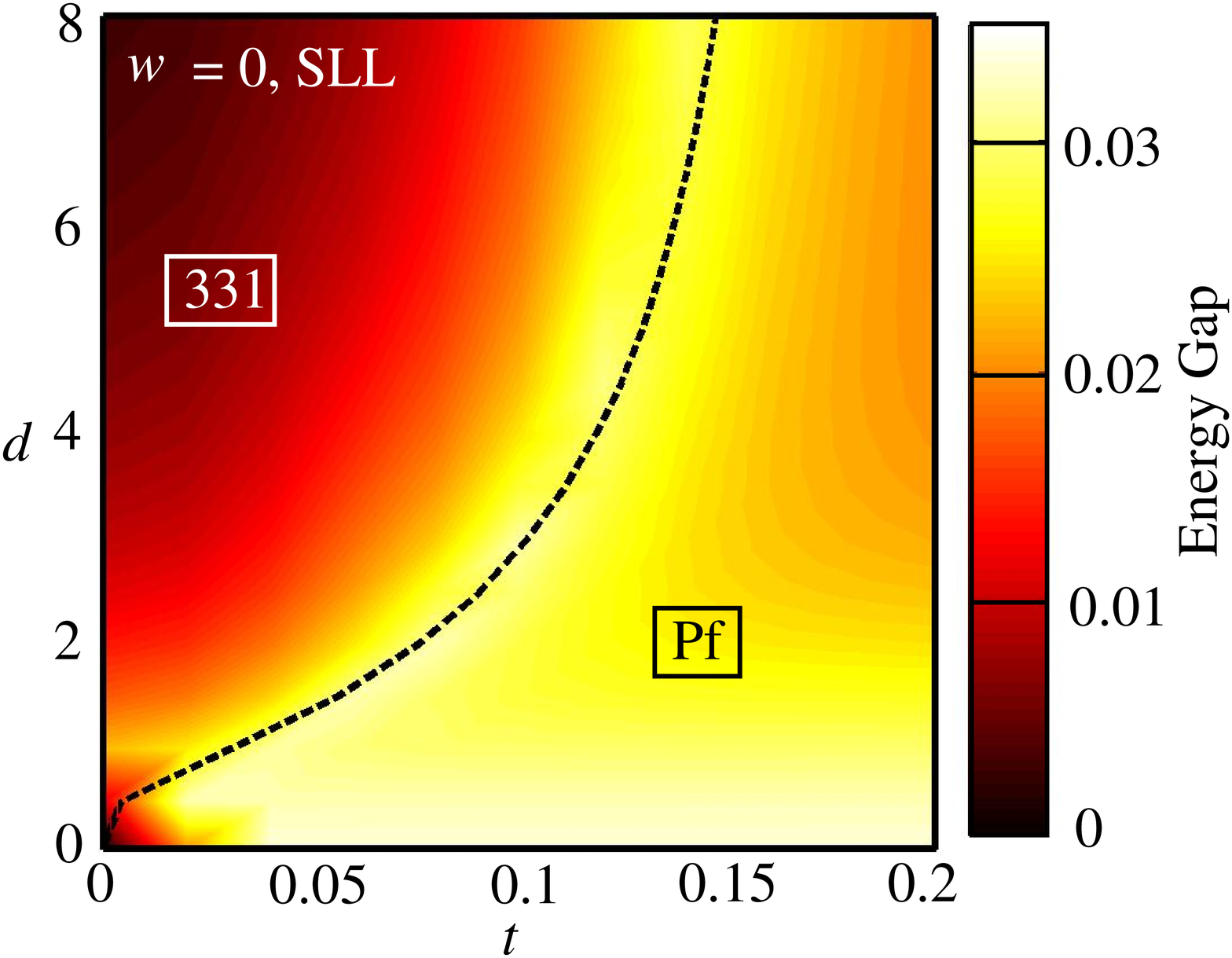}}
\mbox{(b)}
\mbox{\includegraphics[width=7.cm,angle=0]{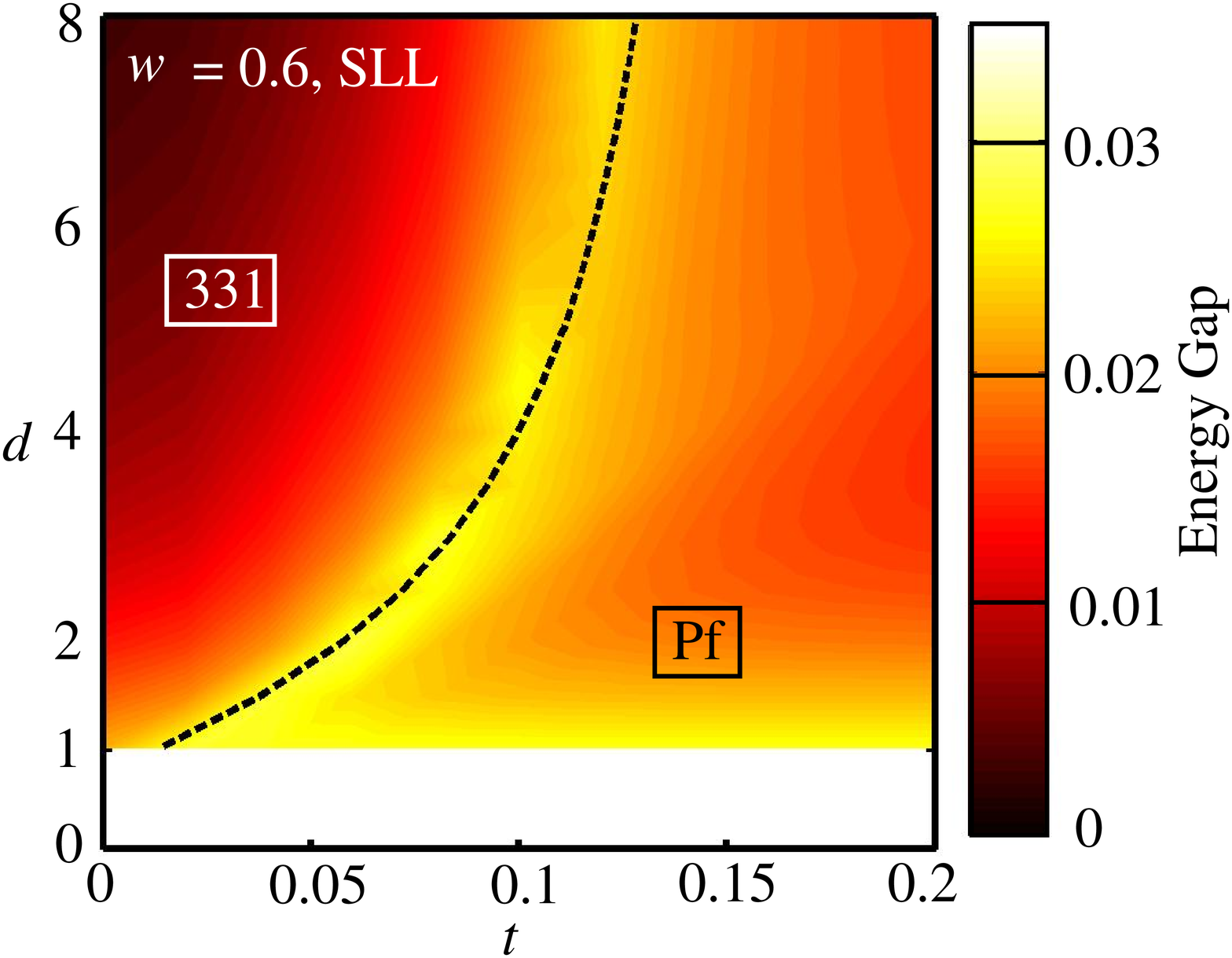}}\\
\mbox{(c)}
\mbox{\includegraphics[width=7.cm,angle=0]{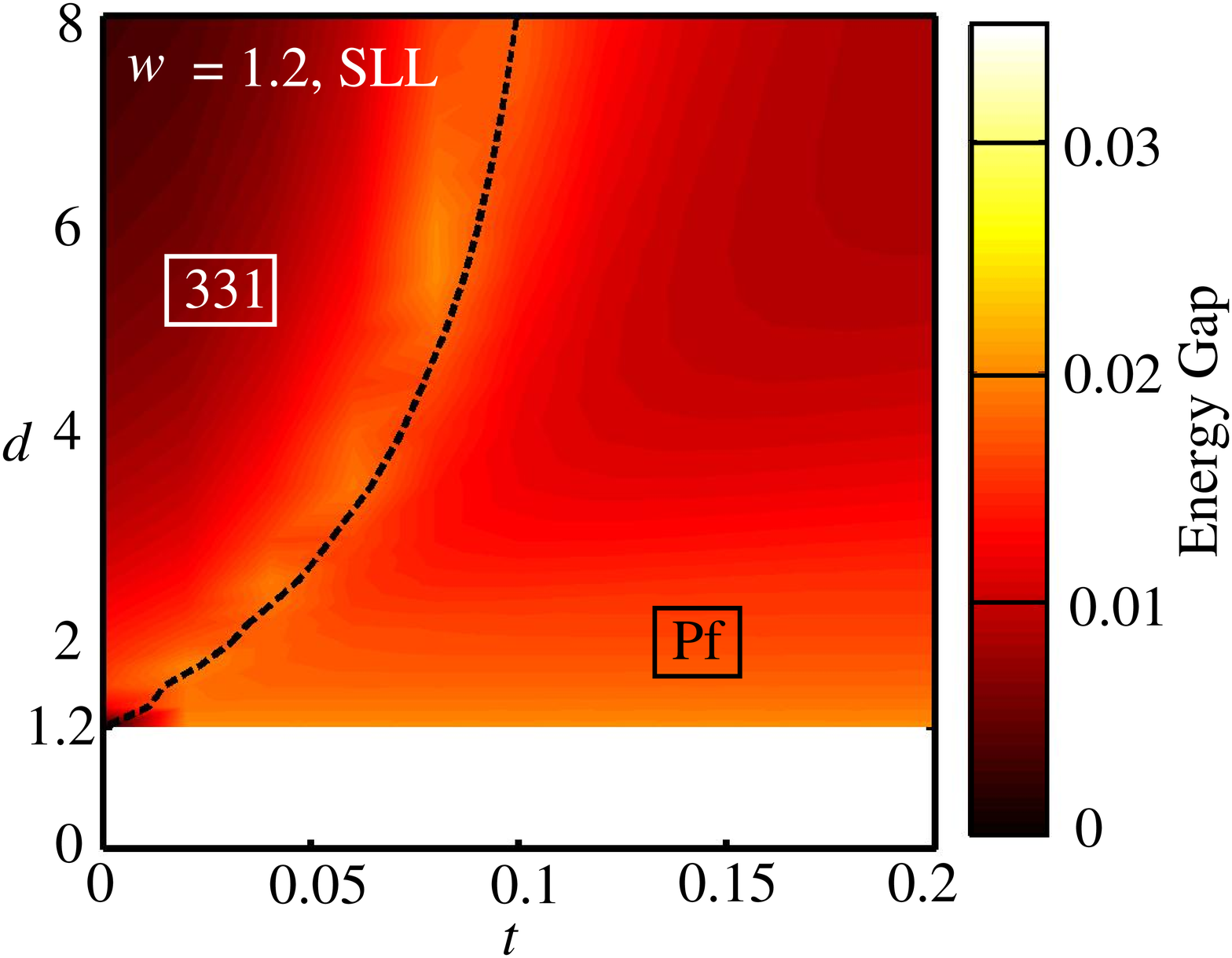}}
\mbox{(d)}
\mbox{\includegraphics[width=7.cm,angle=0]{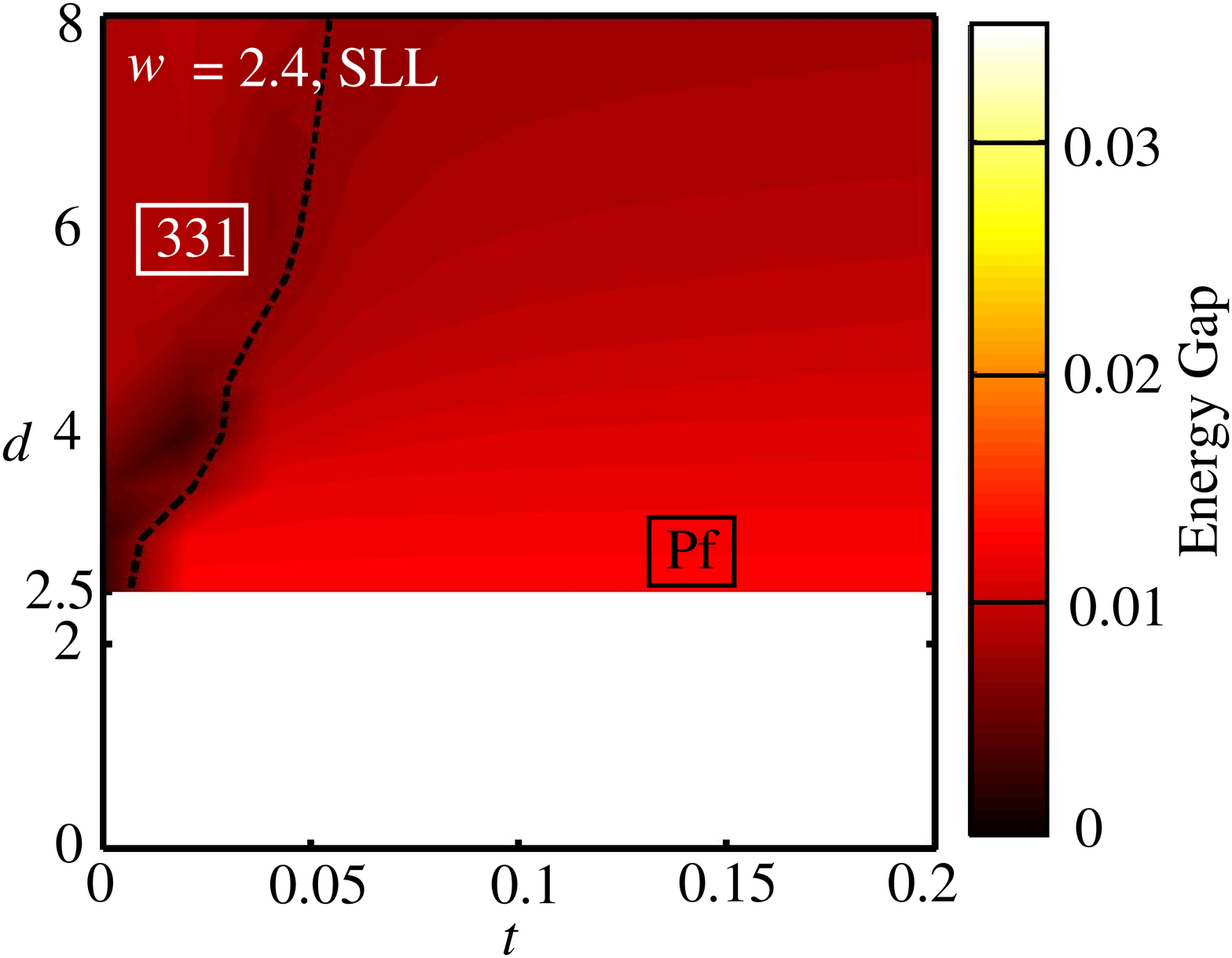}}
\end{center}
\caption{(color online) Quantum phase diagram (QPD) and FQHE gap
(color coded) for the second LL for widths (a) $w=0$, (b) $w=0.6$, (c) $w=1.2$, and (d) $w=2.4$.  
For the QPD, the 331 and Pf phases are labeled appropriately and the gap is given as a
contour plot with color coding given by the color-bar from dark to
light, i.e., white being a largest value of 0.035 and black being value
of 0.}
\label{fig-gap-SLL}
\end{figure*}

In Fig.~\ref{fig-gap3D-SLL}(a)-(d) we show the FQHE energy
gap as a function of $d$ and $t$
for the SLL system for (a) $w=0$, (b) $w=0.6$, (c) $w=1.2$, and (d) $w=2.4$.  
Similar to the LLL, it clear that for finite $d$ and $t$, a 5/2 FQHE with a finite gap (being either 331 
or Pf) exists in a realistic parameter regime.   There is a clear qualitative difference between 
the lowest and second LLs, however.  In the SLL, the largest FQHE energy gap is 
obtained in the zero $d$ limit for finite tunneling.  That is, in the region where the  Moore-Read 
Pfaffian ansatz is the better description of the exact ground state, the energy gap is largest.  With 
an energy gap nearly as large is the ``ridge" region identified in our LLL results--again the 
ridge lies in the region of $d$-$t$ phase space corresponding to the ``quantum phase 
transition" between the Pf phase and 331 phase.  
Finite single-layer width $w$ (Fig.~\ref{fig-gap3D-SLL}(b)-(d)) 
decreases the overall energy gap and moves the ridge area to weaker tunneling.  Note 
that for the largest width considered ($w=2.4$) the energy gap is very small and the 
ridge has essentially become a ``valley" and the overall energy gap is very small.

\begin{figure*}[]
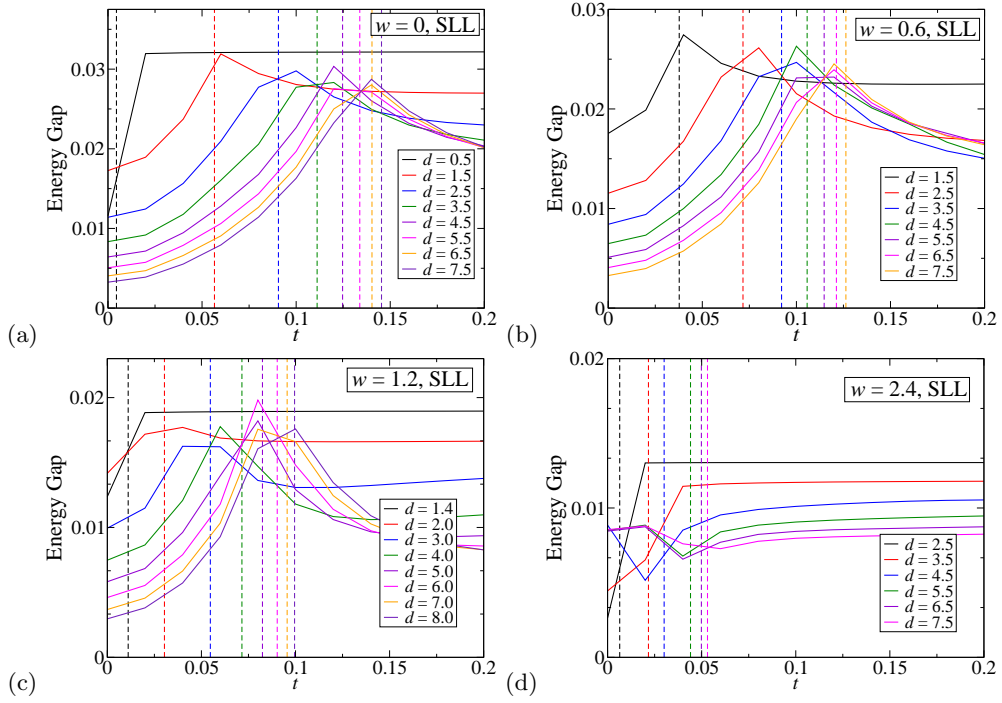

\begin{center}
\mbox{(a)}
\mbox{\includegraphics[width=6.cm,angle=0]{fig13a.eps}}
\mbox{(b)}
\mbox{\includegraphics[width=6.cm,angle=0]{fig13b.eps}}\\
\mbox{(c)}
\mbox{\includegraphics[width=6.cm,angle=0]{fig13c.eps}}
\mbox{(d)}
\mbox{\includegraphics[width=6.cm,angle=0]{fig13d.eps}}
\end{center}
\caption{(color online) FQHE energy gap versus tunneling strength $t$
for a few values of layer separation $d$ for (a) $w=0$, (b) $w=0.6$, (c) $w=1.2$, 
and (d) $w=2.4$.  A dashed
vertical line of the same color corresponds to the boundary
between the Pfaffian phase (right of the line)
and the 331 phase (left of the line).}
\label{fig-gap-vs-D-SLL}
\end{figure*}

\begin{figure*}[]
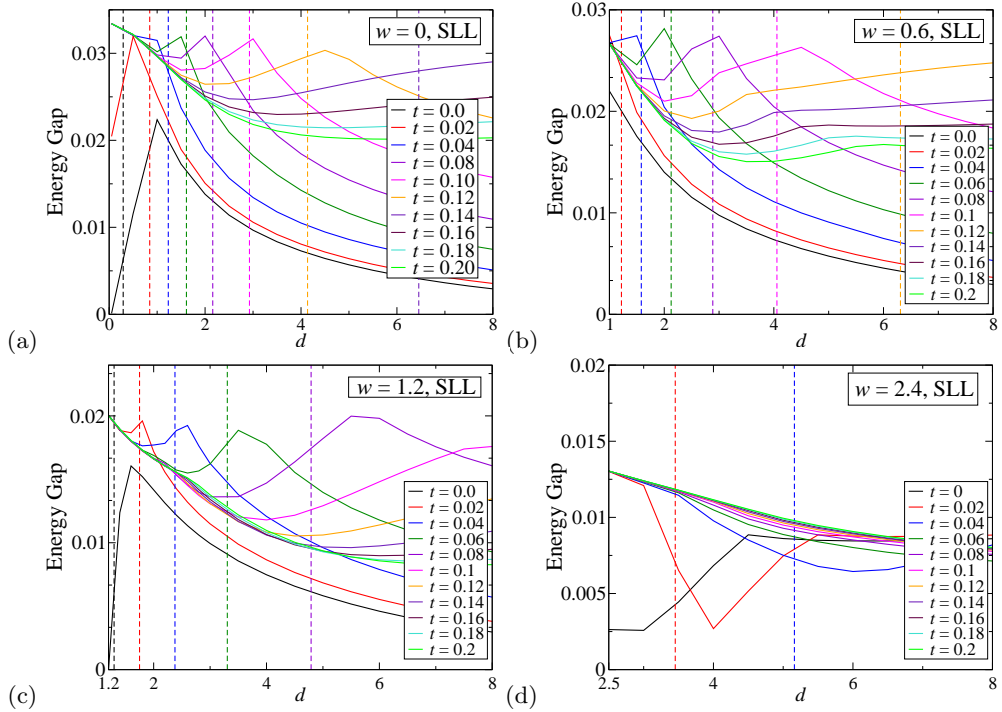

\begin{center}
\mbox{(a)}
\mbox{\includegraphics[width=6.cm,angle=0]{fig14a.eps}}
\mbox{(b)}
\mbox{\includegraphics[width=6.cm,angle=0]{fig14b.eps}}\\
\mbox{(c)}
\mbox{\includegraphics[width=6.cm,angle=0]{fig14c.eps}}
\mbox{(d)}
\mbox{\includegraphics[width=6.cm,angle=0]{fig14d.eps}}
\end{center}
\caption{(color online) FQHE energy gap versus layer separation $d$
for a few values of   tunneling strength $t$ for (a) $w=0$, (b) $w=0.6$, (c) $w=1.2$, 
and (d) $w=2.4$.  A dashed
vertical line of the same color corresponds to the boundary
between the Pfaffian phase (left of the line)
and the 331 phase (right of the line).}
\label{fig-gap-vs-d-SLL}
\end{figure*}

We now discuss the approximate quantum phase diagram for the bilayer system 
for the SLL as shown in Fig.~\ref{fig-gap-SLL} for single-layer 
widths (a) $w=0$, (b) $w=0.6$, (c) $w=1.2$, and (d) $w=2.4$.  This QPD is 
calculated the 
same way as it is for the LLL (cf. Fig.~\ref{fig-gap-LLL})--
identifying the 331 and Pf phases as the regions in the parameter
space where the overlap with $\Psi_0$ is larger (Fig.~\ref{fig-o-SLL}).  In
general, the two-component 331 phase (the upper left region) has a
weaker SLL FQHE than that of the one-component Pf (the
lower right region).  Figure~\ref{fig-gap-SLL}, which is an important
prediction of our work, shows that in the SLL bilayer system, both the
331 and Pf states should be visible, and in a realistic finite
thickness system (Fig.~\ref{fig-gap-SLL}(b) and (c)), the FQHE gap would become very
small, perhaps even zero, in between the two phases.

Again, we ask the 
natural question whether our numerically obtained transition from the large-$d$
(small-$t$) region to the small-$d$ (large-$t$) region is a
quantum phase transition or a crossover.  Of course a finite-size
numerical study cannot definitively answer this question.  Our
results, however, are consistent with the findings~\cite{RG2000} of
Read and Green, and we believe that there would be a QPT 
between 331 and Pf phases in the second LL since a crossover
between topologically trivial (331) and non-trivial (Pf) phases is
difficult to contemplate.    We also note that for the $w=0$ case, the quantum phase 
transition line seems to terminate at $d=0=t$ for the SLL and for $d=0$ and 
\emph{finite} tunneling $t$ in the LLL, i.e., some finite $t$ is 
required to push the system into the Pf phase.  Thus, our SLL results are 
more similar to the QPD of Read and Green~\cite{RG2000}  where a 
multi-critical point exists separating the 331 Abelian phase from the 
non-Abelian Pf phase when $t=0$.  It is perhaps understandable that our SLL results 
would more closely resemble that of Read and Green since they 
were modeling a 1/2-filled bilayer FQHE state as $p_x+ip_y$ superconductor of 
composite fermions which is thought to be the appropriate description for the 
1/2-filled FQHE in the SLL, not necessarily in the  LLL.  
Further, in our SLL QPD, the Pf phase is quite 
strong--as indicated by the FQHE energy gap--along the $d=0$ and 
finite $t$ line, whereas in the LLL, the gap is quite small along that line.
Of course, the parameters in Read and Green's 
analysis are not directly related to our parameters--they have a 
chemical potential $\mu$ in an effective pairing Hamiltonian while 
we have layer separation $d$.  Furthermore, our model is SU(2) symmetric 
at the $d=0=t$ point while their model has explicitly broken SU(2).  More work will be 
needed to understand completely the similarities of our approach and 
that of Read and Green definitively, but the fact that the topology of our numerically
calculated phase diagram of Fig.~\ref{fig-gap-SLL} strongly resembles that of
the phase diagram discussed by Read and Green is
highly suggestive.

In Figs.~\ref{fig-gap-vs-D-SLL}(a)-(d) and~\ref{fig-gap-vs-d-SLL}(a)-(d) we provide a 
detailed view of the SLL FQHE bilayer system with the 
FQHE energy gap given as a function of tunneling energy $t$ for several values of 
the layer separation $d$ and the energy gap as a function of  
layer separation for several values of tunneling energy, respectively.  The FQHE energy gap 
as a function of tunneling is  similar qualitatively to the results in the LLL.  Namely, 
the gap rises to a peak value and then falls off.  Of 
course, for $w=2.4$ (Fig.~\ref{fig-gap-vs-D-SLL}(d)) the 
FQHE gap structure is different than all other cases, lowest or second LL, in that the peak 
has turned into a valley.

In Fig.~\ref{fig-gap-vs-d-SLL} we find that the behavior of the FQHE energy gap as a function 
of layer separation $d$ for constant tunneling energy $t$ is qualitatively different from the 
behavior in the LLL.  The difference is that in the SLL the largest value of the 
energy gap is for $d=0$ and for increasing $d$ the energy gap decreases slightly to 
a minimum (for $t>0.02$) and then rises again to a peak.  For increasing tunneling energy, 
the peak becomes more rounded.  This behavior is qualitatively different from the 
lowest LL in that there are \emph{two} peaks in the energy gap versus $d$ at 
constant $t$ instead of one--note that for $w=0$ and $t\leq 0.02$ there is only 
a single peak not unlike the lowest LL results.   For zero layer width ($w=0$), the phase 
boundary between the one-component Pfaffian phase (to the left of the boundary line) and two-
component  331 phase (to the right of the boundary) is to the left of the \emph{single} peak 
in the energy gap for $t\leq 0.02$, moves to slightly right of the \emph{second} energy gap 
peak for $0.02<t<0.1$, and then finally moves back to being left of the \emph{second} energy 
gap peak.  For $w=0.6$ in Fig.~\ref{fig-gap-vs-d-SLL}(b)  there are 
two energy gap peaks and the phase boundary is slightly to the right of the peak for $t<0.1$ for 
$w=0.6$.  For $t>0.1$ the phase boundary moves to the left of the second peak.  
For $w=1.2$ in Fig.~\ref{fig-gap-vs-d-SLL}(c) (except for $t=0$) 
the phase boundary is always to the left of the second energy gap 
peak.  Lastly, we show results for $w=2.4$ in Fig.~\ref{fig-gap-vs-d-SLL}(d) for sake of 
completeness but it is clear that this case is most likely not a FQHE for any parameter 
values other than for very small $d$ and large $t$.

\subsection{Bilayer FQHE in higher Landau levels}
\label{subsec-bilayer-SLL} 

\begin{figure}[b]
\begin{center}
\mbox{(a)}
\mbox{\includegraphics[width=6.cm,angle=0]{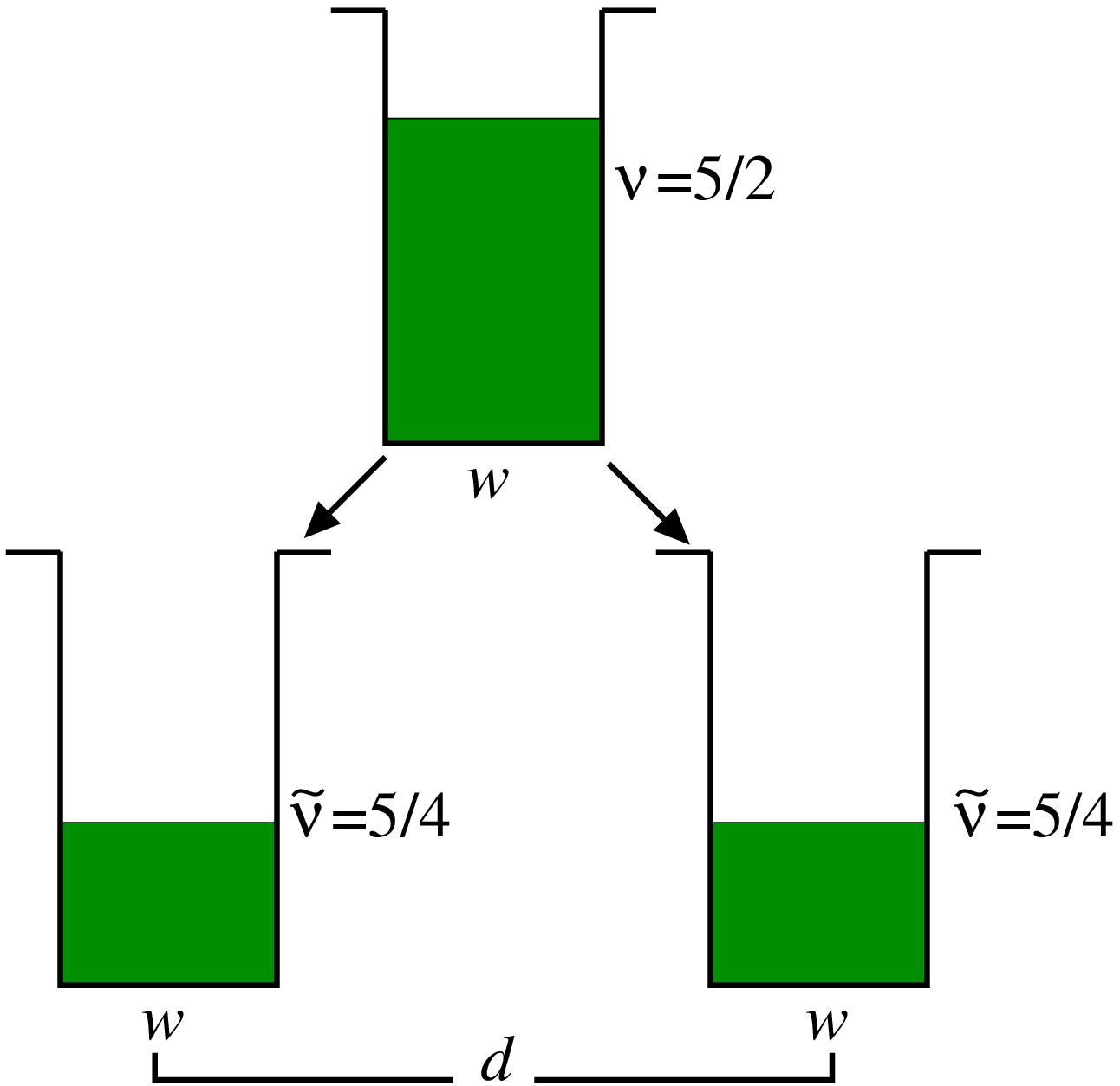}}\\
\mbox{(b)}
\mbox{\includegraphics[width=6.cm,angle=0]{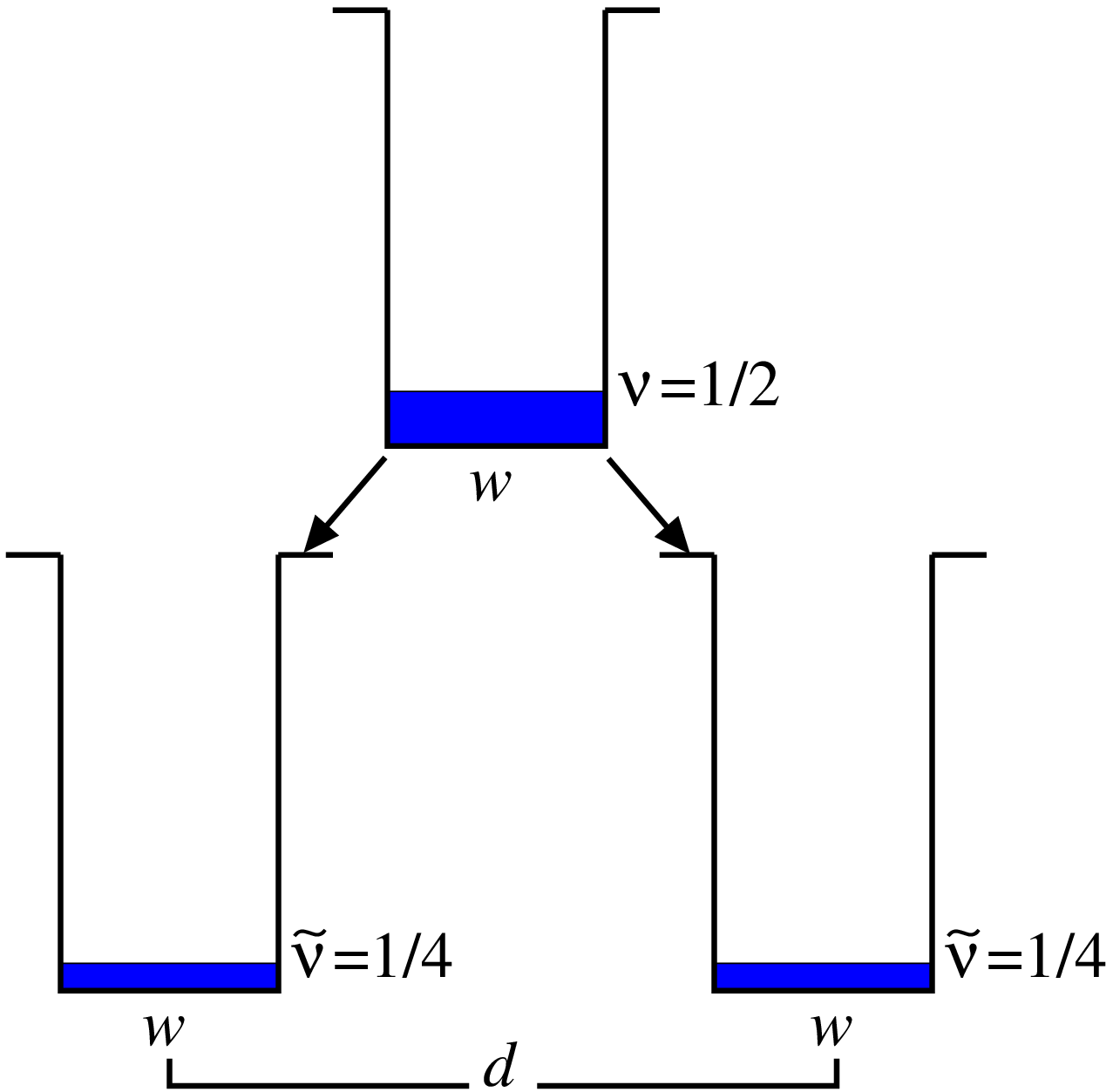}}
\end{center}
\caption{Schematic diagram depicting the FQHE system at 
total filling factor (a) $\nu=5/2$ and (b) $\nu=1/2$ being broken up into 
a bilayer system with each separate layer being at (a) $\tilde\nu=5/4$ and 
(b) $\tilde\nu=1/4$.  The colored shaded section represents a density of electrons 
filling up a quantum well.}
\label{fig-bilayer}
\end{figure}

There is a conceptual difficulty that we postponed when discussing 
bilayer FQHE in the SLL which exists when discussing \emph{any} bilayer 
FQHE in higher LLs.  In fact, we carefully set up the initial SLL bilayer FQHE 
problem as starting with a single-layer, or one-component, 
system in the SLL (presumably the 1/2 filled SLL at $\nu=5/2$) and then systematically driving the 
system into a bilayer by the tuning of model parameters,.   However, 
in our calculation, the electrons fractionally filling the SLL remain in the 
SLL throughout the procedure of tuning the system from one- to two-component.  
A bilayer FQHE at total filling factor $\nu=5/2$ 
would consist of two layers at single layer filling factor $\tilde\nu=(1/2)(5/2)=5/4$, 
see Fig.~\ref{fig-bilayer}.  This is compared to the LLL where the half-filled 
bilayer FQHE consists of $\nu=1/4 + 1/4 = 1/2$, i.e., two layers  at 
$\tilde{\nu}=1/4$, which when 
combined yield total $\nu=1/2=2\tilde\nu$ (Fig.~\ref{fig-bilayer}(b)). 

\begin{figure}[]
\begin{center}
\mbox{\includegraphics[width=6.cm,angle=0]{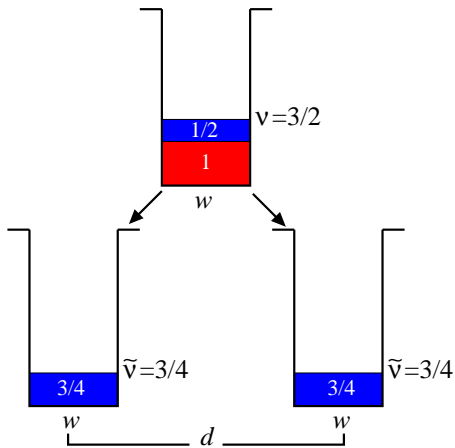}}\\
\end{center}
\caption{Schematic diagram depicting a spin-less FQHE system at 
total filling factor $\nu=3/2$ becoming a bilayer spin-less system with 
each layer at $\tilde\nu=3/4$.  The fraction of the density fractionally filling a 
LL is colored blue (dark grey) while an inert LL (in this case the lowest) is colored 
red (light grey).}
\label{fig-bilayer-32}
\end{figure}

The case of spin-less electrons (we consider the spin-full 
situation below) is depicted in Fig.~\ref{fig-bilayer-32}.  In the one-component 
limit, electrons in the half-filled 
SLL have total filling factor $\nu=1+1/2=3/2$--the LLL is completely 
filled and considered inert.  In the two-component limit, when each of the  
two quantum well layers have been taken far apart from one another, 
the filling factor in each layer (assuming the 
electrons to have equal densities in each layer) will be $\tilde\nu=3/4$.  
Thus, the electrons in individual layers are in the \emph{lowest} LL, not the SLL.  
The key conceptual difficulty is that we started with 
a one-component system at total $\nu=1/2$ in the \emph{second} LL and ended up splitting it 
into a bilayer system of each layer at $\tilde\nu=3/4$ in the \emph{lowest} LL, i.e., the 
electron LL index changed!
 
If one were to study bilayer FQHE in the SLL completely rigorously (necessarily 
numerically) one would have to 
consider--even for spin-less electrons--a system consisting of electrons with a 
layer degree of freedom (pseudo-spin) and at least two LL degrees 
of freedom.  Furthermore, one would 
not necessarily be able to treat the electrons in the LLL (the 1 in the $1+1/2=3/2$) as inert.  For the 
system we studied with $N=8$ electrons half-filling the second LL, 
one would need to fully consider $N=20$ 
interacting electrons with a layer degree of freedom and consider at least 
two LLs.  Hence, the Hilbert space dimension would be extremely large--on the order 
of $10^{11}$ states.  A Hilbert space 
dimension of $10^{11}$ is out of reach for any numerical procedure 
for any computer.

\begin{figure}[t]
\begin{center}
\mbox{(a)}
\mbox{\includegraphics[width=6.cm,angle=0]{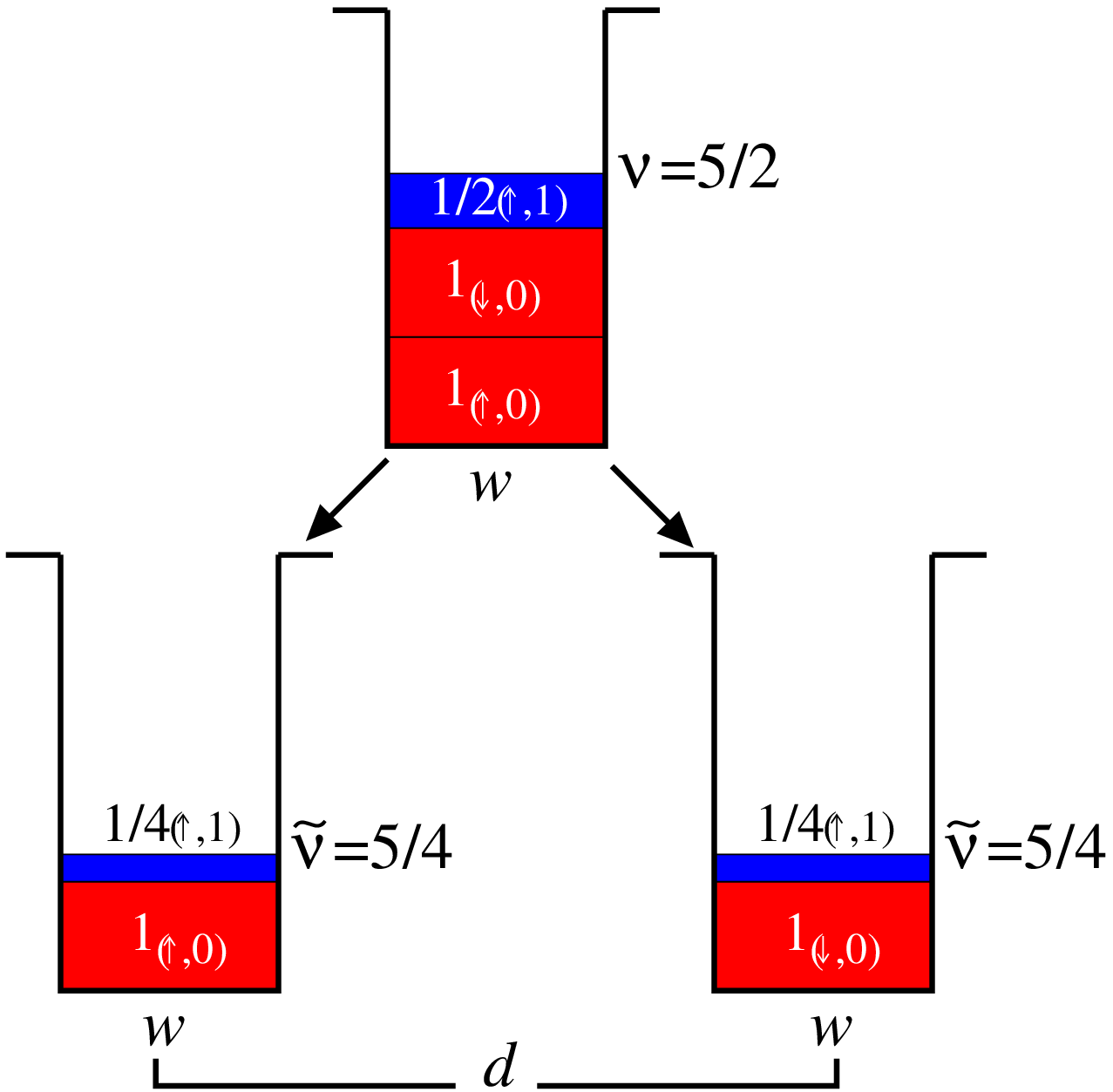}}\\
\mbox{(b)}
\mbox{\includegraphics[width=6.cm,angle=0]{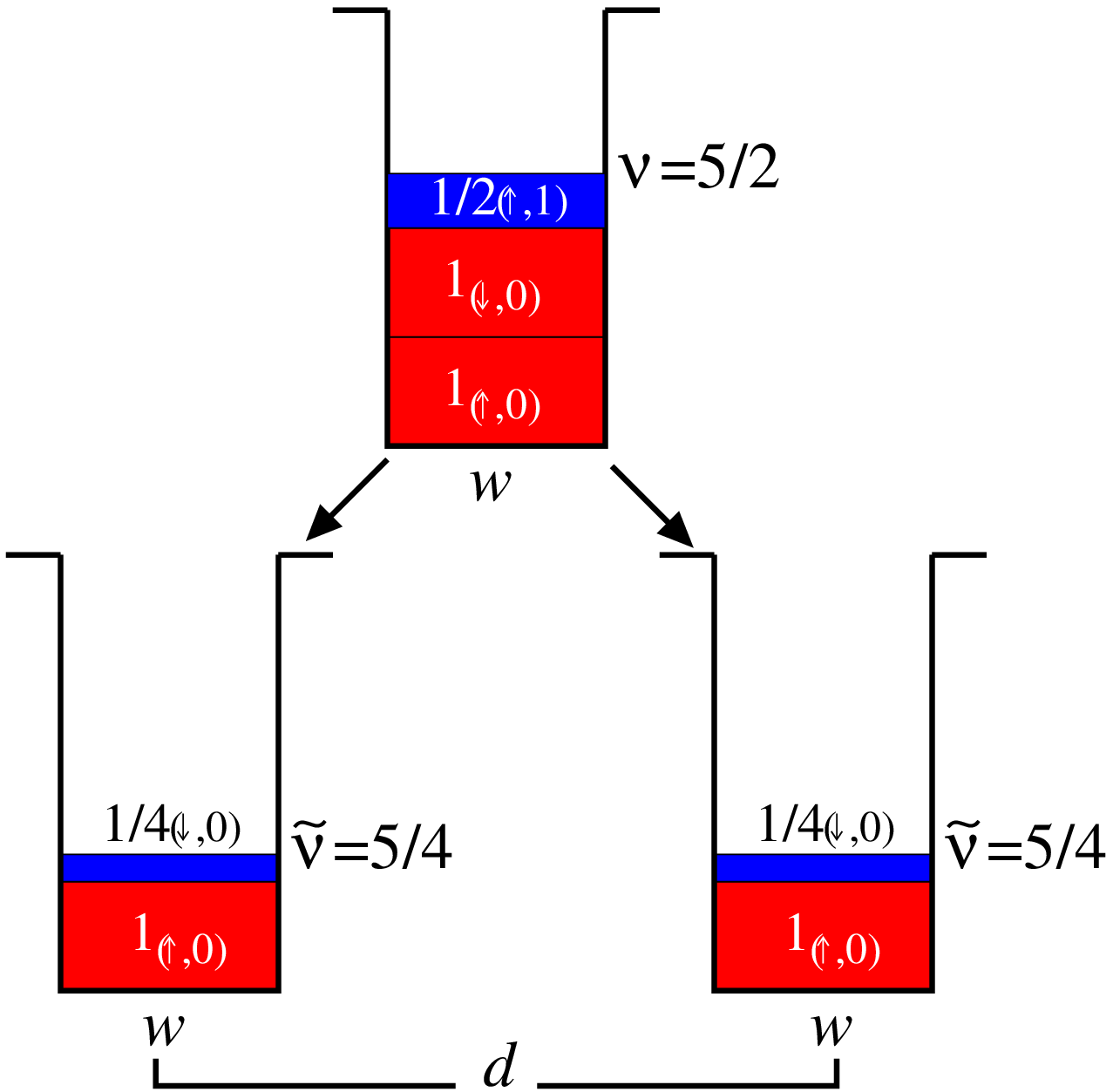}}
\end{center}
\caption{Schematic diagram depicting two possible scenarios for the 
spin-full FQHE system at 
total filling factor $\nu=5/2$ being broken up into 
a bilayer system with each separate layer being at $\tilde\nu=5/4$.  In (a) we 
consider the  small $d$ and strong $t$ limit where the lowest spin-up 
and spin-down Landau levels are completely 
occupied leaving a 1/4 filled spin-up Landau level in each layer.  In (b) we depict the 
large $d$ and weak tunneling $t$ limit with the 
lowest spin-up LLL being filled in each layer leaving a 1/4-filled 
lowest Landau level of spin-down electrons in each layer.  The notation 
is such that $\nu_{(\sigma,n)}$ corresponds to the filling factor $\nu$ for electrons 
with spin-$\sigma$ electrons with orbital Landau level index $n$, 
i.e., $1_{(\uparrow,0)}$ corresponds to filling factor 1 of spin-$\uparrow$ electrons 
with LLL index $0$.}
\label{fig-bilayer-52}
\end{figure}

The experimental reality makes the situation more complicated due to the 
inclusion of spin.  Half-filled 
FQHE in single-layer, presumably, one-component systems has only been observed 
in the SLL, i.e., $\nu=5/2$, which is modeled successfully by treating the completely 
filled spin-up and spin-down LLLs as inert.  Once the 
system is made into a bilayer we are left with, in the 
extreme layer separation limit, two systems at $5/4=1+1/4$ filling.  When the layers are very 
far apart (large $d$ and necessarily weak $t$) it would be expected that the structure of the 5/4 
filled systems would be identical (since they would not interact) 
and the spin-up LLL would be filled and inert and the remaining 
electrons would fill the spin-down LLL up to 1/4 filling.   Such a situation 
is depicted in Fig.~\ref{fig-bilayer-52}(b).  If, on the other hand, the system is barely becoming 
two-component, perhaps the simplest 
possibility is that the two $\tilde\nu=5/4$ layers will consist of a completely filled 
spin-up LLL in one layer and spin-down LL in the other layer (see Fig.~\ref{fig-bilayer-52}(a)).
However, when the two layers are 
close enough that they begin to interact with one another--even 
without any tunneling--the 
situation quickly becomes complicated (as discussed below).  
It is not clear if  \emph{any} of the electrons 
in a bilayer FQHE at total $\nu=5/2$ can be considered inert and one might need 
to consider an interacting electron system where the electrons carry 
spin, pseudo-spin, and two, or possibly three, LLs indices,
making such a problem intractable.

We note that considering each layer to have $\tilde\nu=2+1/4$ filling so that, 
in analogy with the LLL bilayer situation, we have 1/4 filling in each 
SLL layer, does not resolve the ambiguity since this produces a 
total filling of $\nu=(2+1/4)+(2+1/4)=9/2$, taking us to the half-filled \emph{third} 
orbital LL!  Similarly, 
considering each layer to have $\nu=1+1/4$ filling leads to a total 
bilayer filling of $\nu=5/2$, but takes us from the LLL to the SLL in going 
from a one layer to a bilayer system.  The conundrum here is not 
theoretical, but is about how one connects the theoretical results 
with the experimental bilayer 5/2 system.  We emphasize that no such 
ambiguity arises for the $\nu=1/2=1/4+1/4$ bilayer system where 
all of the physics occurs in the LLL, both for individual layers and for 
the total bilayer.  Also, there is no ambiguity in the strong tunneling regime where the 
bilayer effectively acts as a single layer system with $\nu=5/2$ as in a single 2D 
system.

For our fully spin-polarized model, however, there is an easy conceptual (and 
operational) way out of this ambiguity for the bilayer $\nu=5/2$ system.  Since 
we assume the whole system (in both layers and in both orbital Landau levels) 
to have the same spin, the system is effectively spin-less (which is equivalent 
to assuming the Zeeman energy to be much larger than the cyclotron energy).  In 
this spin-less (or spin-polarized) situation, each orbital LL by definition comes 
with just one spin index.  Therefore, the $\nu=5/2$ balanced bilayer system 
is equivalent to a $\tilde\nu=(1/2)(5/2)=5/4=1+1/4$ filling in each layer, 
where the completely filled inert level in each layer is simply the LLL, 
and the 1/4-filled level in each layer is in the SLL.  Since the LLL is 
inert, the incompressible FQH states form entirely in the SLL, and we can 
construct a Halperin 331 SLL state with $\nu=1/2$ from the 
two 1/4-filled SLL states in each individual layer in a manner similar to 
that for the $\nu=1/2$ LLL bilayer state.  We emphasize that this simplicity is 
lost if we include both spin and layer indices on an equal footing since including  
the two spin indices (up and down) and two layer indices (right and left) 
lead to an immediate problem on how to assign individual $\tilde\nu$ values 
which, when combined into the total $\nu=2\tilde\nu$, lead to a 
bilayer SLL $\nu=5/2$ FQHE with \emph{both} the total bilayer (i.e., $\nu$) and 
the individual layer (i.e., $\tilde\nu=\nu/2$) filling factor being in the SLL.  (As emphasized 
already, the situation with two layers, two spins, and two orbital LLs is 
ambiguous unless the numerical diagonalization takes exactly into account 
the full dynamics of all three two-level quantum indices which is impossible 
to do for any computer.) 

We mention that earlier theoretical work (alluded to above) by Zheng \textit{et al.}~\cite{zheng}, 
Das Sarma \textit{et al.},~\cite{SDS-Sachdev-Zheng-PRL,SDS-Sachdev-Zheng-PRB}, 
Demler \textit{et al.}~\cite{Demler-SDS-PRL1998}, 
and Brey \textit{et al.}~\cite{Brey-Demler-SDS-PRL1999} did take into account the 
dynamical interplay 
between layer and spin indices with the conclusion that there should be a novel quantum 
canted antiferromagnetic phase in bilayers for situations where each individual 
layer state has a gap in the spectrum (e.g., $\nu=1+1$ or $1/3+1/3$ 
or $1+1/3+1+1/3$).  For the case we are considering in the current work, i.e., $\nu=5/2$ and 
$\tilde\nu=\nu/2=5/4$, there is not necessarily a gap in the individual layer spectrum, and therefore 
a canted phase is not expected.

\section{Conclusions}
\label{sec-conc}

Our results concerning the $\nu=1/2$ LLL FQHE bilayer system point to 
strong evidence that the dominant FQH phase is the two-component Abelian 
Halperin 331 phase.  We believe that the only chance of observing the $\nu=1/2$ 
non-Abelian Moore-Read Pfaffian FQHE is to look on the Pf side of the 
phase boundary (Fig.~\ref{fig-gap-LLL}) at fairly large values of $d$ and $t$.  
This contrasts the SLL bilayer situation where there are two sharp ridges 
far away from each other in the $d$-$t$ space corresponding to the Pf 
and 331 phases.  We note that for unrealistically large single layer width $w$ 
the Pf phase dominates the 331 phase but the FQHE gap is extremely small.  
If the Pf phase exists at all, it would manifest most strongly in wide samples and 
close to the phase boundary with the 331 phase.

In addition, we predict the existence of both the two-component 331
Abelian (at intermediate to large $d$ and small $t$) and the
one-component Pfaffian non-Abelian (at small $d$ and intermediate to
large $t$) $\nu=5/2$ SLL FQHE phase in bilayer structures.  The
observation of these two topologically distinct phases, one (Pf)
stabilized by large inter-layer tunneling and the other (331)
stabilized by large inter-layer separation would be a spectacular
verification of the theoretical expectation~\cite{RG2000} that bilayer
structures allow quantum phase transitions between topologically
trivial and non-trivial paired even-denominator incompressible FQHE
states.  The direct experimental observation of our predicted ``two
distinct branches'' of two strong SLL FQHE regimes in bilayer structures, as shown in
Fig.~\ref{fig-gap-SLL}(a)-(c), with the
FQHE gap being largest along the two ridges in the phase diagram as
$d$ and $t$ are varied, will be compelling evidence for the existence of the
$\nu=5/2$ non-Abelian Pf state.

Note that for small $d$ and $t$, Pf and 331 phases are
continuously connected, indicating the possible existence of a quantum
phase transition.  We cannot, of course, dismiss a crossover due to
the limitations of finite size calculations, but it is difficult to
contemplate how an Abelian and non-Abelian phase could continuously go into
each other without a quantum phase transition.

Besides our exact diagonalization results for the bilayer $\nu=1/2$ and 5/2 
FQHE, we have raised an important conceptual issue involving the existence 
of Halperin type bilayer FQHE states in higher (i.e., beyond the lowest) Landau 
levels.  In particular, the fact that two well-separated distinct layers have 
individual filling factors $\nu/2$, which necessarily lie in a lower orbital LL, 
make it tricky to define a composite bilayer Halperin wavefunction for $\nu$ which 
is in a higher LL.  (This problem obviously does not arise in the LLL.)  It is 
conceivable that any theoretical study of bilayer FQHE states in higher LLs 
must necessarily include the full dynamics involving layer, orbital LL, and 
spin degrees of freedom.  This is an almost impossible theoretical challenge, and we hope 
that our work will motivate experimental activity in bilayer structures 
at total filling factor 5/2 in order to explore the possible difference 
between the physics of bilayer $\nu=1/2$ and $\nu=5/2$ systems.

\section{Acknowledgements}
We acknowledge support from Microsoft Q and DARPA QuEST.    We also thank Nick 
Read and Kentaro Nomura for helpful discussions, and Bert Halperin for asking a particularly 
probing question.

\section{Appendix: Identifying the Experimental Points in the $\nu=1/2$ Bilayer FQHE 
Quantum Phase Diagram}
\label{sec-app}

We briefly describe our procedure for identifying the experimental data points in our 
calculated $\nu=1/2$ bilayer FQHE quantum phase diagram (Fig.~\ref{fig-gap-LLL}).  In 
other words, we explain how we extracted the parameter values $t$ (tunneling 
energy), $d$ (layer separation), and $w$ (layer width) for each experimental sample point  
shown in Fig.~\ref{fig-gap-LLL}.

\begin{figure}[t]
\begin{center}
\mbox{(a)}
\mbox{\includegraphics[width=5.5cm,angle=0]{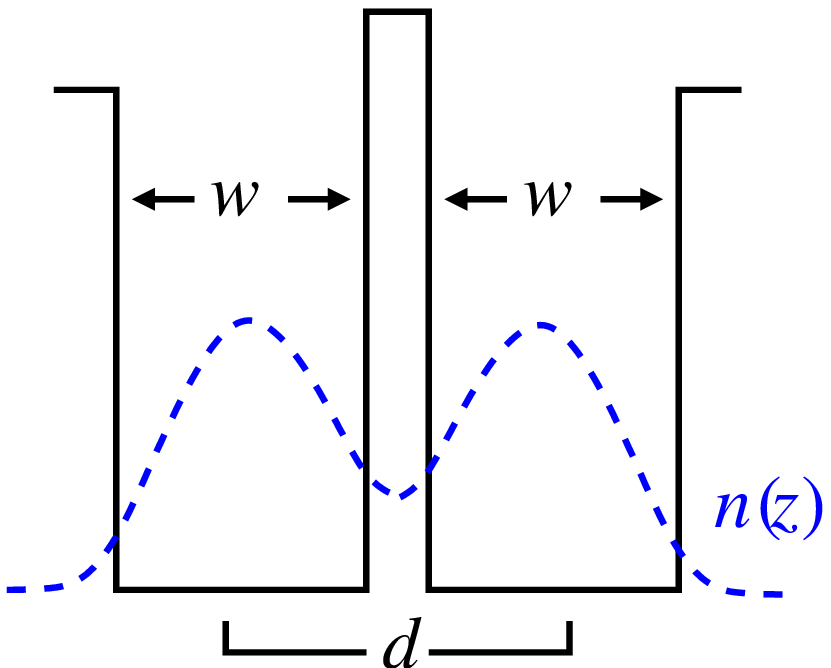}}\\
\mbox{(b)}
\mbox{\includegraphics[width=5.5cm,angle=0]{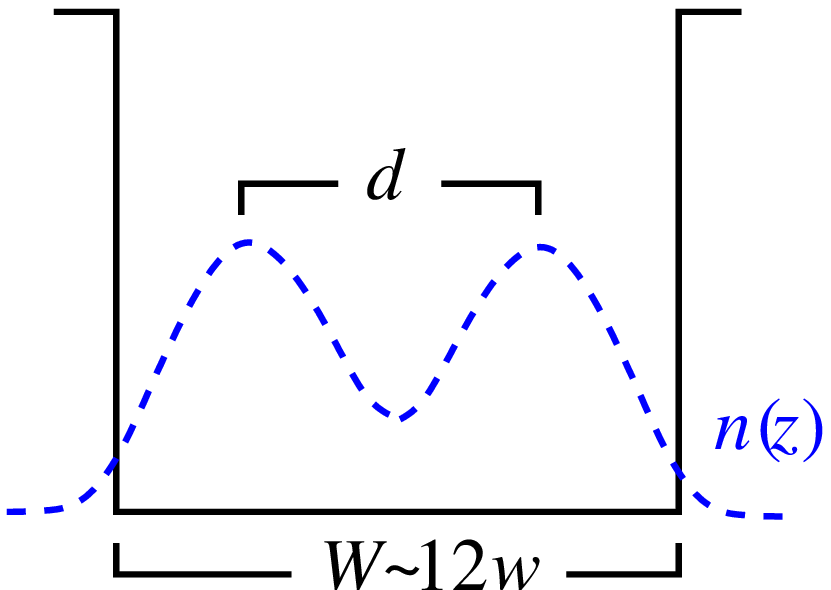}}
\end{center}
\caption{Schematic diagram depicting the two bilayer structures considered 
in this work.  (a) shows a double quantum well structure along with a 
generic density profile in the $z$-direction $n(z)$ (blue dashed line) for 
experiments like Ref.~\onlinecite{eisenstein} while (b) shows a 
single wide-quantum-well structure and generic density profile (blue dashed line) typical of 
experiments like Ref.~\onlinecite{suen}.}
\label{fig-bilayer-dqw-wqw}
\end{figure}

The tunneling energy $t$ is 
connected directly to the symmetric-antisymmetric gap $\Delta_{SAS}=t$ 
in our model and therefore we simply use the quoted value of 
$t$ (or $\Delta_{SAS}$) from the relevant experimental paper.
The layer separation $d$ appears in our interlayer Coulomb interaction as 
$V_\mathrm{inter}=e^2/(\kappa\sqrt{r^2+d^2})$.  For a true bilayer system the 
definition of $d$ is obvious (the well-center to well-center distance between 
the two quantum wells), and we use that value for $d$ from the experimental 
publications (see Fig.~\ref{fig-bilayer-dqw-wqw}(a)).  
For a wide single-well system, where the effective bilayer arises 
as a direct result of the self-consistent density profile in the system, we choose $d$ 
as the distance between the two peaks in the calculated carrier density 
profile for the corresponding wide well sample (see Fig.~\ref{fig-bilayer-dqw-wqw}(b)).  
These self-consistent density 
profiles are given in each experimental paper for each sample 
we show in Fig.~\ref{fig-gap-LLL}.  This procedure is, of course, imperfect--for example, 
the density profiles could be different from the calculated zero-field local density 
approximation (LDA) results since the systems is in the strong field FQHE regime.  We 
do not, however, believe that this is a serious issue since the 2D dynamics of the 
electrons leading to FQHE is separable from the calculated density profile in the 
$z$-direction transverse to the 2D plane.  In any case, the value of $d$, as given by the 
calculated LDA density profile, is the best one can do.

\begin{table}
\begin{tabular}{l c c c} 
\hline\hline
 Experiment &  $d/l$ & $t/(e^2/(\kappa l))$ & $w/l$ \\ \hline
Luhman \textit{et al.}, (squares), Ref.~\onlinecite{luhman-2008} & 5.3 & 0.13 & 0.4 \\
                                                           & 6.9 & 0.07 & 0.3 \\
\hline
Eisenstein \textit{et al.}, (asterisks), Ref.~\onlinecite{eisenstein} & 2.4 & 0.01 & 0.3 \\
                                                                & 2.7 & 0.01 & 0.3 \\
                                                                & 3.6 & 0.01 & 0.3 \\
\hline
Suen \textit{et al.}, (triangles), Ref.~\onlinecite{suen,suen-1} & 5.3 & 0.08 & 0.4 \\
                                                                         & 4.8 & 0.09 & 0.4 \\
                                                                         & 3.9 & 0.14 & 0.4 \\
\hline
Shabani \textit{et al.}, (circles), Ref.~\onlinecite{shayegan-new} & 5.9 & 0.11 & 0.25 \\
                                                               & 6.0 & 0.13 & 0.25 \\
\hline\hline
\end{tabular}
\caption{Values of layer separation $d$, quantum well width $w$, and 
tunneling energy $t$ in units of magnetic length $l$ and Coulomb
energy $e^2/(\kappa l)$, respectively, taken from Refs.~\onlinecite{luhman-2008},
~\onlinecite{eisenstein},~\onlinecite{suen},~\onlinecite{suen-1}, and~\onlinecite{shayegan-new} as described 
in Section~\ref{sec-app}.
\label{tab-1}}
\end{table}

The most problematic parameter for our model is the finite width parameter $w$ which, 
in the Zhang-Das Sarma model (Ref.~\onlinecite{zds}), does not correspond 
\emph{at all} to the physical layer width.  This is why we have shown the experimental 
points on all four ``width" values in Fig.~\ref{fig-gap-LLL}.  It is known from earlier 
work~\cite{he-1990,mrp-tj-sds-prb,morf,dabrumenil,dassarmaprb,park-jain,park-meskini-jain} 
that the Zhang-Das Sarma 
parameter $w$ has the following, very approximate, correspondence with the 
quantum well width parameter $w_\mathrm{QW}$:  $w\simeq \frac{w_\mathrm{QW}}{6}$.
Given that a balanced, symmetric, wide-quantum-well structure with a total width 
of $W$ has $w_\mathrm{QW}=W/2$ for each individual layer, we conclude 
$w_\mathrm{ZDS}\approx w_\mathrm{QW}/12$.  This is, however, a 
very qualitative and crude estimate.  Because of all the approximations involved 
in our depiction of the experimental data, points in Fig.~\ref{fig-gap-LLL} should be taken 
as a qualitative comparison, rather than a quantitative one.

In Table~\ref{tab-1} we provide the values for $t$, $d$, and $w$ we 
get from individual experimental papers whose data points show 
up in Fig.~\ref{fig-gap-LLL}

Lastly, we schematically show in Fig.~\ref{fig-bilayer-dqw-wqw} 
how $d$ and $w$ in our model correspond 
to those in the experimental sample.  The tunneling energy $t$ corresponds directly 
to the symmetric-antisymmetric splitting, and the density profile in the wide well is obtained 
from LDA calculations for the experimental samples.  Fig.~\ref{fig-bilayer-dqw-wqw}(a) 
corresponds to double quantum well structure and typical density profile in 
the experiments by Eisenstein \textit{et al.}~\cite{eisenstein} and 
Fig.~\ref{fig-bilayer-dqw-wqw}(b) matches the 
single wide-quantum-well structure 
and typical density profile in experiments by Suen \textit{et al.}~\cite{suen,suen-1}, Luhman 
\textit{et al.}~\cite{luhman-2008}, and Shabani \textit{et al.}~\cite{shabani}.


\begin{thebibliography}{72}
\bibitem[{foo({\natexlab{b}})}]{foot-note-1}
\bibinfo{note}{The relation between the number of electrons $N$ and the total
  magnetic flux $N_\phi$ for $\Psi_{222}$ is $N_\phi=2N-2$, while the relation
  for $\Psi_{440}$ is $N_\phi=2N-4$. Thus, they cannot be compared directly to
  $\Psi_{331}$ or $\Psi_\mathrm{Pf}$ which have $N_\phi=2N-3$.}
\end{thebibliography}

\end{document}